%% file: ms_x1.tex
\documentclass[twocolumn]{aastex62}

\hypersetup{linkcolor=blue,citecolor=blue,filecolor=magenta,urlcolor=cyan}

\usepackage{natbib}
\def\editR{}

\def\editW{}

\shorttitle{How Galaxies Get Their Gas}
\shortauthors{Ho et al.}

\begin{document}
\input setdef.tex

\title{How Gas Accretion Feeds Galactic Disks}

%% Use \affiliation for affiliation information. The old \affil is now aliased
%% to \affiliation. AASTeX v6.2 will automatically index these in the header.
%% When a duplicate is found its index will be the same as its previous entry.
%%
%% Note that \altaffilmark and \altaffiltext have been removed and thus 
%% can not be used to document secondary affiliations. If they are used latex
%% will issue a specific error message and quit. Please use multiple 
%% \affiliation calls for to document more than one affiliation.
%%
%% The new \altaffiliation can be used to indicate some secondary information
%% such as fellowships. This command produces a non-numeric footnote that is
%% set away from the numeric \affiliation footnotes.  NOTE that if an
%% \altaffiliation command is used it must come BEFORE the \affiliation call,
%% right after the \author command, in order to place the footnotes in
%% the proper location.
%%
%% If done correctly the peer review system will be able to
%% automatically put the author and affiliation information from the manuscript
%% and save the corresponding author the trouble of entering it by hand.

%\correspondingauthor{Stephanie H. Ho, Crystal L. Martin}
\email{shho@physics.ucsb.edu, cmartin@physics.ucsb.edu}

%\author[0000-0002-9607-7365]{Stephanie H. Ho}
\author{Stephanie H. Ho}
\affiliation{Department of Physics, University of California, Santa Barbara, CA 93106, USA}

%\author[0000-0001-9189-7818]{Crystal L. Martin}
\author{Crystal L. Martin}
\affiliation{Department of Physics, University of California, Santa Barbara, CA 93106, USA}

%\author[0000-0001-5389-6312]{Monica L. Turner}
\author{Monica L. Turner}
\affiliation{Las Cumbres Observatory, 6740 Cortona Drive, Goleta, CA 93117, USA}

%%%%%%%%%%%%%%%%%%%%%%%%%%%%%%%%%%%%%%%%%%%%%%%%%%%%%%%%%%%%%%%%%%%%%%%%%%%%%%%%

%%%%%%%%%%%%%%%%%%%%%%%%%%%%%%%%%%%%%%%%
% ABSTRACT
%%%%%%%%%%%%%%%%%%%%%%%%%%%%%%%%%%%%%%%%
\begin{abstract}

Numerous observations indicate that
galaxies need a continuous gas supply to fuel star formation
and explain the star formation history.
However, direct observational evidence of gas accretion remains rare.
Using the \EAGLE\ cosmological hydrodynamic simulation suite, 
we study cold gas accretion onto galaxies 
and the observational signatures of the cold gas kinematics.  
For \EAGLE\ galaxies at $z=0.27$, 
we find that cold gas accretes onto galaxies anisotropically
with typical inflow speeds between 20\kms\ and 60 \kms.
Most of these galaxies have comparable mass inflow rates 
and star formation rates, implying that 
\editR{the cold inflowing gas plausibly accounts for
sustaining the star-forming activities 
of the galaxies.}
As motivation for future work 
to compare the cold gas kinematics 
with measurements from quasar sightline observations, 
we select an \EAGLE\ galaxy with an extended cold gas disk,
and we probe the cold gas using mock quasar sightlines.
We demonstrate that by viewing the disk edge-on, 
sightlines at azimuthal angles below 10\deg\ 
and impact parameters out to 60 pkpc 
can detect cold gas that corotates with the galaxy disk.  
This example suggests 
cold gas disks that extend beyond the optical disks
possibly explain the sightline observations that 
detect corotating cold gas near galaxy major axes.

\end{abstract}


%%%%%%%%%%%%%%%%%%%%%%%%%%%%%%%%%%%%%%%%
% INTRODUCTION
%%%%%%%%%%%%%%%%%%%%%%%%%%%%%%%%%%%%%%%%
\section{Introduction}
\label{sec:intro}

Observations have indicated the need for
gas accretion onto galaxies.  
Infalling gas prolongs the gas consumption times of galaxies, 
or else the star formation of galaxies would exhaust the gas
within a few gigayears (Gyrs) 
\citep{Bigiel2008,Bigiel2011, Leroy2008,Leroy2013,Rahman2012}.  
Gas infall helps regulate galaxy star formation 
and is responsible for   
the color of galaxy disks along the Hubble sequence
\citep{Kennicutt1998}. 
Continuous accretion of metal-poor gas explains
the relative paucity of low metallicity stars in the disk,
known as the G-dwarf problem in the solar neighborhood
\citep{vandenBergh1962,Schmidt1963,SommerLarsen1991}
and is also observed in other galaxies
(e.g., \citealt{Worthey1996}).
These observations strongly suggest the need for
galaxy gas accretion.

However,
direct observation of gas accretion onto galaxies remains sparse
\citep{Putman2012}. 
Although the Milky Way (MW) is accreting gas
(e.g., the Magellanic Stream, \citealt{Fox2014}), 
our location in the MW makes it difficult 
to detect inflowing gas besides high velocity clouds
\citep{Zheng2015}.
Beyond the local universe, 
\editR{the detection rate of net inflow 
identified from galaxy spectra
stays low at roughly 5\% \citep{Martin2012,Rubin2012}.} 
These down-the-barrel galaxy observations have been challenged 
by the fact that
inflowing gas can be identified only if the Doppler shift
can be distinguished from the velocity dispersion 
of the interstellar medium (ISM).  
Otherwise, the inflowing gas will produce absorption 
that overlaps with the dense ISM in wavelength (or velocity) space,
and the true inflow mass flux will be underestimated.

In contrast to down-the-barrel observations,
transverse sightlines through 
the circumgalactic medium (CGM, \citealt{Tumlinson2017}) 
eliminate the problem of overlapping absorption from 
the intervening CGM and the ISM.  
Transverse sightlines probe the CGM against 
bright background sources such as quasars.  
Recent quasar sightline studies 
use a series of metal absorption-line systems
to provide better constraints on 
the significant cool ($\sim 10^{4-5}$ K) gas mass in the CGM.  
Together with the warm-hot phase,
the CGM potentially accounts for at least half and up to all
of the missing baryons associated with galaxy halos
(\citealt{Werk2014,Prochaska2017}, also see \citealt{Stocke2013}).

Recent measurements of circumgalactic absorption 
along quasar sightlines draw attention to
the inhomogeneous distribution of baryons in the CGM.  
Sightlines along the galaxy major or minor axes
frequently detect absorption systems 
with large equivalent widths and broad velocity ranges, 
but these strong absorbers
are largely absent from sightlines that 
do not align with either of the two axes
\citep{Bouche2012,Kacprzak2012,Kacprzak2015,Nielsen2015}.  
This bi-modality in spatial geometry suggests
the position of the sightline relative to the galactic disk 
potentially distinguishes the origin 
of the circumgalactic absorption.

Sightlines along galaxy major axes 
often detect circumgalactic absorption 
with the Doppler shift sharing the same sign as 
the galactic disk.  
This implies the CGM corotates with 
the galaxy disks out to large radii
\editR{\citep{Steidel2002,Kacprzak2010,Kacprzak2011ApJ,
Bouche2013,Bouche2016,DiamondStanic2016,Ho2017,Martin2019}}.  
However, a simple rotating disk poorly reproduces 
the broad velocity ranges spanned by the absorption
\citep{Steidel2002,Kacprzak2010,Kacprzak2011ApJ,Ho2017}.
Some studies even demonstrate that the corotation 
can be modeled as inflowing gas with a disk-like geometry
\editR{\citep{Bouche2016,Bowen2016,Ho2017}}.  
Hence, probing the CGM along galaxy major axes 
provides a promising strategy to explore 
how galaxies obtain their gas.

From the theoretical perspective, 
circumgalactic gas, especially for gas accreted in `cold-mode', 
has significant angular momentum 
which can lead to corotation. 
In contrast to `hot-mode', where shock-heated gas 
cools and accretes onto the central galaxies isotropically
\citep{Fall1980,Mo1998}, 
`cold-mode' gas has a cooling time shorter than
the time needed to establish a stable shock
\citep{Dekel2006,Keres2005}.
Recent hydrodynamical simulations emphasize the 
importance of `cold-mode' accretion.
In addition to accreting along filamentary streams,
cold-mode gas has higher specific angular momentum
than its dark matter and `hot-mode' gas counterparts
\editR{\citep{Keres2009,Brook2011,Kimm2011,Stewart2011ApJ,
Stewart2013,Teklu2015,Stewart2017,Stevens2017}}.

In hydrodynamical simulations, 
galactic disks grow by 
accreting cooling, high angular momentum gas from the CGM.
As gas streams fall toward a galaxy, 
torques generated by the disk align the infalling gas with the pre-existing disk
\citep{Danovich2012,Danovich2015}. 
The newly accreted gas forms an extended cold flow disk,
which corotates with the galaxy out to large radii
\citep{Stewart2011ApJ,Stewart2013}. 
With gas accreted at later times having higher specific angular momentum, 
galaxy disks thereby grow inside-out
\editR{\citep{Kimm2011,Pichon2011,Lagos2017,ElBadry2018}}.

This paper presents results from 
the \EAGLE\ simulation suite \citep{Schaye2015,Crain2015, McAlpine2016}.
\EAGLE\ has been found to 
produce a realistic galaxy population 
and broadly reproduce a number of observations.
These include the $z\sim0$ galaxy stellar mass function
and the Tully-Fisher relation \citep{Schaye2015},
the evolution of galaxy masses \citep{Furlong2015}, 
the color bimodality of galaxies \citep{Trayford2015,Trayford2016},
\editR{and the atomic \citep{Bahe2016,Crain2017}
and molecular gas \citep{Lagos2015} content of galaxies.  
Similar to other hydrodynamical simulations
that show cold gas with high specific angular momentum, 
\citet{Stevens2017} have demonstrated that 
both cooling gas and hot gas in \EAGLE\ 
have higher specific angular momentum 
than the dark matter halo.} 
Previous work using the \EAGLE\ simulations 
(and \OWLS, \citealt{Schaye2010}) has also
demonstrated the importance of cold gas accretion onto galaxies
\citep{vandeVoort2011,Correa2018a,Correa2018b}.
\citet{Turner2017} have compared 
\EAGLE\ mock spectra with metal-line absorption data 
of $z\approx2$ star-forming galaxies \citep{Turner2014}
in the Keck Baryonic Structure Survey (KBSS, \citealt{Rudie2012,Steidel2014}).
The comparison has found evidence of infalling gas 
that explains the observed redshift-space distortions.

Using the \EAGLE\ simulations,
we examine how cold gas accretes onto galaxies
and relate the gas kinematics to measurements in
quasar sightline observations.  
We identify the inflowing gas particles using two methods:
(i) the analytical ballistic approximation that 
    predicts the motion of particles under the influence of gravity,
    and 
(ii) tracking particles through time in \EAGLE\
     which includes full hydrodynamic calculations. 
We study and compare the inflow properties 
from identifying inflow particles using the two methods,
and we gain insight into the factors 
that affect whether a cold gas particle reaches the inner galaxy 
within a disk rotation period.
To motivate the use of \EAGLE\ simulations to explain 
the CGM observations in the future, 
we use mock sightlines to probe the cold gas around  
one of the \EAGLE\ galaxies.  
We identify the structural features that can 
reproduce the corotation signature and  
the broad velocity ranges detected
in quasar sightline observations.
We defer the spectral analysis and 
the use of a large galaxy sample to a future paper.

We present the paper as follows.
Section~\ref{sec:sample_simulation} describes the 
\EAGLE\ galaxy selection at $z=0.27$, 
and we demonstrate that the short gas consumption time demands an
external gas supply.  
Then we identify the cold inflowing gas that feeds the inner galaxies   
and examine the inflow properties.  
We show the results of using the ballistic approximation 
and particle tracking
in Section~\ref{sec:identify_inflow} and \ref{sec:test_ballistic} 
respectively.
In Section~\ref{sec:observations}, 
we probe the cold gas around 
an \EAGLE\ galaxy using mock quasar sightlines,
and we focus on the gas structures that
corotate with the disk and span broad velocity ranges.  
Finally, we summarize our results in Section~\ref{sec:summary}.

%%%%%%%%%%%%%%%%%%%%%%%%%%%%%%%%%%%%%%%%
% EAGLE/SAMPLE SELECTION
%%%%%%%%%%%%%%%%%%%%%%%%%%%%%%%%%%%%%%%%
\section{Galaxy Selection from the EAGLE Simulation}
\label{sec:sample_simulation}

%%%%%%%%%%%%%%%%%%%%%%%%%
% Simulation Overview
%%%%%%%%%%%%%%%%%%%%%%%%%
\subsection{Simulation Overview}
\label{ssec:simulation}

The \EAGLE\ simulation suite consists of a large number of 
cosmological, hydrodynamic simulations 
\citep{Schaye2015,Crain2015,McAlpine2016}.  
\EAGLE\ was run on a modified version of the $N$-Body Tree-PM 
smoothed particle hydrodynamics (SPH) code \GADGET3\
(last described in \citealt{Springel2005}).
State-of-the-art subgrid models were implemented to capture unresolved physics, 
including radiative cooling and photoheating, star formation, 
stellar evolution and enrichment, stellar feedback, 
black hole growth, and feedback from active galactic nuclei.
The simulations also varied in 
cosmological volumes, resolutions, and subgrid physics,
and the stellar feedback was
calibrated to reasonably reproduce sizes of disk galaxies, 
and the galaxy stellar mass function at $z\sim0$.  

\EAGLE\ defines galaxies as gravitationally bound subhalos 
identified by the \texttt{SUBFIND} algorithm \citep{Springel2001, Dolag2009}.  
In brief, first the friends-of-friends (FoF) algorithm \citep{Davis1985}
places dark matter particles into the same group 
if the particle separation is below 0.2 times the average particle separation.
Baryons are associated with the same FoF halo (if any) 
as their closest dark matter particle.
Then, within FoF halos, 
\texttt{SUBFIND} defines self-bound overdensity substructures as subhalos, 
and each subhalo represents a galaxy.  
Within each FoF halo, 
the subhalo that contains the particle with the lowest value 
of gravitational potential is the central galaxy.  
The remaining subhalos are classified as satellite galaxies.  

In this pilot study, we use the simulation Ref-L012N0188 
% which has a box size of 12.5 cMpc.
with a box size of 12.5 cMpc.  
In the future, we will expand our study to 
large simulations and with higher resolutions.
We summarize the simulation parameters in Table~\ref{tb:simulation}.  
We use the particle data output\footnote{
    Particle data from snapshots can be downloaded from
    \texttt{\url{http://icc.dur.ac.uk/Eagle/database.php}}
    }
and mainly focus on galaxies at a single `snapshot' of $z=0.271$;
this redshift is comparable to the galaxy redshifts in 
quasar absorption line studies  
that measure the CGM kinematics of low redshift galaxies
(e.g., \citealt{Ho2017,Martin2019}).
In Section~\ref{sec:test_ballistic}, 
we also show results from particle tracking 
at finer timesteps than consecutive snapshots;
we use the reduced set of particle properties in `snipshots'.

% =================================
% TABLE with EAGLE box info
%\begin{center}
\input{table1.tex}
%\end{center}
% =================================

%%%%%%%%%%%%%%%%%%%%%%%%%
% Defining
%%%%%%%%%%%%%%%%%%%%%%%%%
\subsection{Defining Cold Gas}
\label{ssec:define_cold_gas}

In the simulation, 
we identify cold gas particles using temperature cutoffs.
For most of this work, 
especially when we focus on cold inflowing gas
(Section~\ref{sec:identify_inflow} and \ref{sec:test_ballistic}), 
we select cold gas using a temperature cutoff of $2.5\times10^5$ K.  
This cutoff has been commonly used to distinguish between 
cold-mode and hot-mode accretion
\citep{Keres2005,Keres2009, Stewart2013,Stewart2017}.  
However, when we compare cold gas kinematics to 
quasar sightline observations  (Section~\ref{sec:observations}),
which detect absorption from low ionization ions 
(e.g., \mgII, \siII, \feII),
we redefine the cutoff as $3\times10^4$ K.  
This is because low ions such as
\mgII\ do not exist at $\sim 10^5$ K
\citep{Oppenheimer2013, Tumlinson2017}.

In addition, we assign all star-forming gas as `cold' gas, 
because in \EAGLE, 
the temperature of the star-forming (i.e., interstellar) gas
is artificially increased to reflect
the effective pressure.
\EAGLE\ lacks the resolution to 
resolve the interstellar gas phase of $T_\mathrm{gas} \ll 10^4$ K.  
Hence, \EAGLE\ imposes a temperature floor, 
such that the corresponding effective equation of state  
prevents numerical Jeans fragmentation due to the finite resolution
(\citealt{Schaye2008}, also see \citealt{Robertson2008} and \citealt{Schaye2008}).

%%%%%%%%%%%%%%%%%%%%%%%%%
% Galaxy Selection
%%%%%%%%%%%%%%%%%%%%%%%%%
\subsection{Galaxy Selection}
\label{ssec:galaxy_selection}

\begin{figure}[thb]
    \centering
    \includegraphics[width=0.85\linewidth]{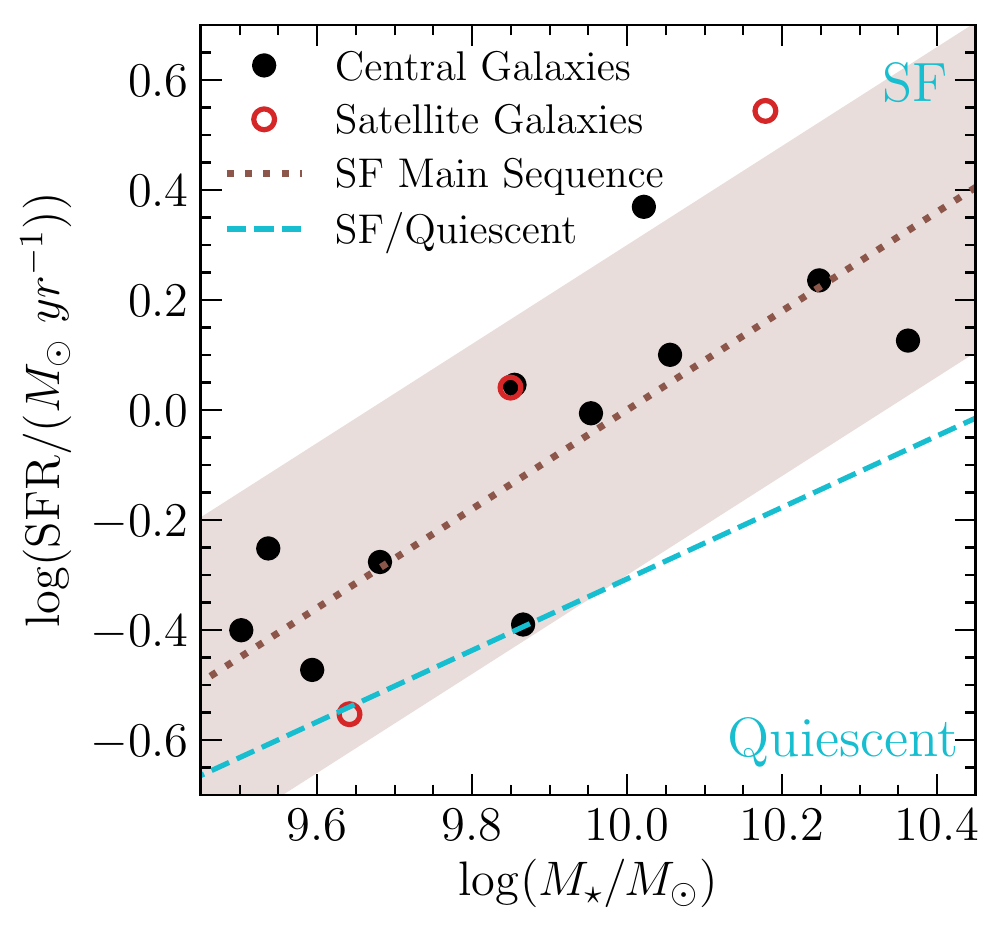}
    \caption{
        Galaxies on the SFR--$M_\star$ plane.  
        Black filled and red empty circles 
        represent central and satellite galaxies, respectively.  
        The brown dotted line and the shaded region 
        show the star-forming main sequence 
        and the 0.3 dex scatter from \citet{Peng2010},
        which fit the SDSS star-forming galaxies 
        in \citet{Brinchmann2004}.
        The cyan dashed line divides the sample into 
        star-forming or quiescent, 
        depending on whether the galaxies lie above or below 
        the line \citep{Moustakas2013}.  
        The selected galaxies lie along the main sequence, 
        and most galaxies are star-forming.
        }
    \label{fig:sfr_mstar} 
\end{figure}

We select galaxies at $z=0.271$ with a stellar mass range of
$\log(M_\star / M_\odot)$ between 9.5 and 10.5, 
comparable to that of the galaxy samples in 
\citet{Ho2017} and \citet{Martin2019}
that study the CGM kinematics of low-redshift galaxies. 
To measure the stellar mass of each galaxy, 
we use a 3D aperture with a radius of 30 pkpc from the galaxy center\footnote{
    The galaxy center is defined as the location 
    of the most bound particle of the subhalo.}, 
and sum over the masses of star particles that belong to the subhalo 
(as defined in \citealt{Schaye2015}).
Figure~\ref{fig:sfr_mstar} shows the selected galaxies 
on the SFR--$M_\star$ plane, 
for which we use the same 30-pkpc aperture to measure the SFR.  
We show the SFR--$M_\star$ main sequence from \citet{Peng2010},
fitted to the star-forming galaxies 
in the Sloan Digital Sky Survey (SDSS) from \citet{Brinchmann2004}.
We also plot the line that divides star-forming and quiescent galaxies 
\citep{Moustakas2013};
the line comes from a redshift dependent relation 
using $\sim$120,000 galaxies with spectroscopic redshifts
from the PRism MUlti-object Survey 
(PRIMUS, \citealt{Coil2011,Cool2013}) and SDSS.
Figure~\ref{fig:sfr_mstar} shows that our selected galaxies lie 
along the main sequence.  
Star-forming galaxies dominate our sample, 
with most of them being central galaxies.

The halo virial masses of our sample
range from $\log(M_\mathrm{vir} / M_\odot)$ of 11.5 to 12.6.
We define the virial radius $r_\mathrm{vir}$ as 
the radius that encloses an average density of
$\Delta_{vir}(z) \rho_{c}(z)$, 
where $\rho_{c}(z)$ represents the critical density at redshift $z$, 
and the overdensity $\Delta_{vir}(z)$ 
follows the top-hat spherical collapse calculation in \citet{Bryan1998}.\footnote{
    At $z=0.271$, $\Delta_{vir}(z) = 124$.}

%%%%%%%%%%%% FIG: SFR--Mstar plane %%%%%%%%%%%%
%%%%%%%%%%%%%%%%%%%%%%%%%%%%%%%%%%%%%%%%%%%%%%%%%%%%%%%%%%%%

%%%%%%%%%%%%%%%%%%%%%%%%%
% Gas Consumption Timescale
%%%%%%%%%%%%%%%%%%%%%%%%%
\subsection{Global Gas Consumption Timescale}
\label{ssec:gas_consumption_time}

The gas consumption timescale 
is sometimes referred as the `Roberts time' \citep{Roberts1963},
which is defined using the 
following relation (\citealt{Kennicutt1994}, hereafter K94),
\begin{equation}
    \tau_{R} = \frac{M_\mathrm{gas}/SFR}{1-R}
    \label{eq:roberts_time}
\end{equation}
where $R$ is the returned gas fraction.  
The $(1-R)^{-1}$ correction factor accounts for gas recycling 
and future time dependence on SFR.  
K94 have analyzed how $(1-R)^{-1}$ changes with different parameters, 
such as initial mass functions, star formation laws, and gas surface density etc.
From their time-dependent modeling of gas return rate (from stars), 
they suggest that recyling gas extends $\tau_R$ 
of typical star-forming disks by 1.5 to 4 times.

For our selected \EAGLE\ galaxies, 
we calculate the gas consumption time $\tau_R$
using Equation~(\ref{eq:roberts_time}).
Including gas beyond the star-forming region
will overestimate the actual gas depletion time in the inner galaxy regions,
where star formation takes place.
To avoid this overestimation, 
first we define the star-forming radius $r_\mathrm{SFR90}$ as the radius 
that encloses 90\% of the galaxy SFR, 
and we
round it up to the closest 5 pkpc\footnote{
    We round up to the closest 1 pkpc, 
    if the star-forming radius is below 5 pkpc.
    }.
Then, we calculate $\tau_R$ using
the gas mass and SFR within a 3D aperture of radius $r_\mathrm{SFR90}$.
In addition, we impose limits on gas temperature $T_\mathrm{gas}$ 
while calculating the gas mass: 
all $T_\mathrm{gas}$, $T_\mathrm{gas} \leq 2.5\times10^5$ K, 
and $T_\mathrm{gas} \leq 3\times10^4$ K.

%%%%%%%%%%%% FIG: Compare Gas Consumption Time %%%%%%%%%%%%
\begin{figure}[htb]
    \centering
        \includegraphics[width=1.0\linewidth]{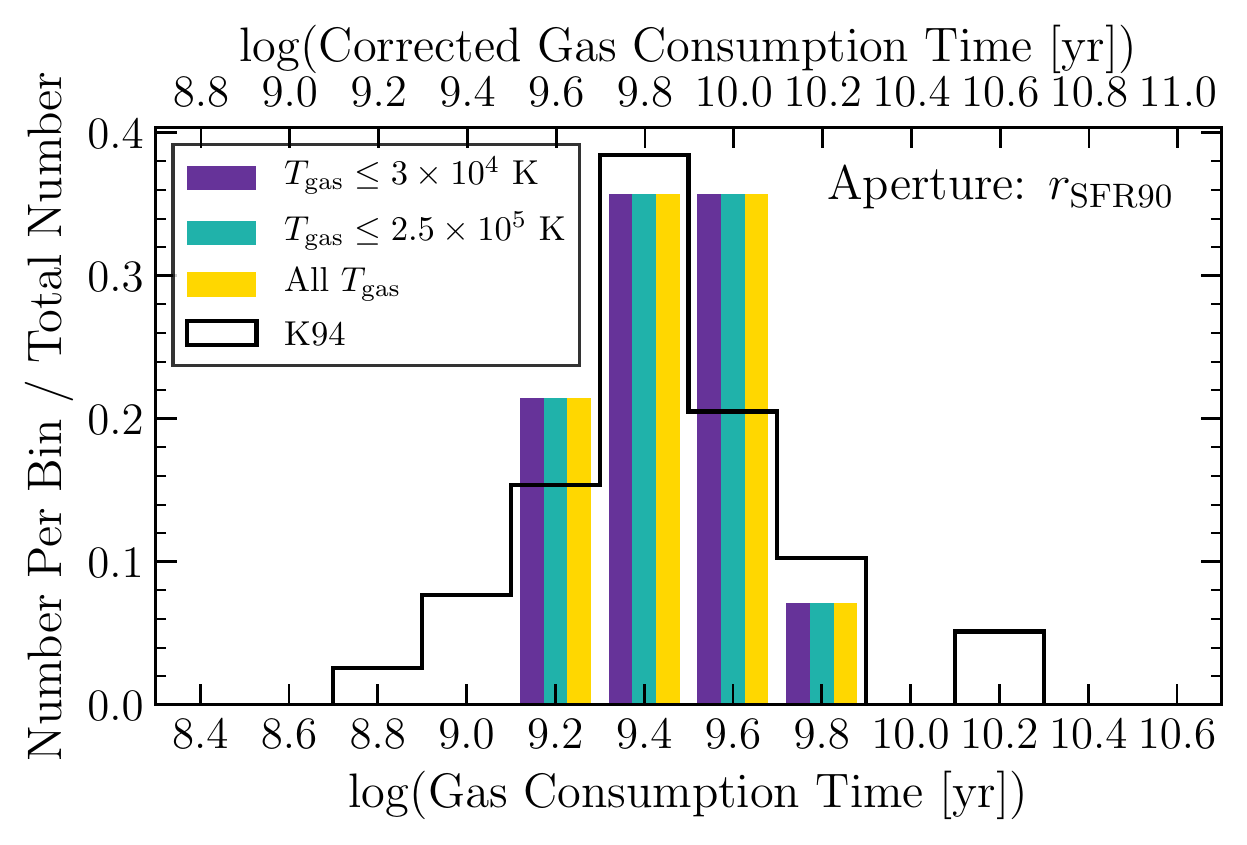}
    \caption{
        Distribution of gas consumption time.  
        We calculate the gas consumption time 
        using the gas mass and SFR within 
        the star-forming radius $r_{\mathrm{SFR90}}$.  
        Within each histogram bin,
        different colors of the filled histograms distinguish
        the use of gas masses with different temperature limits. 
        The bottom and top axis show the timescale before and after 
        applying the gas recycling correction 
        factor, i.e., $(1-R)^{-1} = 1$ or 2.5.  
        The black histogram shows the gas consumption time distribution 
        of nearby galaxies in \citet{Kennicutt1994}, 
        and their calculation includes only the gas mass 
        within the star-forming disk.  
        Even with the $(1-R)^{-1}$ correction factor,
        at least half of the \EAGLE\ galaxies 
        have gas consumption timescales shorter than the Hubble time.
        }
    \label{fig:gas_consumption} 
\end{figure}
%%%%%%%%%%%%%%%%%%%%%%%%%%%%%%%%%%%%%%%%%%%%%%%%

Figure~\ref{fig:gas_consumption} 
shows the distribution of gas consumption time $\tau_R$ of the galaxies.  
We show $\tau_R$ before and after applying the 
$(1-R)^{-1}$ correction factor, 
i.e., set $(1-R)^{-1} = 1$ or use an assumed value of 2.5.
The latter is consistent with K94,\footnote{
    $(1-R)^{-1}=2.5$ is also adopted in other publications, 
    e.g., \citet{Boselli2001}.
    }
and Figure~\ref{fig:gas_consumption}  shows that 
their nearby galaxy sample (black histogram) and our selected galaxies
have comparable gas consumption times.
If we adopt a fixed 30-pkpc aperture, 
i.e., the same aperture 
used to measure the stellar mass and SFR of a galaxy, 
then $\tau_R$ of each galaxy increases by over 50\% (0.2 dex).  
This is because star-forming regions 
are generally smaller than 30 pkpc, 
and hence a 30-pkpc aperture includes a larger gas reservior
without increasing the SFR.  
This phenomenon agrees with K94, who have also
shown that including gas 
outside the star-forming radius increases $\tau_R$.

\editR{The gas consumption times of several Gyrs for 
our galaxies 
are also comparable to the atomic and molecular
consumption timescales (uncorrected)
of galaxies in other observed samples.
For example, the nearby disk galaxies 
in the HERACLE survey have 
$\mathrm{H}_2$ consumption timescales
of $\sim$2.4 Gyr \citep{Bigiel2008,Bigiel2011},
and the low-redshift, 
$M_\star \sim 10^{10} M_\odot$ galaxies
in the COLD GASS survey
have \hI\ and $\mathrm{H}_2$ consumption timescales 
of $\sim$3 Gyr and $\sim$1 Gyr respectively
\citep{Schiminovich2010,Saintonge2011}; 
also see Section 4 of \citet{Lagos2015} 
for a detailed comparison 
of $\mathrm{H}_2$ consumption timescales
between \EAGLE\ and COLD GASS galaxies.}

Furthermore, while Figure~\ref{fig:gas_consumption}
shows that $\tau_R$ of the \EAGLE\ galaxies 
do not vary significantly with the temperature cuts,
$\tau_R$ shows a stronger temperature dependence 
if we adopt a fixed 30-pkpc aperture.  
This suggests cold gas with temperature 
below $3\times10^4$ K
dominates the total gas mass within $r_\mathrm{SFR90}$.

Regardless of our gas temperature limit, 
and even if we apply the $(1-R)^{-1}$ correction factor, 
Figure~\ref{fig:gas_consumption} shows that
gas consumption timescales of at least half of the galaxies are 
significantly shorter than the Hubble time.
This implies that in order to sustain the star-forming activity,
the galaxies have to replenish their gas supply.

%%%%%%%%%%%%%%%%%%%%%%%%%%%%%%%%%%%%%%%%
% Radial Inflow Model
%%%%%%%%%%%%%%%%%%%%%%%%%%%%%%%%%%%%%%%%
\section{Identification of Inflowing Cold Gas}
\label{sec:identify_inflow}

We aim to understand how galaxies get the gas to fuel star formation.
Hot, virialized gas does not directly accrete onto galaxies.
Even in the hot-mode accretion scenario, 
the virialized gas has to cool and condense, 
before being accreted to form stars.
In the cold-mode scenario, 
the accreted gas remains cold during accretion.
Therefore, we focus on the accretion of cold gas onto galaxies.  
In Section~\ref{sec:test_ballistic}, 
by identifying inflowing gas through tracking particles,
we will verify that, 
at least for our selected central galaxies in \EAGLE,
inflow from hot gas is negligible.  
That will justify our focus on 
considering cold gas only in this section.

We want to identify the cold gas particles that will fall into 
the galactic disk within a rotation period.
In principle, we can do so by tracking particles 
through time.
However, the simulation only produces snapshot output 
at coarse redshift (i.e., time) intervals 
typically larger than a rotation period.   
And in general, 
tracking the particle positions at finer timesteps 
than the default simulation output requires re-running the simulation, 
which may be impossible if one only has access to output
at preset time intervals.
\editR{Running future cosmological simulations 
will also become more computationally expensive,
making it hard or even impossible to 
save the full simulation output at fine timesteps.  
As a result, the reconstruction of gas infall rates 
by tracking particles through time will be difficult.}
Therefore, in this section,
we introduce a procedure to estimate the infall from the simulation
output at a single time step.

We neglect pressure forces on the cold gas
and calculate the orbits of
gas particles in the gravitational potential.  
Orbits that reach pericenter beyond the galactic disk cannot feed the disk.  
We are interested in the orbits that intersect the galactic disk, 
because the collision of the cloud with the gas disk 
or the dense star-forming region
will dissipate energy.
We assume such gas will be incorporated into the disk as fuel.  
We then estimate a mean accretion rate and the average inflow speed 
over a disk rotation period.  
We also examine the distribution of the inflowing gas
and its angular momentum.
\editR{Then in Section~\ref{sec:test_ballistic}, 
we will discuss the validity and the caveats 
of \editR{this simple, analytical calculation
of the ballistic approximation,
and we will explore whether the ballistic approximation
can reasonably reproduce the inflow properties.}  
For the analysis in this Section 
and Section~\ref{sec:test_ballistic},
we will only focus on the central galaxies and 
exclude the satellite galaxies.}

%%%%%%%%%%%%%%%%%%%%%%%%%
% Ballistic Approximation
%%%%%%%%%%%%%%%%%%%%%%%%%
\subsection{Ballistic Approximation}
\label{ssec:ballistic}

To identify the inflowing cold gas 
($T_\mathrm{gas} \leq 2.5\times10^5$ K),
we predict whether individual cold gas particles can 
reach the star-forming region within a rotation period.  
We define the star-forming region   
as a spherical region of radius $r_\mathrm{SFR90}$ 
(defined in Section~\ref{ssec:gas_consumption_time})
from the galaxy center.  
We set the rotation period as
$4\times\sqrt{3 \pi/(16 G \langle\rho\rangle})$,
where $\langle\rho\rangle$ represents 
the average density of the star-forming region.  
We calculate each particle orbit 
within a rotation period of the star-forming region.  
Each particle conserves energy and angular momentum, 
and the particle
moves in a centrally directed gravitational field characterized 
by the dark matter.
In the absence of hydrodynamical interactions between particles,
the gravitational force determines the particle trajectory. 
Hence, we characterize the radial motion of each particle by
\begin{eqnarray}
    \ddot{r} & = & f_\mathrm{grav}(r) + \frac{j^2}{r^3},\\
    \varepsilon & = & \frac{1}{2}\dot{r}^2 + \frac{j^2}{2 r^2} +  
                    \Phi_\mathrm{grav}(r),
\label{eq:ballistic_motion}
\end{eqnarray}
where $f_\mathrm{grav}(r) = -\partial \Phi_\mathrm{grav}(r)/\partial r$
represents the
gravitational force per unit mass at radius $r$, 
and $j$ and $\varepsilon$ represent the 
constant specific angular momentum and energy of the particle.  
If the particle orbit ever intersects the star-forming region, 
then we consider the particle as inflowing.

We justify our assumption 
of gravitational force 
dominating over the pressure force on the cold gas as follows.  
Analogous to a multi-phase interstellar medium (ISM) model, 
gas of different phases, 
e.g., cold ($\sim$$10^4$ K) and hot ($\sim$$10^6$ K) gas, 
reach pressure equilibrium and thereby 
share a similar pressure gradient,  
i.e., $\nabla P_\mathrm{cold} \approx \nabla P_\mathrm{hot}$.
The cold gas has a higher density, 
i.e., $\rho_\mathrm{cold} \gg \rho_\mathrm{hot}$, 
since gas density scales inversely with temperature.
As a result,
the cold gas experiences a smaller pressure force 
because
$|\nabla P_\mathrm{cold}/\rho_\mathrm{cold}| \ll 
|\nabla P_\mathrm{hot}/\rho_\mathrm{hot}|$.  
For the virialized hot gas, 
the pressure force balances the gravitational pull
and prevents the gas from collapsing by its own gravity.
Since the cold gas experiences a smaller pressure force 
than the hot gas, 
the pressure force on the cold gas is negligible 
compared to gravity.
Hence, for the cold gas, the gravitational force dominates.

We expect the ballistic approximation to break down 
under the following circumstances.
First, if the cold gas clouds intercept hot gas 
with comparable mass column density, 
then the cold gas will be suspended by the hot gas, 
which prevents the cold clouds from falling in.  
Because this scenario violates the assumption 
of the ballistic approximation that
the gravitational force dominates over the pressure force,
the ballistic approximation becomes invalid.  
Second,
although we assume any gas that reaches the star-forming region
will be incorporated into the disk, 
this neglects the presence of galactic winds.  
Not only can winds remove 
gas that has once accreted onto the star-forming disk, 
they can also push out gas that would otherwise be accreted.
Consequently, whether the gas remains in the disk as fuel 
is sensitive to feedback, 
which the ballistic approximation ignores.  
\editR{Third, 
gas particles may lose angular momentum 
due to the interactions and collisions between them, 
e.g., with other cold infalling gas particles
or with the pre-existing disk
\citep{Danovich2015,Stevens2017}\footnote{
    \editR{
    Specifically, \citet{Stevens2017} use \EAGLE\ and 
    study the change of angular momentum 
    of the \textit{cooling} gas, 
    i.e., gas that transitioned from hot to cold phase,
    and find non-negligible angular momentum loss}.}.
The angular momentum dissipation thereby depends 
on the distribution of the gas particles 
and the frequency of particle collision,
which the ballistic approximation neglects.
}
Keeping these caveats in mind, 
we will identify the inflowing gas
and
predict the inflow properties using this analytical calculation.
Section~\ref{sec:test_ballistic} will explore the 
validity of the ballistic approximation.  
\editR{ 
We will show that 
although the ballistic approximation
can estimate the mass inflow rate
to within a factor of two compared to 
that from particle tracking,
the prediction is less accurate for the average radial inflow speed.  
Our analysis will thereby suggest that generally, 
the reconstruction of gas inflow properties, 
including inflow rates and inflow speeds, 
will benefit from simulation output at higher time cadence,
e.g., at the order of the rotation timescale of $\sim100$ Myr.
}

%%%%%%%%%%%%%%%%%%%%%%%%%
% Geometry
%%%%%%%%%%%%%%%%%%%%%%%%%
\subsection{Geometry of the Inflowing Cold Gas}
\label{ssec:geometry}

%%%%%%%%%%%% FIG: Example Galaxy Plot Set %%%%%%%%%%%%
\begin{figure*}[ht]
    \centering
    \includegraphics[width=0.88\linewidth]{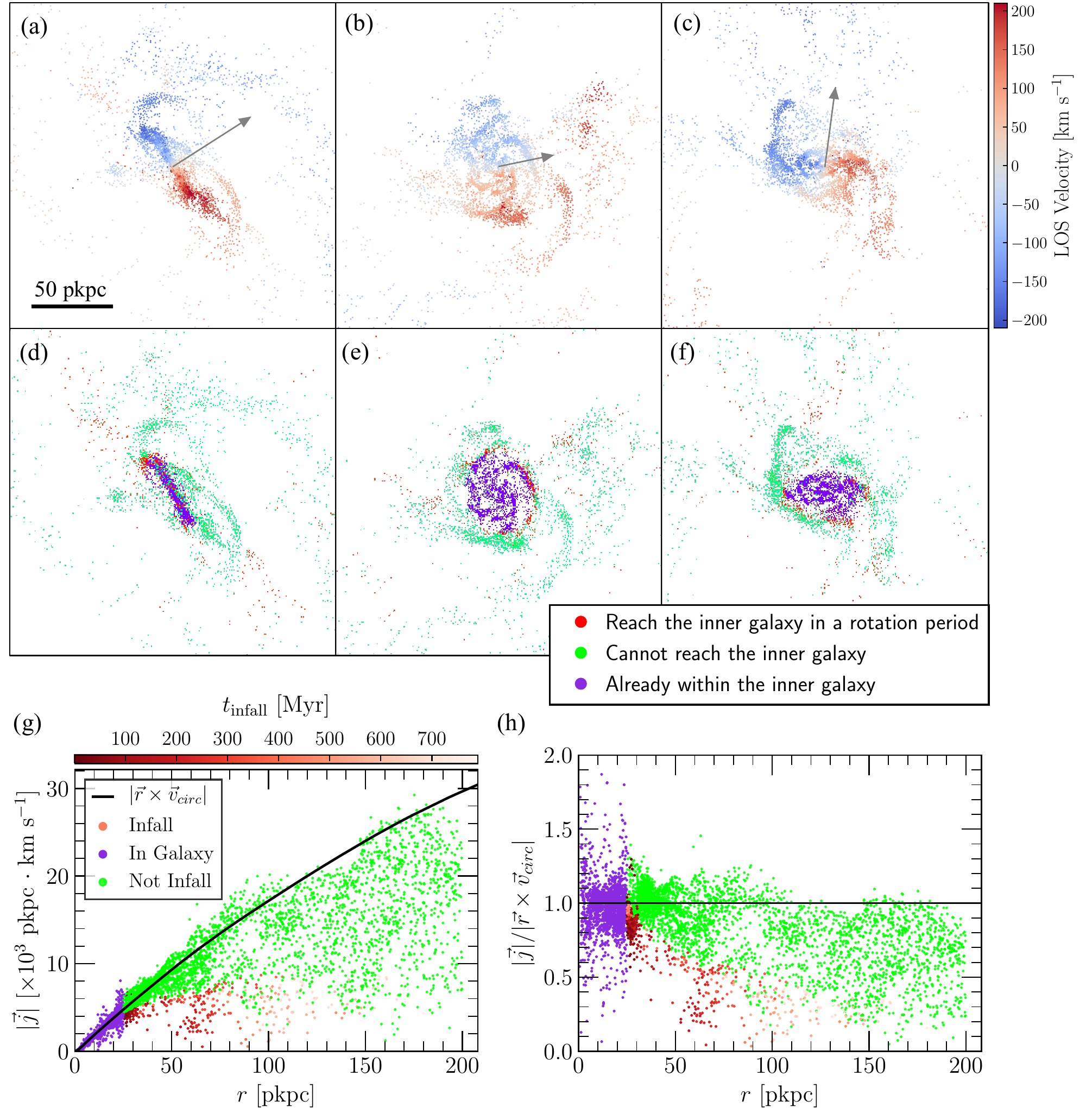}
    \caption{
        Cold gas with $T_\mathrm{gas} \leq 2.5\times10^5$ K of a disk galaxy 
        (galaxy ID: 37448).  
        The top two rows (panels a to f) 
        show the projection plots of cold gas particles, 
        and each column shows a cut in $x$, $y$ and $z$ plane respectively.  
        Each panel is 200 pkpc $\times$ 200 pkpc.  
        \textit{(Top row, panels a to c)}
        Particles are color-coded by their projected LOS velocities, 
        which clearly show a disk-like rotating structure.  
        The gray arrows show the direction of the net angular momentum 
        (projected onto each 2D plane) of the inner cold gas disk.  
        The galaxy has an approaching (blueshifted) 
        and a receding (redshifted) side, 
        which indicates the presence of a rotating structure.  
        \textit{(Middle row, panels d to f)}
        Particles are color-coded by whether 
        they will reach the star-forming region 
        in our ballistic approximation.  
        Most inflowing gas particles (red) reside at the outer spiral arms.
        \textit{(Bottom row, panels g and h)}
        The left panel shows the specific angular momentum $|\vec{j}|$ 
        of each particle.  
        The black line shows the specific angular momentum 
        required for the particle 
        to be rotationally supported,
        i.e., $|\vec{r}\times \vec{v}_\mathrm{circ}|$, 
        where $\vec{v}_\mathrm{circ}$ is the circular velocity at radius $r$.  
        The right panel normalizes each particle $|\vec{j}|$
        by $|\vec{r}\times \vec{v}_\mathrm{circ}|$.  
        Most infalling gas particles lack sufficient angular momentum 
        to maintain circular orbits,
        since $|\vec{j}| / |\vec{r}\times \vec{v}_\mathrm{circ}| < 1$
        for most infalling particles. 
        }
    \label{fig:ppj_jr_example_2p5e5} 
\end{figure*}
%%%%%%%%%%%%%%%%%%%%%%%%%%%%%%%%%%%%%%%%%%%%%%%%%%%%%%%%%%%%

In this section, 
we examine the distribution 
and the morphological structure of the inflowing cold gas.  
As an example, we show a galaxy (ID: 37448) in 
Figure~\ref{fig:ppj_jr_example_2p5e5},
and we project the cold gas particles 
($T_\mathrm{gas} \leq 2.5\times10^5$ K) onto 2D planes.
First, panels (a) to (c)
indicate that the galaxy has a rotating structure.  
The blueshifted and redshifted 
particle projected velocities clearly show that the galaxy has 
an approaching and a receding side.
Second, panel (a) also illustrates that
most particles reside in a thin structure, 
and the galaxy resembles a disk morphology.  
To visualize the distribution of the inflowing gas,
in panels (d) to (f), 
the colors of the gas particles show whether 
the particles will reach the star-forming region, i.e., inflowing.  
Most inflowing gas particles (red dots) 
reside in the outer spiral arms.

%%%%%%%%%%%% FIG: Particle Plots (color by infall) for 10 centrals %%%%%%%%%%%%
\begin{figure*}[ht]
    \centering
    \includegraphics[width=0.95\linewidth]{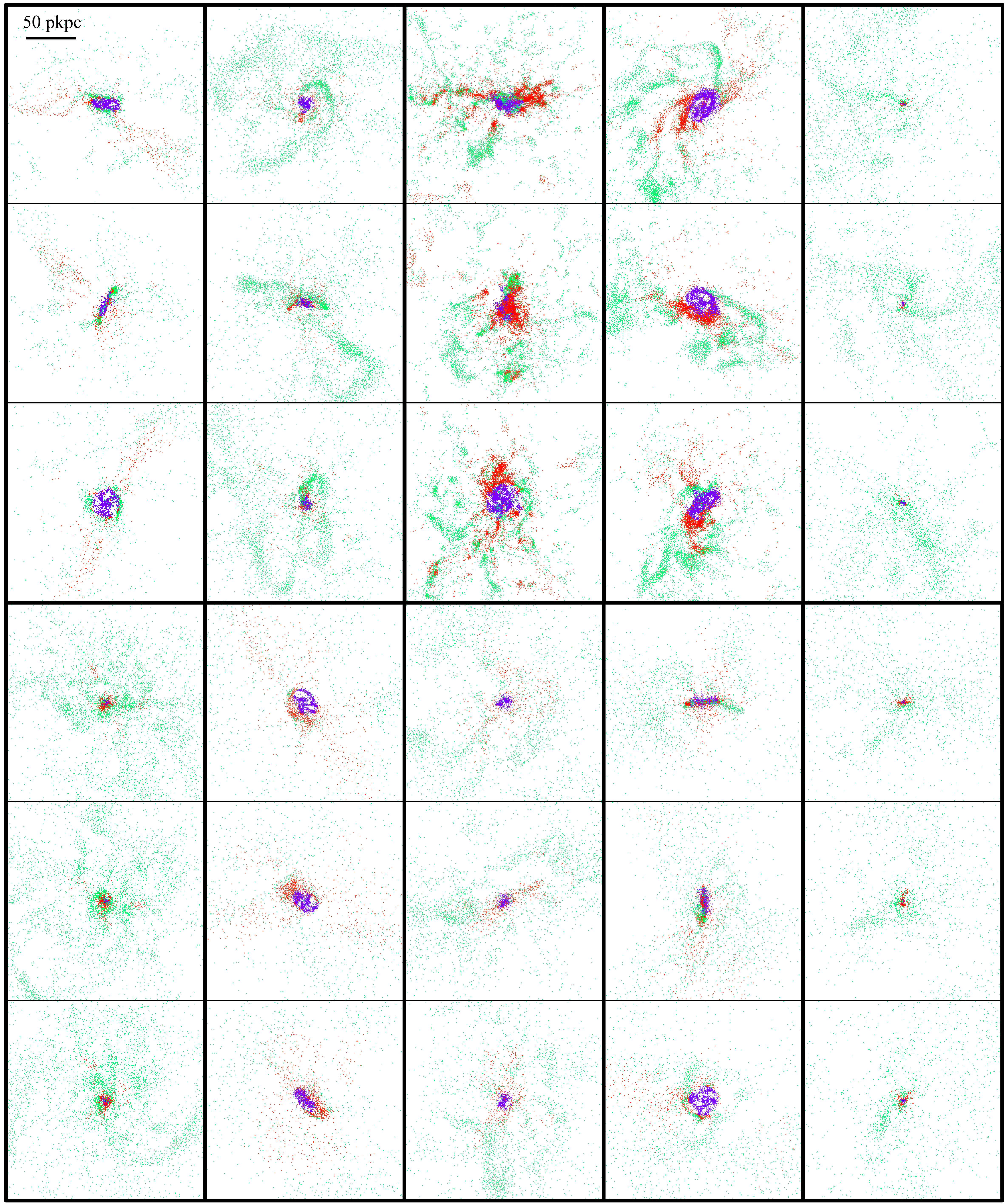}
    \caption{
        Particle projection plots 
        showing whether the cold gas particles can reach 
        the star-forming region within a rotation period.  
        The particles have the same color-coding as in 
        panel (d) to (f) of
        Figure~\ref{fig:ppj_jr_example_2p5e5}.
        Each galaxy occupies 3 rows $\times$ 1 column, 
        enclosed by the thick black lines.  
        For individual galaxies, from top to bottom, each panel
        represents the cut in $x$, $y$ and $z$ plane respectively, 
        and has a side length of 200 pkpc.
        The plotted central galaxies 
        are ordered in decreasing stellar masses, 
        from left to right, and top to bottom.
        }
    \label{fig:particle_pjplot_2p5e5}
\end{figure*}
%%%%%%%%%%%%%%%%%%%%%%%%%%%%%%%%%%%%%%%%%

Analogous to Figure~\ref{fig:ppj_jr_example_2p5e5}, 
Figure~\ref{fig:particle_pjplot_2p5e5} shows 
the remaining 10 central galaxies,
and we use to same color scheme as in 
Figures~\ref{fig:ppj_jr_example_2p5e5}(d) to (f) 
to distinguish the inflowing particles.
A similar figure with particles color-coded 
by the particle projected velocities 
can be found at the Appendix.  
The particle projection plots in 
Figure~\ref{fig:ppj_jr_example_2p5e5} and \ref{fig:particle_pjplot_2p5e5}
(and the Appendix)
show that despite the different morphologies among galaxies,
most galaxies have rotating structures 
The infalling gas particles are distributed anisotropically
and form a variety of structures, e.g.,
particles are concentrated along streams or 
are located near the thin, `disk-like' structures 
as in Figure~\ref{fig:ppj_jr_example_2p5e5}.

%%%%%%%%%%%% FIG: Example Particle Plots %%%%%%%%%%%%
\begin{figure}[thb]
    \centering
    \includegraphics[width=1.0\linewidth]{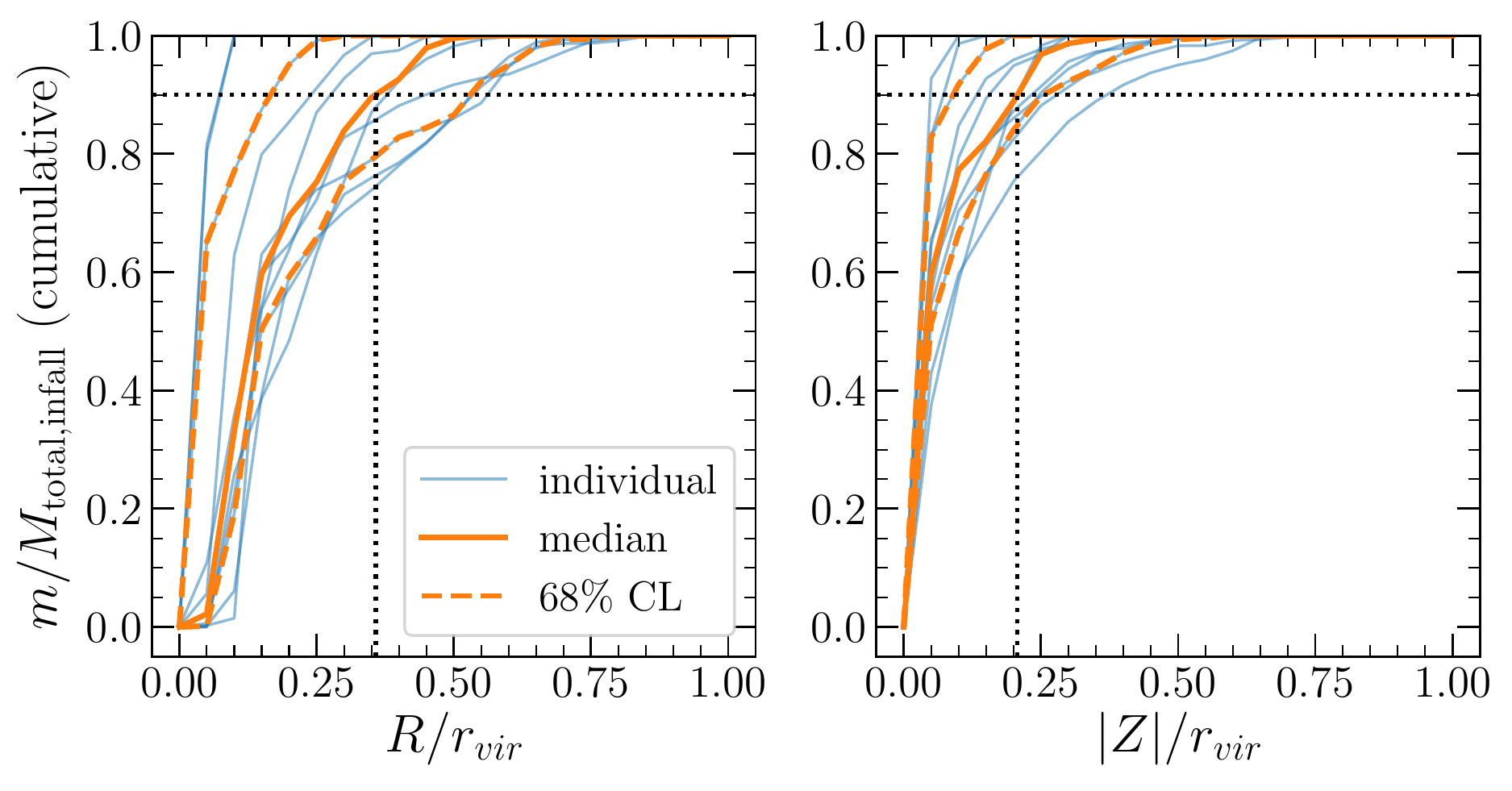}
    \caption{
        Spatial distribution of cold inflowing gas particles 
        relative to warped disk planes.  
        The left and the right panels show the cumulative mass profiles 
        of the inflowing gas in radial $R$, and perpendicular $|Z|$ direction 
        relative to the disk planes respectively
        (i.e., in cylindrical coordinates). 
        Both $R$ and $|Z|$ are normalized by the 
        galaxy virial radius $r_\mathrm{vir}$,
        and the mass profiles 
        are normalized by the total mass of the infalling gas. 
        Thin, cyan lines represent individual central galaxies, 
        and the thick, orange solid and dashed lines show 
        the median and the 68\% confidence level. 
        As measured from the median profiles,
        enclosing 90\% of the inflowing gas mass
        requires $R \approx 0.35r_\mathrm{vir}$ and 
        $|Z| \approx 0.2 r_\mathrm{vir}$ (black dotted lines).
        The inflowing gas particles extend further
        in the $R$-direction compared to the $|Z|$-direction.
        }
    \label{fig:particle_infall_geom} 
\end{figure}
%%%%%%%%%%%%%%%%%%%%%%%%%%%%%%%%%%%%%%%%%

To quantify the spatial distribution of inflowing gas particles,  
we calculate their positions
relative to the plane of the cold gas disk.  
For each galaxy,
we define disk plane using the net angular momentum vector of cold gas.  
We allow the disk plane to change with radius, 
since \hI\ observations often find warped gas disks
(e.g., \citealt{Heald2011,Zschaechner2012}), 
and we measure the net angular momentum vector in 
concentric shells with thickness of $\Delta r = 10$ pkpc.
Figure~\ref{fig:particle_infall_geom} shows the cumulative mass profiles 
of the inflowing gas relative to the warped disk planes.  
The thin, cyan lines represent individual galaxies,
and the thick, orange solid and dashed lines represent 
the median and the 68\% confident levels of the sample.  
The profiles show that the inflowing particles 
extend further in the 
$R$-direction on the disk plane 
than in the $|Z|$-direction perpendicular to the disk plane.
As an illustration, the black dotted lines 
show that from the median profiles, 
$R \approx 0.35r_\mathrm{vir}$ encloses 90\% of the inflowing gas mass,
in contrast to $0.2 r_\mathrm{vir}$ in the $|Z|$-direction.
On the one hand, 0.2$r_\mathrm{vir}$ or tens of pkpc 
(Table~\ref{tb:inflow_avgv_result} shows $r_\mathrm{vir}$ of each galaxy)
along the $|Z|$-direction
implies that the infalling particles do not just lie on a thin disk. 
But on the other hand, 
as the inflowing gas particles extend further in 
the $R$-direction than in the $|Z|$-direction, 
the gas tends to have a cylindrical structure,
instead of an isotropic distribution.

% =================================
% TABLE with mass flow rate, average velocity info
%\begin{center}
\input{table2.tex}
%\end{center}
% =================================

%%%%%%%%%%%%%%%%%%%%%%%%%
% Angular Momentum
%%%%%%%%%%%%%%%%%%%%%%%%%
\subsection{Angular Momentum of Inflowing Gas}
\label{ssec:angular_momentum}

Figures~\ref{fig:ppj_jr_example_2p5e5}(g) and (h)
show the specific angular momentum $|\vec{j}|$ of each cold gas particle.  
To be rotationally supported, 
each particle requires a specific angular momentum of 
$|j_\mathrm{circ}| = |\vec{r}\times \vec{v}_\mathrm{circ}|$ 
(solid line), % in panel (g)),
where $\vec{v}_\mathrm{circ}$ is the circular velocity at radius $r$.  
Except for the inflowing gas at $r \approx r_\mathrm{SFR90}$, 
most inflowing particles have $|j|/|j_\mathrm{circ}| < 1$.  
Thus, these particles lack sufficient angular momentum 
to be on circular orbits.
Inflowing gas of other central galaxies 
also show the same characteristics.

Without angular momentum, 
a particle will fall radially towards the galaxy center 
due the gravity.
Angular momentum provides rotational support for a particle.  
Under the assumption of angular momentum conservation, 
the particle $|\vec{j}|$ defines the pericenter of the particle orbit,
and only particles with pericenters smaller than 
the star-forming radius $r_\mathrm{SFR90}$
can possibly be considered as inflowing.  
In reality, however, 
a particle may lose angular momentum 
due to tidal interactions with the existing disk.  
Hence, the predicted pericenter represents an upper limit,
and more particles than predicted 
may reach $r_\mathrm{SFR90}$ within a rotation period.

%%%%%%%%%%%%%%%%%%%%%%%%%
% Mass Inflow Rate
%%%%%%%%%%%%%%%%%%%%%%%%%
\subsection{Average Mass Inflow Rate and SFR}
\label{ssec:ballistic_inflow_rate}

\begin{figure}[thb]
    \centering
    \includegraphics[width=0.82\linewidth]{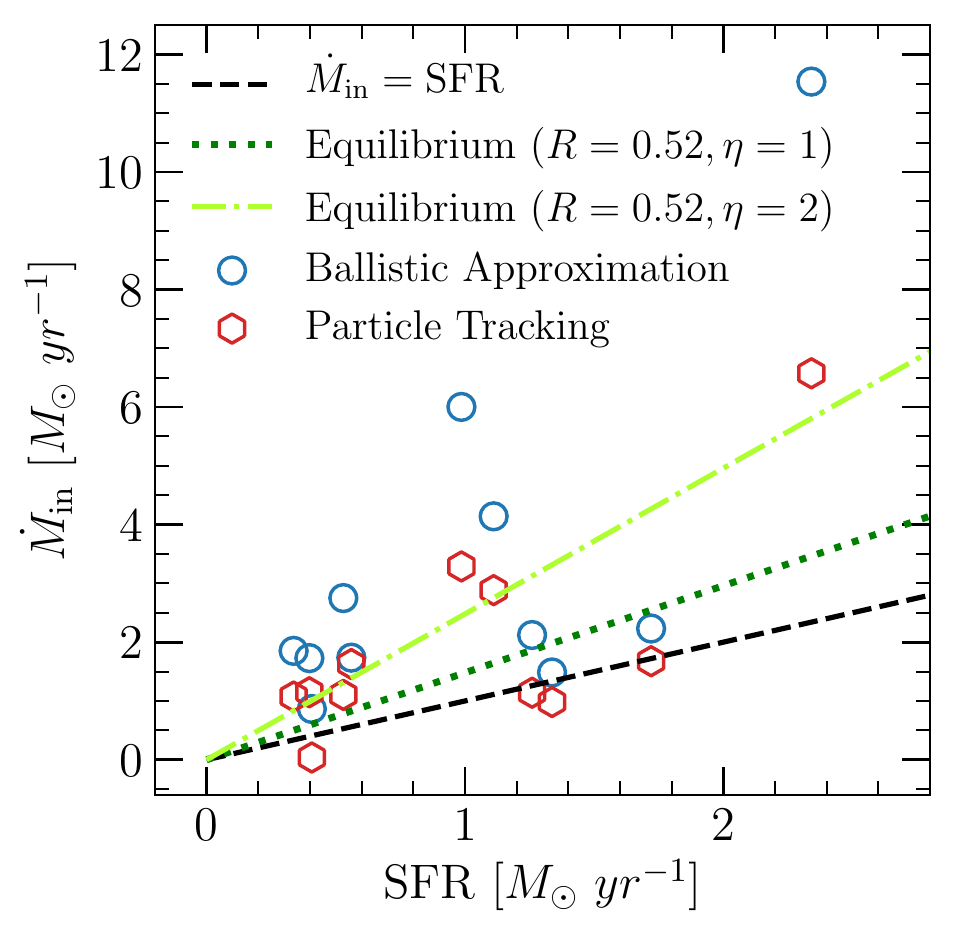}
    \caption{
        \editR{
        Comparison between average mass inflow rates
        and SFRs of central galaxies. 
        The ballistic approximation (blue circles) 
        generally overestimates
        the cold gas inflow rate
        compared to particle tracking (red hexagons).  
        The inflow rates are comparable to that
        suggested by the equilibrium model;
        we show the case for
        $R=0.52$ and $\eta = 1$ (dotted)
        and $\eta = 2$ (dot-dashed)
        in Equation~\ref{eq:equilibrium}.
        The cold gas inflow plausibly
        accounts for sustaining the galaxy star formation activities.
        }
        }
    \label{fig:ballistic_inflow_rate_2p5e5} 
\end{figure}

Unless galaxies replenish their gas, 
galaxies will eventually quench due to the lack of fuel.  
The inflow rate provides an important clue 
of whether a galaxy can sustain its star formation rate.
To calculate the average mass inflow rate 
$\dot{M}_\mathrm{in}^\mathrm{ballistic}$, 
we sum over the mass of the inflowing cold gas 
and divide it by the rotation period of each galaxy.  
We list $\dot{M}_\mathrm{in}^\mathrm{ballistic}$ for 
individual galaxies in Table~\ref{tb:inflow_avgv_result}.   
Figure~\ref{fig:ballistic_inflow_rate_2p5e5}
compares $\dot{M}_\mathrm{in}^\mathrm{ballistic}$ to the galaxy SFR
\editR{(blue circles)},
and the figure shows that
$\dot{M}_\mathrm{in}^\mathrm{ballistic}$ 
exceeds the SFR for all central galaxies.

Equality between the average mass inflow rate 
and the galaxy SFR 
does not imply a sustainable star formation rate.  
Gas outflows remove gas from the galaxy
and thereby demand a higher inflow rate to support star formation, 
whereas the recycling stellar gas reduces the 
need for new gas supply.
In a self-sustainable evolution model, i.e., equilibrium model, 
a galaxy reaches equilibrium
when the gas inflow rate is equal to a linear combination of 
the SFR and the gas outflow rate \citep{Bouche2010, Dave2012}.
Under the equilibrium condition, 
we express the inflow rate $\dot{M}_\mathrm{gas,in}$ as
\begin{equation}
    \dot{M}_\mathrm{gas,in} = (1-R)\dot{M}_\star + \dot{M}_\mathrm{gas,out},
    \label{eq:equilibrium}
\end{equation}
where $\dot{M}_\mathrm{gas,out}$ and $\dot{M}_\star$
represent the mass outflow rate and SFR respectively.  
The $(1-R)$ factor corrects for the gas recycling fraction, 
and hence $(1-R)\dot{M}_\star$ represents the corrected net SFR.  
The mass outflow rate scales with the SFR 
according to the mass loading factor, 
$\eta = \dot{M}_\mathrm{gas,out} / \mathrm{SFR}$.
For simplicity, we use $R=0.52$ as in \citet{Bouche2010}\footnote{
    Adopting $R=0.6$ as in Section~\ref{ssec:gas_consumption_time} 
    will have a negligible effect on
    the equilibrium expectation 
    in Figure~\ref{fig:ballistic_inflow_rate_2p5e5}.
    }.
\editR{
While $\eta$ is not well-constrained in observations 
due to the uncertainties in mass outflow rates, 
we adopt $\eta \sim 1$ and 2
(dotted and dot-dashed lines respectively in 
Figure~\ref{fig:ballistic_inflow_rate_2p5e5}), 
typically inferred from observations
for galaxies with similar stellar masses 
as our \EAGLE\ galaxies \citep{Martin2012,Kacprzak2014,Heckman2015},
as well as dwarf galaxies \citep{Martin1999}
and infrared-luminous galaxies \citep{Rupke2005}.
Figure~\ref{fig:ballistic_inflow_rate_2p5e5} shows that 
the inflow rates are comparable to that 
expected by the equilibrium models,
suggesting that
the cold inflowing gas plausibly accounts 
for sustaining the galaxy star formation activities.
We further discuss the comparison between 
inflow rates and SFRs and the implication 
in Section~\ref{ssec:tracking_inflow_rate}.
}

%%%%%%%%%%%% FIG: Mass Flow Rate %%%%%%%%%%%%

%%%%%%%%%%%%%%%%%%%%%%%%%
% Ballistic: Infall Speed
%%%%%%%%%%%%%%%%%%%%%%%%%
\subsection{Mass-weighted Average Inflow Speed}
\label{ssec:ballistic_inflow_speed}

To estimate the average inflow speed over the rotation period, 
we take the mass-weighted average 
of the inflow speeds of individual inflowing cold gas particles.
First, for each inflowing cold gas particle, 
we calculate its inflow speed from the change in radial distance 
from the galaxy center, $\Delta r$, and divide it by the rotation period.  
However, the ballistic approximation breaks down when the particle 
enters the star-forming region.  
Once the gas particle collides with the dense gas clouds 
in the star-forming region and dissipates energy, 
our assumption of energy conservation no longer holds.  
Therefore, 
we estimate $\Delta r$ by differencing
the initial particle radial position and 
the star-forming radius $r_\mathrm{SFR90}$; we assume
the particle halts once it reaches $r_\mathrm{SFR90}$.  
Then, we weight the inflow speeds of all inflowing gas particles
by their particle masses, 
and we obtain the mass-weighted average inflow speed 
$\langle v_r \rangle_\mathrm{in}^\mathrm{ballistic}$ for each galaxy.

Table~\ref{tb:inflow_avgv_result} lists 
the mass-weighted average inflow speed 
$\langle v_r \rangle_\mathrm{in}^\mathrm{ballistic}$
for the central galaxies, 
which ranges from 17\kms\ to 62\kms.
Although most particles have low inflow speeds, 
resulting in
low $\langle v_r \rangle_\mathrm{in}^\mathrm{ballistic}$, 
the particle inflow speeds often span a large range, 
and the distribution has a high velocity tail.
To characterize the spread in individual inflow speeds, 
we also estimate the mass-weighted average inflow speed 
using the subset of particles with the highest inflow speeds;
this particle subset accounts for 
10\% of the total inflowing gas mass.
We show this average inflow speed from the `high speed tail' as 
$\langle v_r \rangle_\mathrm{in, top 10\%}^\mathrm{ballistic}$ in 
Table~\ref{tb:inflow_avgv_result}.
For most galaxies, 
$\langle v_r \rangle_\mathrm{in, top 10\%}^\mathrm{ballistic}$
exceeds 100\kms.  
The significant difference between 
$\langle v_r \rangle_\mathrm{in}^\mathrm{ballistic}$ and 
$\langle v_r \rangle_\mathrm{in, top 10\%}^\mathrm{ballistic}$ 
demonstrates 
the large spread in particle inflow speeds
despite the low mass-weighted average.

%%%%%%%%%%%%%%%%%%%%%%%%%%%%%%%%%%%%%
% Tracking Particles [SECTION]: Ballistic vs. Hydrodynamics
%%%%%%%%%%%%%%%%%%%%%%%%%%%%%%%%%%%%%
\section{Ballistic Approximation vs. Hydrodynamic Calculations}
\label{sec:test_ballistic}

Instead of tracking particles through continuous time,
the ballistic approximation 
provides an alternate method to predict the properties of gas inflows.
To understand how the gas inflow properties
from the ballistic approximation compare to 
that from the hydrodynamic calculations in \EAGLE, 
we track the gas particles through `snipshots'.  
Snipshots are output from the same \EAGLE\ run, 
sampled at finer time intervals 
but with less information per particle 
compared to the full `snapshot' output.  
This allows
us to select the time slice
around the rotation period of a galaxy.  
In Section~\ref{ssec:tracking}, 
we explain how we track particles to define inflowing cold gas.   
Section~\ref{ssec:tracking_inflow_rate} 
compares the average mass inflow rate obtained from 
both the ballistic approximation and particle tracking,
whereas Section~\ref{ssec:tracking_inflow_speed}  
compares the mass-weighted average inflow speed using the two methods.
In Section~\ref{ssec:become_stars},
we explore how the newly accreted gas contributes to 
recent star formation.
Section~\ref{ssec:ballistic_limitation} discusses 
the limitations of the ballistic approximation
and the caveats of comparing its inflow prediction
to the outcome of \EAGLE's hydrodynamical calculations.

%%%%%%%%%%%%%%%%%%%%%%%%%
% Tracking Gas Particles
%%%%%%%%%%%%%%%%%%%%%%%%%
\subsection{Tracking Gas Particles to Identify Inflowing Gas}
\label{ssec:tracking}

We follow each gas particle through \EAGLE\ snipshots, 
sampled at every $\sim$70 Myr around $z\sim$0.271.  
To identify the inflowing cold gas particles for each galaxy, 
first we choose the snipshot for which
the time evolved from $z = 0.271$ 
is closest to the galaxy rotation period.
Hence, we use different snipshots for different galaxies.  
We define cold gas using the same criteria in 
Section~\ref{ssec:define_cold_gas}.  
To classify a gas particle as inflowing, 
the particle has to reside beyond $r_\mathrm{SFR90}$ at $z=0.271$
and have reached
$r_\mathrm{SFR90}(z=0.271)$ or deeper
after the rotation period in the selected snipshots.  
If this particle has been converted into a star,
we still classify it as an inflowing gas particle.  
Since only a tiny fraction of inflowing gas particles 
have turned into stars after a rotation period,
our conclusions will not change even if we exclude 
this inflowing particle subset.

\begin{figure}[htb]
    \centering
    \includegraphics[width=0.85\linewidth]{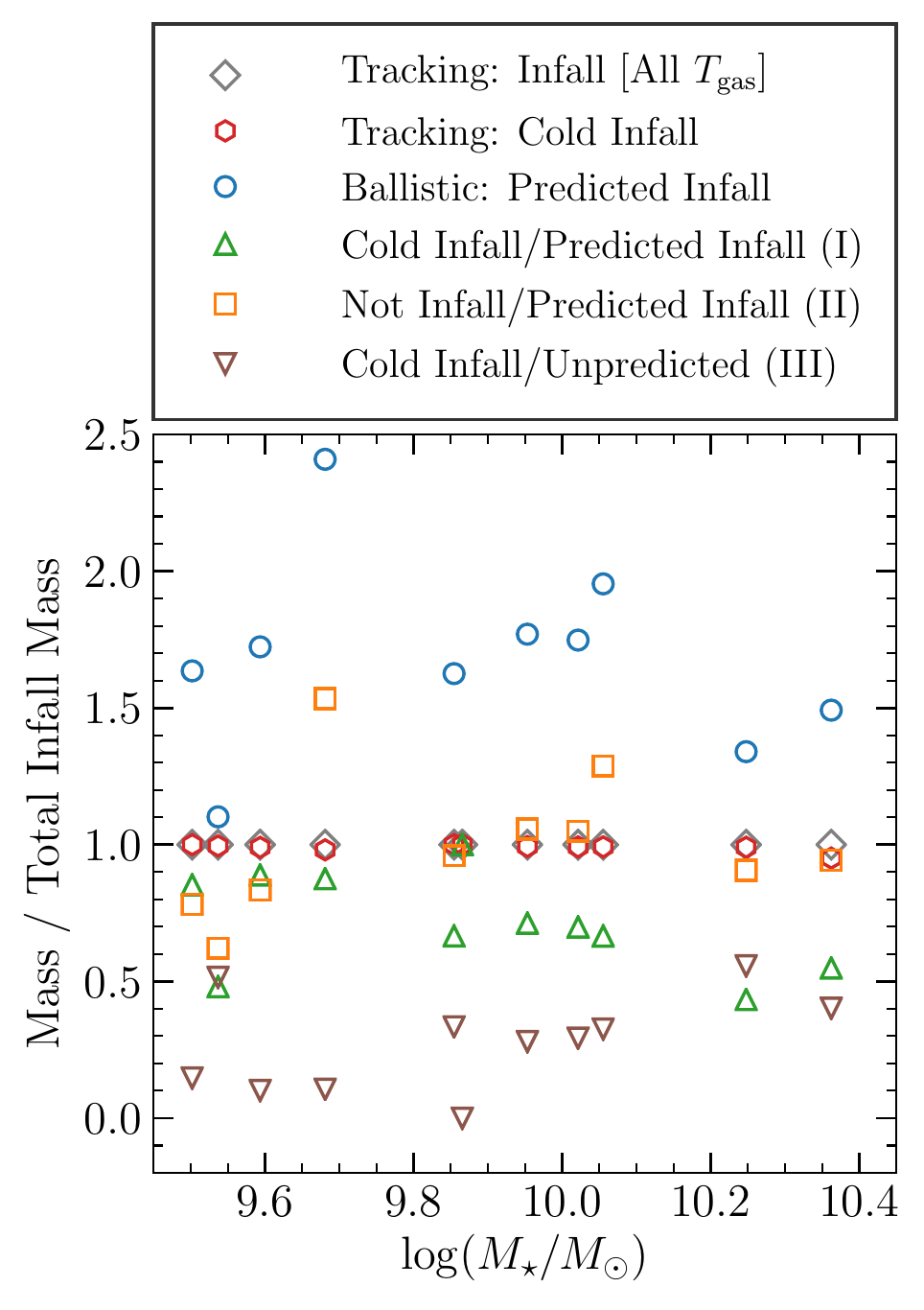}
    \caption{
        Comparison of inflowing gas mass 
        between ballistic approximation 
        and particle tracking. 
        The masses of inflowing gas of different subsets 
        (see legend) are normalized by 
        the total mass of inflowing gas at all temperatures.  
        Compared to the `true' inflow mass 
        that has reached the star-forming region,
        our ballistic approximation generally overestimates 
        the inflowing mass by a factor of 1.5 to 2.
        See Section~\ref{ssec:tracking_inflow_rate} for details.
        }
    \label{fig:tracking_2p5e5} 
\end{figure}

In our ballistic approximation,
we have assumed that inflow from hot gas is negligible
and only focused on cold gas 
with $T_\mathrm{gas} \leq 2.5\times10^5$ K. 
Figure~\ref{fig:tracking_2p5e5} justifies this assumption.  
Through particle tracking,
comparing the inflow mass with and without applying
a temperature cutoff $T_\mathrm{\mathrm{gas}} (z=0.271)$ 
(red hexagons and gray diamonds)
shows that over 95\% of the 
tracked inflowing gas (by mass) is cold.  
This implies that for our selected \EAGLE\ galaxies,
inflow from cold gas dominates.

Using the results of particle tracking and 
the prediction of the ballistic approximation, 
Figure~\ref{fig:tracking_2p5e5} compares 
the inflowing gas mass from both methods.
The inflow mass from the ballistic approximation 
generally exceeds the traced inflow mass by 
a factor of 1.5 to 2 (blue circles).  
To understand what contributes to this factor of two overestimation,
we sub-divide the gas particles into three categories:
(I)   predicted cold inflow from the ballistic approximation, 
      and the particles have reached $r_\mathrm{SFR90}$ (green triangles),
(II)  predicted cold inflow from the ballistic approximation, 
      but the particles still reside outside of 
      $r_\mathrm{SFR90}$ (orange squares), 
      and
(III) not predicted as cold inflow, 
      but the particles have reached $r_\mathrm{SFR90}$ 
      (inverted brown triangles).
While over 50\% of the predicted inflowing gas 
really reaches the star-forming region (type I), 
a significant fraction of the predicted inflow fails 
to be accreted (type II).
In contrast, 
the amount of unpredicted inflowing gas (type III) 
stays relatively low. 
Therefore, the factor of two difference in inflowing mass comes from 
the correct inflow prediction, 
as well as the 
compensation between the unpredicted inflow and 
the particles incorrectly predicted as inflow.  
As we will discuss in Section~\ref{ssec:ballistic_limitation},
this discrepancy is mainly caused by feedback.

%%%%%%%%%%%% FIG: Inflow rate: Ballistic vs Tracking (2.5e5) %%%%%%%%%%%%
%%%%%%%%%%%%%%%%%%%%%%%%%%%%%%%%%%%%%%%%%%%%%%%%

%%%%%%%%%%%%%%%%%%%%%%%%%
% Tracking: Average Mass Inflow Rate
%%%%%%%%%%%%%%%%%%%%%%%%%
\subsection{Average Mass Inflow Rate from Particle Tracking}
\label{ssec:tracking_inflow_rate}

We calculate the average mass inflow rate by summing 
the masses of identified inflow gas particles
and dividing it by the time evolved from $z=0.271$ 
to the selected snipshot.  
Table~\ref{tb:inflow_avgv_result} lists the inflow rate 
in column $\dot{M}^\mathrm{tracking}_\mathrm{in}$.

\begin{figure}[htb]
    \centering
    \includegraphics[width=0.82\linewidth]{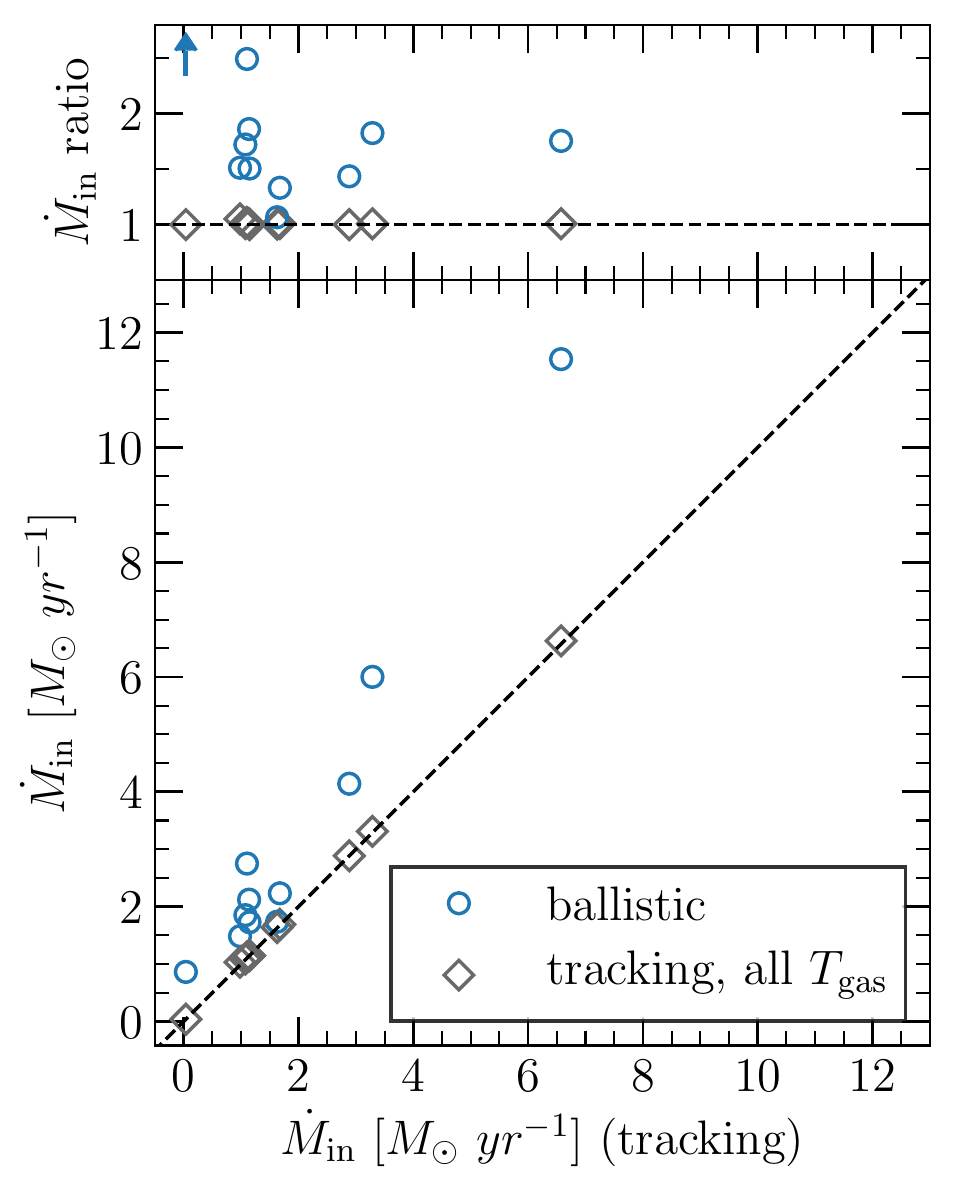}
    \caption{
        Average mass inflow rate from ballistic approximation 
        and particle tracking.
        The horizontal axis shows the tracked inflow rate from cold gas 
        ($T_\mathrm{gas}\leq 2.5\times10^5$ K) of the 11 central galaxies.  
        The cyan circles show the 
        predicted inflow rates from cold gas using 
        the ballistic approximation, 
        and the gray diamonds show the tracked inflow rates 
        from gas at all temperatures.  
        The top panel shows the ratio between each of 
        these quantities and the tracked inflow rate from cold gas.  
        The up arrow indicates that 
        the value exceeds the axis limit.
        }
    \label{fig:inflow_flux_tracking} 
\end{figure}

Figure~\ref{fig:inflow_flux_tracking} shows 
the tracked inflow rate from cold gas particles,
and we compare that with the tracked inflow rate 
from gas at all temperatures
and the predicted inflow rate 
from the ballistic approximation.  
The tracked inflow rate 
from cold gas matches that from gas at all temperatures.
This suggests that most inflowing gas is cold
(as deduced from Figure~\ref{fig:tracking_2p5e5}).  
\editR{Figure~\ref{fig:inflow_flux_tracking} also shows 
that the predicted inflow rates from the ballistic approximation 
correlate with the tracked inflow rates.  
All predicted inflow rates 
exceed the tracked inflow rates, however,}
and with the exception of two galaxies,
the overestimations are within a factor of two.  
For the galaxy that shows the largest discrepancy (ID: 48386), 
feedback has removed most of the gas from the star-forming region, 
resulting in a low average inflow rate.
We will discuss how feedback affects the average mass inflow rate in 
Section~\ref{ssec:ballistic_limitation}.

\editR{We also compare the tracked inflow rates 
to the galaxy SFRs in Figure~\ref{fig:ballistic_inflow_rate_2p5e5}
(red hexagons).
Although the inflow rates from particle tracking
are lower than those from the ballistic approximation,
Figure~\ref{fig:ballistic_inflow_rate_2p5e5} still shows that
most galaxies have comparable 
inflow rates and SFRs.  
This suggests that
even without other types of inflows, 
such as satellite accretion and hot gas accretion 
(negligible as deduced in Figure~\ref{fig:tracking_2p5e5}),
the cold gas infall alone can account for most of the fuel needed 
to sustain the star formation in galaxies.}

%%%%%%%%%%%% FIG: Compare Mdot (from ballistic and tracking) %%%%%%%%%%%%

%%%%%%%%%%%%%%%%%%%%%%%%%
% TRACKING: Infall Speed
%%%%%%%%%%%%%%%%%%%%%%%%%
\subsection{Mass-weighted Average Inflow Speed from Particle Tracking}
\label{ssec:tracking_inflow_speed}

\begin{figure}[htb]
    \centering
    \includegraphics[width=1.0\linewidth]{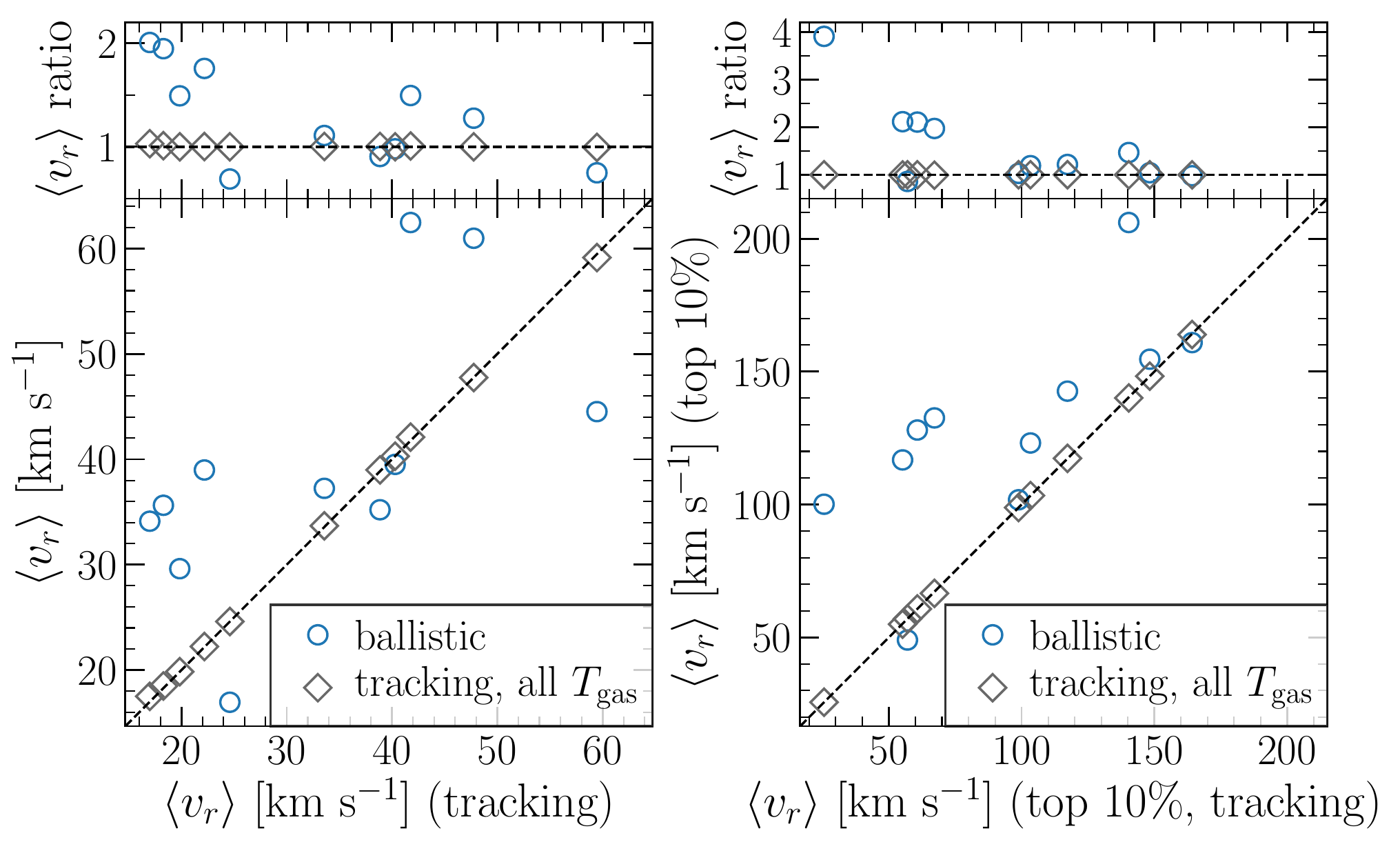}
    \caption{
        Mass-weighted average inflow speed from ballistic approximation 
        and particle tracking.  
        The left panel shows the mass-weighted average inflow speed, 
        and the right panel shows the 
        mass-weighted average calculated from 
        10\% of the particles by mass that
        have the highest particle inflow speeds.
        Each horizontal axis shows the average inflow speed of the cold gas 
        ($T_\mathrm{gas}\leq 2.5\times10^5$ K) in particle tracking.  
        The cyan circles show the mass-weighted average 
        from cold inflowing gas using 
        the ballistic approximation, 
        and the gray diamonds show the mass-weighted average 
        of the tracked inflow gas at all temperatures.  
        The top panels show the ratio between each of these quantities to the 
        mass-weighted average inflow speed of the tracked cold gas 
        (i.e., the quantity plotted on the horizontal axis).
        }
    \label{fig:compare_inflow_speed} 
\end{figure}

We calculate the average inflow speed
as in Section~\ref{ssec:ballistic_inflow_speed}, 
but with two differences.
First, instead of assuming the inflowing gas particles 
halt at $r_\mathrm{SFR90}$, 
we calculate $\Delta r$ as the change in radial distance 
relative to the galaxy center.
Second, we divide $\Delta r$ by the time evolved from $z=0.271$ 
to the selected snipshot, instead of the galaxy rotation period.  
In Table~\ref{tb:inflow_avgv_result}, 
$\langle v_r \rangle_\mathrm{in}^\mathrm{tracking}$ 
shows the mass-weighted average inflow speed, 
and $\langle v_r \rangle_\mathrm{in, top 10\%}^\mathrm{tracking}$ 
shows the mass-weighted average from only the 10\% particles by mass
that have the highest inflow speeds.
Figure~\ref{fig:compare_inflow_speed} shows these two sets of  
average inflow speeds calculated from the cold gas, 
and compare them with the predicted inflow speeds 
from the ballistic approximation.  

The average inflow speeds calculated from particle tracking 
share similar characteristics as those from the ballistic approximation: 
(1) low mass-weighted average inflow speed of $\lesssim$60\kms,
(2) large range of individual particle inflow speeds 
with a `high speed tail', and
(3) if we only consider 10\% of the particles by mass with 
    the highest individual inflow speeds, 
    then the mass-weighted average 
    ($\langle v_r \rangle_\mathrm{in, top 10\%}^\mathrm{tracking}$) 
    often exceeds 100\kms.
In addition, 
for both sets of average inflow speeds,
i.e., either from all cold gas or
only cold gas particles at the high end of inflow speeds, 
the ballistic approximation overestimates the average inflow speeds
for most galaxies.  
The differences in inflow speeds from 
the ballistic approximation and particle tracking 
mostly stay within a factor of two.  
The only exception is the galaxy (ID: 48386) with feedback 
removing most of the gas from the star-forming region 
(see Section~\ref{ssec:tracking_inflow_rate} and 
\ref{ssec:ballistic_limitation}).

\editR{However, 
unlike the mass inflow rate 
that shows a correlation
between the prediction from the ballistic approximation 
and the calculation from particle tracking, 
inflow speeds do not reveal the same characteristic.
While one possible explanation is the mismatch 
of inflowing particles identified by the two methods,
another reason is due to the assumption of the ballistic approximation.
When we calculate the average radial inflow speed 
for the ballistic approximation,
because the approximation breaks down 
when a particle interacts with the dense clouds in the star-forming region,  
we assume the particle halts when it reaches the star-forming region.
However, in reality, 
where the particle ends up depends on the density and 
the distribution of cold gas within the star-forming region,
both of which the ballistic approximation ignores.
Therefore, 
even though the average radial inflow speeds from
ballistic approximation and particle tracking
agree within a factor of two,
the lack of correlation suggests 
the ballistic approximation does not provide 
a satisfactory estimate on the inflow speed.
Hence, this suggests that
inflow speed estimation, 
and generally, inflow properties reconstruction,
require tracking particles at series of simulation output
with higher time candence, 
preferably of the order of $\sim100$ Myr,
i.e., around (or shorter than) the rotation period of a galaxy.
}

%%%%%%%%%%%% FIG: Compare Inflow speeds (from ballistic and tracking) %%%%%%%%%%%%

%%%%%%%%%%%%%%%%%%%%%%%%%
% Tracking: Particles that become stars
%%%%%%%%%%%%%%%%%%%%%%%%%
\subsection{Star Formation from Gas Accreted onto the Inner Galaxy}
\label{ssec:become_stars}

Gas accreted onto galaxies eventually forms stars.  
For each galaxy, we identify the gas at $z=0.271$ that 
has turned into stars at the selected snipshot after a rotation period.
Although most of these particles reside within $r_\mathrm{SFR90}(z=0.271)$ 
at both times, 
on average, around 10\% of the new stars are formed 
from the newly accreted gas
that originally resided outside of $r_\mathrm{SFR90}$.

\begin{figure}[htb]
    \centering
    \includegraphics[width=0.9\linewidth]{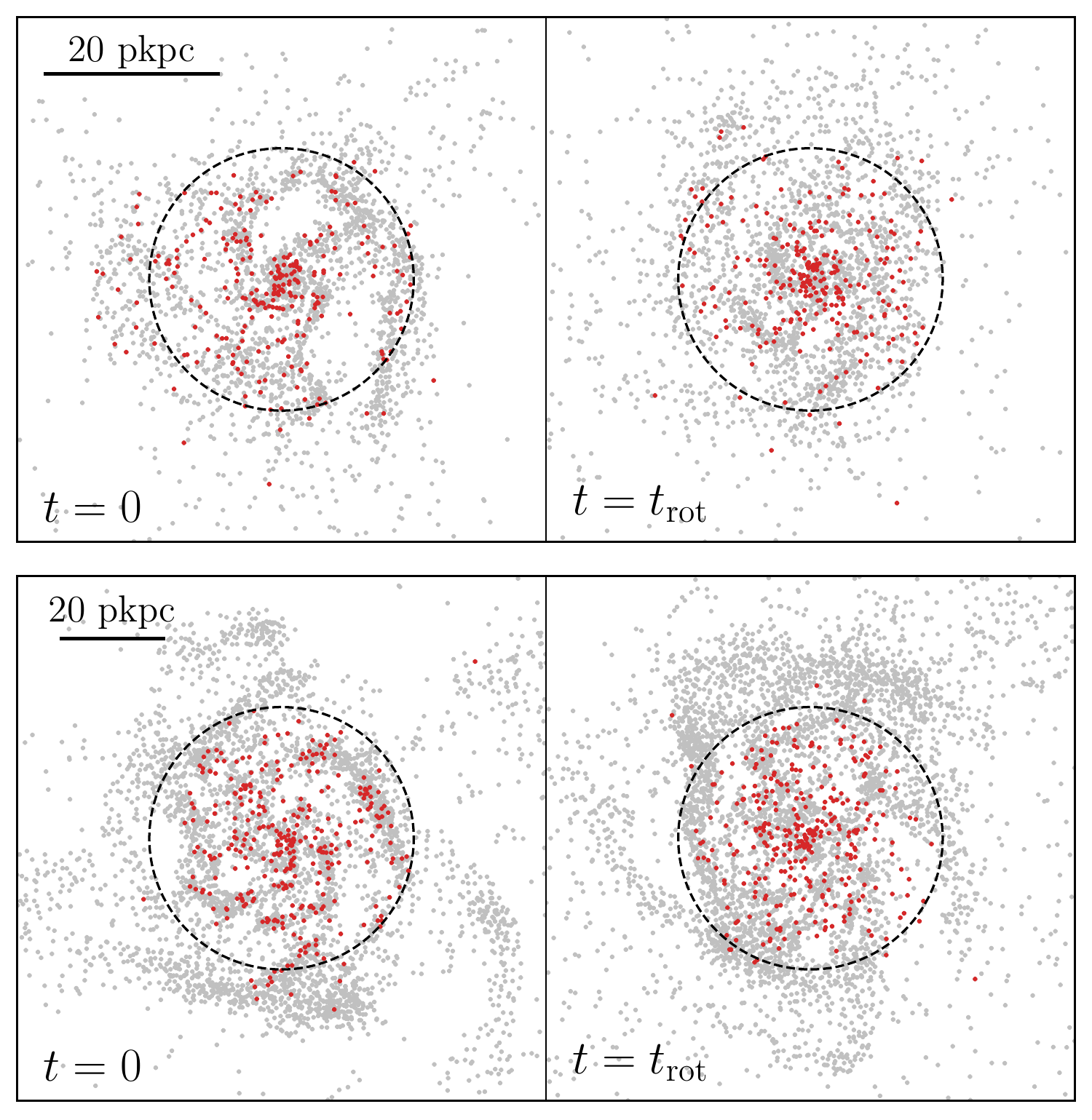}
    \caption{
        Cold gas moves towards inner galaxy to form stars.  
        By projecting cold gas particles onto a 2D plane, 
        each row shows a different galaxy: 
        43166 \textit{(top)} and 37448 \textit{(bottom)}.
        The left and right column show the particle distribution 
        before and after a rotation period respectively,
        i.e., $t=0$ at $z=0.271$, and $t=t_\mathrm{rot}$.  
        Each gray point represents a cold gas particle at both times. 
        Each red point respresents a cold gas particle 
        at $z=0.271$ \textit{(left)}
        that has turned into a star after a rotation period \textit{(right)}.  
        Each dashed circle represents $r_\mathrm{SFR90}$ of each galaxy, 
        defined at $z=0.271$.
        }
    \label{fig:tracking_bePT45_example} 
\end{figure}

Figure~\ref{fig:tracking_bePT45_example} shows 
two galaxy examples, 
and the red points illustrate the gas particles 
that have turned into stars after a rotation period.  
Most of these particles originally resided along the spiral arms.
After a rotation period, 
these particles become more concentrated near each galaxy center,
where some of these particles are newly accreted from 
outside of  $r_\mathrm{SFR90}$.  
This demonstrates 
the fueling of star-forming activity from newly accreted gas.
Moreover, some stars are formed outside the star-forming radius, 
which indicates the growth of the star-forming disk and the stellar disk.  
This qualitatively supports the picture of inside-out galaxy growth. 

%%%%%%%%%%%% FIG: Example of where particles turn into stars %%%%%%%%%%%%
%%%%%%%%%%%%%%%%%%%%%%%%%%%%%%%%%%%%%%%%%%%%%%%%

%%%%%%%%%%%%%%%%%%%%%%%%%
% Tracking: Limitation of Ballistic Approximation
%%%%%%%%%%%%%%%%%%%%%%%%%
\subsection{Limitations of Ballistic Approximation}
\label{ssec:ballistic_limitation}

Using the ballistic approximation, 
both predicted inflow rate and mass-weighted average velocity 
generally agree with those from particle tracking to 
within a factor of two.  
In addition to the breakdown conditions of 
the ballistic approximation
stated in Section~\ref{ssec:ballistic}, 
here we describe another cause of discrepancies 
between the two sets of inflow properties.  
We also discuss two caveats of comparing 
predictions from ballistic approximation to 
the outcome from the hydrodynamical calculations in \EAGLE.

By following the gas particles in continuous timesteps, 
i.e., at multiple times for each galaxy,
we find that feedback plays an important role.  
The ballistic approximation
ignores the effect of feedback 
and predicts each particle trajectory by
considering only gravity.  
But within a rotation period, feedback and winds 
expel some recently accreted gas particles to large radii.
In the most extreme case (galaxy ID: 48386), 
not only does the feedback expel the newly accreted gas,
but the feedback also disrupts the star-forming region 
and disperses all of the cold gas.

\editR{The temperature increase 
of the cold gas particles also provides evidence 
for feedback;
\EAGLE\ implements the stellar feedback by
stochastically heating particles to high temperature, 
known as the stochastic thermal feedback 
\editW{\citep{DallaVecchia2008,DallaVecchia2012,Schaye2015}}.
We focus on the change of temperature 
for particles that are predicted to reach the galaxy
within a rotation period (by the ballistic approximation)
but failed to do so (according to particle tracking).
For the 11 central galaxies, 
around 50\% of the particles that fail to be accreted
have at least doubled their temperatures after a rotation period;
the fraction varies among galaxies, from around 30\% to 75\%.
For the most extreme case (galaxy ID: 48386), 
the particles have increased their temperatures by 
over 100 times to $\sim10^7$ K, 
indicating that the particles are strongly affected by feedback.}

Because we identify inflowing gas
using the location of particles after a rotation period, 
we have tracked the net inflow as
the difference between the newly accreted gas 
and the gas expelled from the galaxy due to outflows.  
As some particles have once reached 
the star-forming region but have been expelled at later times
\editR{due to feedback and winds},
these particles explain the non-negligible fraction 
of predicted inflow particles that fail to be in 
the star-forming region after a rotation period
(orange squares in Figure~\ref{fig:tracking_2p5e5}).  
In other words,
whether or not a particle can accrete onto a galaxy
and stay at the star-forming region without being expelled
depends strongly on feedback.
This phenomenon 
also qualitatively agrees with the results 
in \citet{Nelson2015} with \AREPO\ and 
\citet{Correa2018b} with \EAGLE; 
at $z\leq 5$ and $z\leq2$, feedback suppresses 
the accretion rates onto galaxies with
halo masses below $10^{12}$\msun.

We emphasize that the comparison between the
ballistic approximation and particle tracking
\editR{only explores whether or how well the former 
resembles the hydrodynamical calculations in \EAGLE.}   
This comparison, however, has two caveats: 
there exist at least two factors that 
will result in different inflow properties 
deduced from different hydrodynamical simulations.
First, 
because whether the gas can accrete onto galaxies
is sensitive to feedback, 
this means the 
method of feedback implementation 
and the feedback strength 
will alter the inflow properties deduced from 
hydrodynamical simulations.  
Second, 
the inflow properties may be code-dependent.  
\citet{Nelson2013} (also see \citealt{Nelson2015})
study the cosmological gas accretion
of $\simeq 10^{11}\ M_\odot$ halos at $z=2$.
They find that compared to the SPH code \GADGET3,
\AREPO, a moving mesh code,
produces an order of magnitude higher in hot accretion rate,
but a factor of two lower in cold accretion rate.
Therefore, this demonstrates that 
the properties of inflowing gas
are affected by the 
nature of the numerical approach.

%%%%%%%%%%%%%%%%%%%%%%%%%%%%%%%%%%%%%%%%
% Observations
%%%%%%%%%%%%%%%%%%%%%%%%%%%%%%%%%%%%%%%%
\section{Interpreting Cold Gas Kinematics from Observations}
\label{sec:observations}

Previous sections have demonstrated 
the ubiquitousness of gas accretion onto the \EAGLE\ galaxies.  
In contrast, observations rarely detect inflowing gas directly, 
as down-the-barrel galaxy observations 
can identify inflowing gas only if 
its Doppler shift can be distinguished from the ISM velocity dispersion.  
As a result, 
quasar sightline observations have provided 
the best observational probe of gas inflows,
and these sightline observations have advanced our understanding 
of the gas kinematics of the CGM.

Quasar sightline observations have revealed important 
results regarding the cold gas kinematics of the CGM.
First, through quasar sightlines that probe the CGM 
near galaxy major axes, 
observations often show that the cold CGM ($\sim$$10^4$ K), 
which is traced by low-ionization-state ions such as \mgII, 
corotates with the galaxy disks.  
Combining galaxy rotation curves and \mgII\ absorption profiles 
from quasar spectra,
studies find cold absorbing gas that corotates with the disks 
at various redshifts:
from low redshifts of $z\sim0.1$ \citep{Kacprzak2011ApJ} 
and $z\sim0.2$ \citep{Ho2017,Martin2019}, 
to intermediate redshifts of $z~\sim0.5$ \citep{Kacprzak2010,Steidel2002}.
While similar types of data rarely exist at higher redshifts,   
there exists evidence suggesting that the cold gas shows 
similar kinematic properties as the lower redshift counterparts.
Using other low-ionization-state ions (e.g., \znII, \crII), 
\citet{Bouche2013,Bouche2016} study two galaxies at  
$z = 2.3283$ and $z=0.9096$ respectively
and detect corotating cold gas out to 26 pkpc and 12 pkpc.
Second, for the observed corotating gas, 
the sightline impact parameters range 
from around 10 pkpc to even $\sim$100 pkpc.
\editR{This exceeds the typical gas disk sizes
of nearby galaxies
measured from deep \hI\ 21-cm observations,\footnote{
    \editR{The observed \hI\ disk size (diameter) 
    depends on the depth of the observation;
    deeper observations measure \hI\ emission down to 
    a lower column density limit, 
    resulting in a larger \hI\ disk size.
    As an example, for NGC 5023,
    the \hI\ diameter is 19 kpc 
    at a column density limit of $10^{20}$ \percm2,
    but the disk extends to about 27 kpc in diameter 
    at $2\times10^{19}$ \percm2 \citep{Kamphuis2013}.}
    }
e.g., 
$\sim20$ pkpc in radius 
down to \hI\ column density of $\sim5\times10^{19}$ \percm2
for NGC 5023 and UGC 2082
from the HALOGAS survey \citep{Heald2011,Kamphuis2013},
and $\sim30$ pkpc in radius for NGC 891
down to $10^{19}$ \percm2 \citep{Sancisi1979,Oosterloo2007}}.
Third, the low-ion absorption typically spans over 100\kms.
Despite the corotation, 
a thin rotating disk fails to reproduce the broad velocity range.
This problem demonstrates the need of additional components 
to describe the observed cold gas kinematics.

Inspired by these observations, 
we now observe the cold CGM kinematics in \EAGLE\ 
using mock quasar sightlines.
In this pilot study,
limited by the galaxy sample size and resolution, 
our following discussion aims to encourage future work
and focuses on one edge-on ($i=90$\deg) galaxy.  
We will show that with this single galaxy, 
we can detect cold gas kinematics with 
characteristics similar to those detected in
real quasar sightline observations.  
Therefore, 
this case study serves as motivation for future studies
to `observe' the cold CGM using
\EAGLE\ simulations with larger box sizes and a higher resolution,
and to include the analysis of mock absorption line spectra.

%%%%%%%%%%%%FIG: Example Particle Plots %%%%%%%%%%%%
\begin{figure}[htb]
    \centering
    \includegraphics[width=0.86\linewidth]{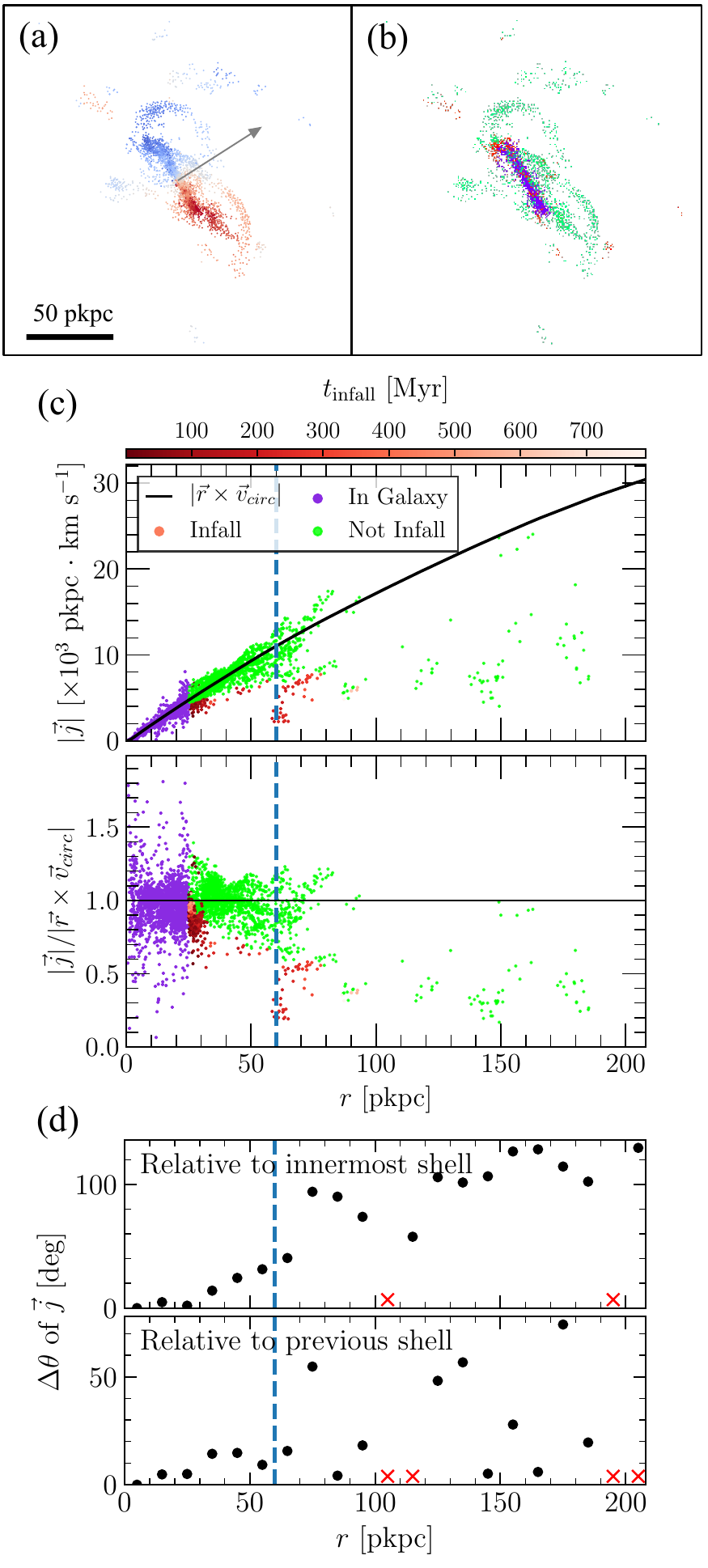}
    \caption{
        Distribution of cold gas particles 
        ($T_\mathrm{gas} \leq 3\times10^4$ K) 
        of a disk galaxy (ID: 37448)
        and the particle angular momenta.  
        Panels (a) and (b) 
        show cold gas particles projected onto 2D planes.
        Each particle is color-coded as in  
        the top and middle row, respectively, of
        Figure~\ref{fig:ppj_jr_example_2p5e5}.
        The gray arrow in panel (a) shows the 
        direction of the net angular momentum of cold gas at the 
        innermost 10-pkpc disk.
        Similar to the bottom row of Figure~\ref{fig:ppj_jr_example_2p5e5}, 
        panel (c) shows the specific angular momentum 
        of each cold gas particle.
        By first finding the net angular momentum vector of each 
        concentric 10-pkpc shell, 
        panel (d) shows the angle between the net vector at 
        each shell ($i$-th shell), and that at the innermost shell, 
        or at the previous ($(i-1)$-th) shell.
        Red crosses indicate cold gas particles 
        do not exist at the $i$-th shell 
        (or at the $(i-1)$-th shell).
        The verticle dashed lines in panel (c) and (d) mark the
        radius of the cold gas disk of 60 pkpc.
        }
    \label{fig:ppj_jr_example_3e4} 
\end{figure}
%%%%%%%%%%%%%%%%%%%%%%%%%%%%%%%%%%%%%%%%%%%%%%%%

Figure~\ref{fig:ppj_jr_example_3e4} shows 
the example galaxy to be studied in this Section.
This Figure is similar to Figure~\ref{fig:ppj_jr_example_2p5e5},  
but in order to compare with cold gas traced by 
low-ionization-state ions, 
here we only show cold gas particles with 
$T_\mathrm{gas}\leq3\times10^4$ K.  
Not only do the particles form a structure with 
a redshifted and a blueshifted side, which thereby indicates rotation,
but the more extended particle distribution 
in the radial than in the $Z$-direction 
implies the rotating structure morphologically 
represents a disk (panel (a)).
Furthermore, out to around 60 pkpc, 
the majority of the cold gas particles 
have angular momenta comparable to those 
required to be on circular orbits
(panel (c)).  
This suggests a rotating disk radius of around 60 pkpc.  
\editR{
Therefore, 
from both morphological and kinematical perspectives,  
this selected galaxy clearly has a giant, rotating cold gas disk.
}

\editR{Panel (d) shows that this extended gas disk is warped.}  
By finding the net angular momentum vector
at each concentric 10-pkpc shell, 
we calculate the angle between the net vector at the $i$-th shell
and that at the innermost shell, or at the $(i-1)$-th shell.  
The small but non-zero variations of both angles 
out to 60 pkpc implies that 
the disk plane changes orientation with radius, 
i.e., the gas disk is warped.  
\editR{
Warped gas disks are also commonly found in 
hydrodynamical simulations, 
a result from cooling gas accreting onto galaxies
and aligning its angular momentum vector
with the central disks
(e.g., \citealt{Roskar2010,Stewart2011ApJ}).
In particular with the \EAGLE\ simulations,
\citet{Stevens2017} show that
the angular momentum directions of both hot and cooling gas 
differ from that of the cooled gas by tens of degrees.  
This means that the gas particles precess as they cool, 
and their angular momenta
align with the pre-existing cold gas in the galaxy.  
This leads to the formation of an extended, warped gas disk, 
similar to our example galaxy 
in Figure~\ref{fig:ppj_jr_example_3e4}.
}

In this section,
first we describe how we generate the mock sightlines 
in Section~\ref{ssec:sightline_obs}.
In Section~\ref{ssec:vlos_offset}, 
by measuring the mean LOS velocities along sightlines,
we explore which sightlines can detect
cold gas that corotates with the galaxy disk.
Then in Section~\ref{ssec:vlos_range}, 
we study how LOS velocity varies along sightlines,
and we discuss the galaxy structures 
that can also produce velocity ranges comparable to those 
measured in real quasar sightline observations.

%%%%%%%%%%%%%%%%%%%%%%%%%
% Mock Sightline Observations
%%%%%%%%%%%%%%%%%%%%%%%%%
\subsection{Creating Mock Quasar Sightlines}
\label{ssec:sightline_obs}

\begin{figure}[htb]
    \centering
    \includegraphics[width=0.75\linewidth]{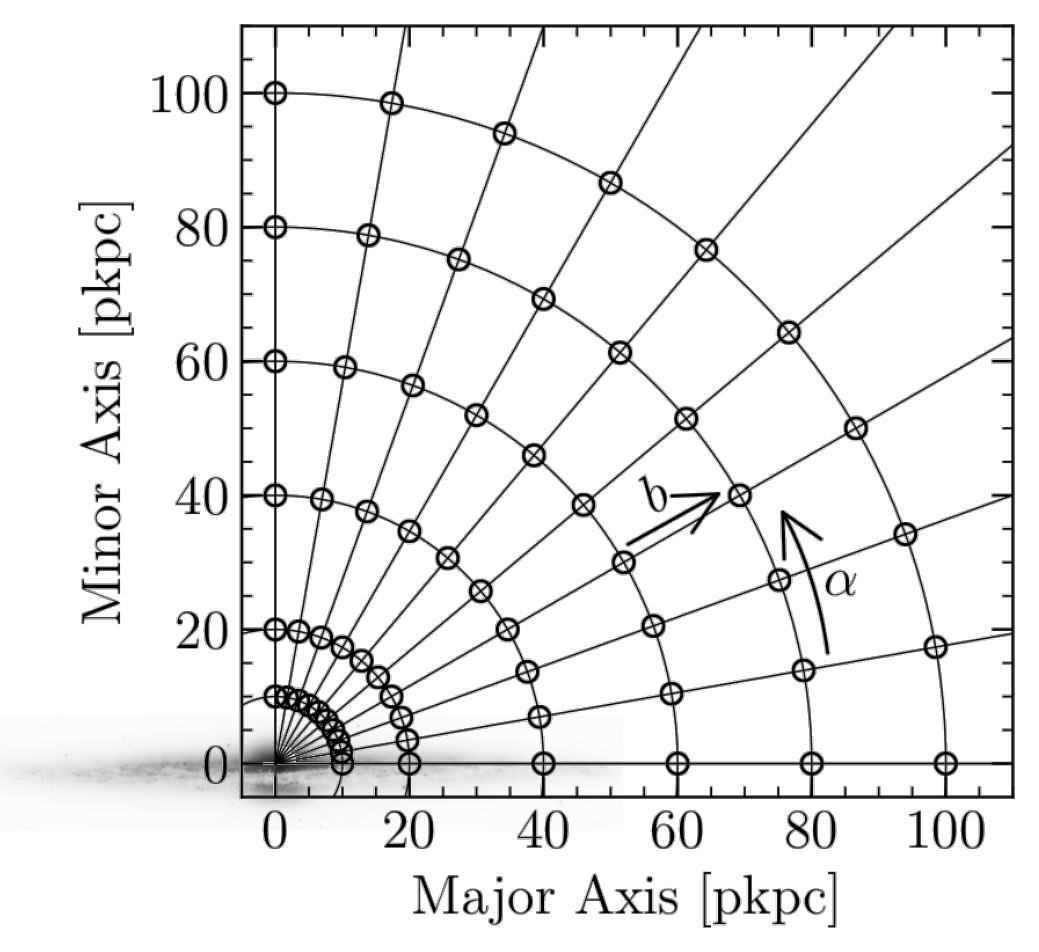}
    \caption{
        Grid of mock sightlines relative to galaxy major axis.
        The origin of the coordinate grid 
        represents the center of the galaxy. 
        Each circle represents the position of a sightline.
        }
    \label{fig:los_grid} 
\end{figure}

To create the mock quasar sightlines, 
first we define the disk plane of the galaxy.
Within the innermost 10-pkpc of the galaxy, 
we calculate the net angular momentum vector of the cold gas
and use this vector to define the disk plane.
For a fixed galaxy inclination angle $i$ (e.g., $i = 90$\deg), 
we generate sightlines with 
impact parameter $b$ between 10 pkpc and 100 pkpc 
and azimuthal angle\footnote{
    Measured from the galaxy center, 
    azimuthal angle $\alpha$ is the angular separation 
    of the sightline from the galaxy major axis.  
    A sightline at $\alpha = 0$\deg\ lies on the major axis, 
    whereas a sightline at $\alpha = 90$\deg\ lies on the minor axis.
    }
$\alpha$ from 0\deg\ to 90\deg. 
Figure~\ref{fig:los_grid} illustrates the grid 
of sightlines relative to the galaxy major axis.  
Then for each combination of $i$, $b$, and $\alpha$, 
we produce 12 LOSs by changing the location of the observer;
each sightline runs along a different direction 
relatively to the simulation box,
but as viewed from the `relocated' observer, 
the sightlines still intersect the disk 
at the given $i$, $b$ and $\alpha$.
Moreover, because in this work,
we are interested in the kinematic signatures of cold gas within the halo, 
we create each LOS to extend only to the 
galaxy virial radius $r_\mathrm{vir}$.

Analogous to observations, 
which trace cold gas using low ions (e.g., \mgII, \siII, \feII), 
we redefine the cold gas temperature cutoff as $3\times10^4$ K
(see Section~\ref{ssec:define_cold_gas}).  
We use \texttt{yt} \citep{Turk2011} to convert 
the cold gas particle fields to grid-based fluid quantities.  
We extract the cold gas density and velocity
at cells intersected by each sightline.  
Then we calculate the mean LOS velocity, 
and we explore how the LOS velocity changes along sightlines
in Section~\ref{ssec:vlos_offset} and \ref{ssec:vlos_range} respectively.

%%%%%%%%%%%% FIG: LOS GRID %%%%%%%%%%%%
%%%%%%%%%%%%%%%%%%%%%%%%%%%%%%%%%%%%

%%%%%%%%%%%%%%%%%%%%%%%%%
% Velocity Offset/Corotation
%%%%%%%%%%%%%%%%%%%%%%%%%
\subsection{CGM Corotation with Galactic Disk}
\label{ssec:vlos_offset}

We pose the question of 
which sightlines can detect
cold gas that corotates with the inner galaxy disk.
Using the mock sightlines that probe the edge-on galaxy, 
for each of the 12 LOS at different $b$ and $\alpha$, 
we calculate the mean LOS velocity (column density weighted).  
If this mean LOS velocity has the same sign as the disk rotation
and is at least 10\kms\ away 
from the galaxy systemic velocity, 
then we classify the LOS as detecting corotating gas.
The latter criterion prevents misassigning cold gas
that moves randomly within thermal and turbulent velocity
as corotating.

\begin{figure}[htb]
    \centering
    \includegraphics[width=0.9\linewidth]{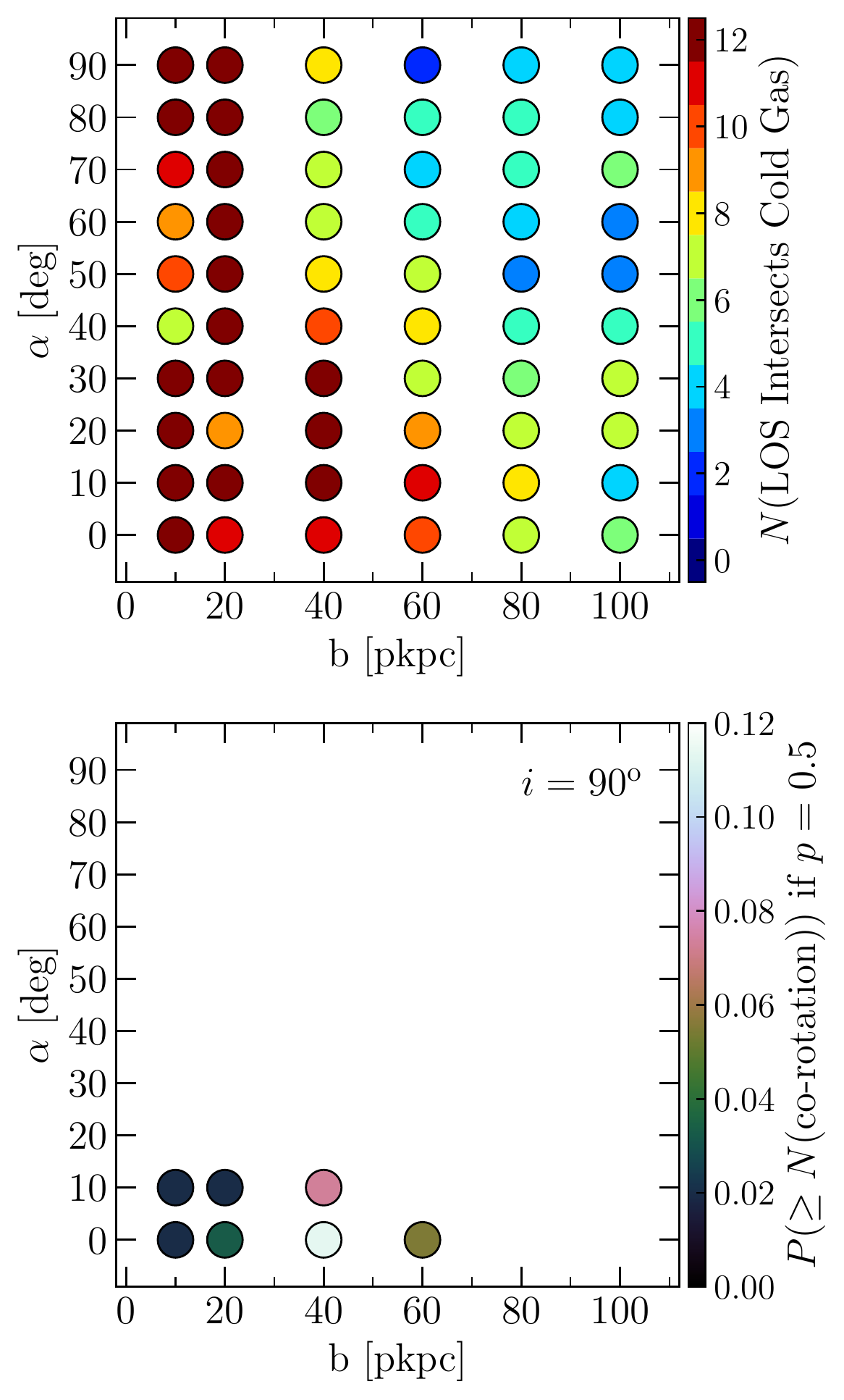}
    \caption{
        Locations of sightlines that detect 
        corotating cold gas around an edge-on ($i = 90$\deg)
        galaxy (galaxy ID: 37448).
        \textit{(top)} 
        Number of LOS (out of 12)
        that intersects cold gas 
        with $T_\mathrm{gas} \leq 3 \times10^4$ K.  
        Fewer LOS at large $b$ and high $\alpha$ 
        intersect cold gas within the galaxy virial radius.
        \textit{(bottom)}
        This panel only shows the sightlines 
        along which we rule out random gas motion
        at $1\sigma$-level of confidence.    
        Sightlines at $\alpha=0$\deg\ and 10\deg\ 
        detect corotating cold gas out to 60 pkpc and 40 pkpc 
        respectively.
        In particular, 
        \textit{if} the gas had moved randomly, then 
        the probabilities of detecting $\geq N_\mathrm{cr}$ 
        LOS with corotating cold gas
        at $\alpha = 0$\deg\ and 10\deg\ within $b=20$ pkpc 
        would lie below 5\% (indicated by the colors).  
        This implies that these sightlines have detected corotating cold gas.
        }
    \label{fig:los_corotation} 
\end{figure}

Before discussing the detection of cold corotating gas,
we note that not all LOS intersect cold gas.  
The top panel of Figure~\ref{fig:los_corotation} 
shows that among the 12 LOS per $b$--$\alpha$ combination,
the number of LOS that intersects cold gas 
decreases with increasing $b$ and $\alpha$.  
This can be explained by the decreasing 
cold gas density when the radial separation 
from the galaxy center increases 
(e.g., see the cold gas distribution in 
Figure~\ref{fig:ppj_jr_example_3e4}(a)).

%%%%%%%%%%%% FIG: Corotation %%%%%%%%%%%%
%%%%%%%%%%%%%%%%%%%%%%%%%%%%%%%%%%%%%%%%%%%%%%%%%%%%%%%%%%%%

Using the sightlines that have intersected cold gas, 
we explore what ranges of $b$ and $\alpha$ can
detect corotating gas.  
At each $b$--$\alpha$ combination, 
first we find the number of LOS 
that intersects cold gas $N_\mathrm{cold}$ 
and the number of LOS that 
detects corotating cold gas $N_\mathrm{cr}$.  
With binomial statistics \citep{Gehrels1986},
we calculate the $1\sigma$ lower limit 
of the rate of detecting corotating cold gas $p_{cr,1l}$. 
The bottom panel of Figure~\ref{fig:los_corotation} 
isolates the sightlines in the  $b$--$\alpha$ plane
with $p_{cr,1l} > 0.5$; within
$1\sigma$-level of confidence,
it is unlikely that these sightlines have 
detected randomly moving cold gas.  
This is because with random gas motion, 
sightlines will detect corotation at half of the times, 
and the probability of detecting corotating gas 
at one sightline will be $p = 0.5$.  
This probability, however, 
is excluded with $1\sigma$-level of confidence
if $p_{cr,1l} > 0.5$.
Then, among these sightlines in the $b$--$\alpha$ plane, 
we also ask the following question: 
\textit{if} the gas moves randomly,
what is the probability of having 
$\geq N_\mathrm{cr}$ LOS that detect
corotating gas among the $N_\mathrm{cold}$ LOS.  
We show this probability using the color scale 
in the bottom panel of Figure~\ref{fig:los_corotation}.  
Among the selected $b$--$\alpha$ locations,
the chances of detecting $\geq N_\mathrm{cr}$ sightlines 
with corotating gas lie below 12\%.  
In particular, at $b=10$ pkpc  and 20 pkpc, 
the probabilities at both $\alpha = 0$\deg\ and 10\deg\ lie below 5\%.
Thus, the calculation suggests that sightlines 
shown in the lower panel of Figure~\ref{fig:los_corotation}, 
i.e., sightlines at $\alpha=0$\deg\ and 10\deg\ 
out to 60 pkpc and 40 pkpc, respectively, 
have unlikely intersected randomly moving gas.
In other words, these sightlines can detect 
corotating cold gas.

Our edge-on galaxy example demonstrates that
quasar sightlines are more likely to detect 
corotating cold gas near the galaxy disk plane:
at sightlines with $b \leq 60$ pkpc 
and low $\alpha$ of $\leq10$\deg.  
This can be naturally explained by sightlines 
intersecting the rotating gas disk 
as shown in Figure~\ref{fig:ppj_jr_example_3e4}(a).
Since real quasar sightline observations 
often detect corotating cold gas 
along sightlines near the galaxy major axes, 
our example suggests that these sightlines 
have likely intersected cold gas disks 
that extend beyond the optical disks.

\begin{figure*}[htb]
    \centering
    \includegraphics[width=1.0\linewidth]{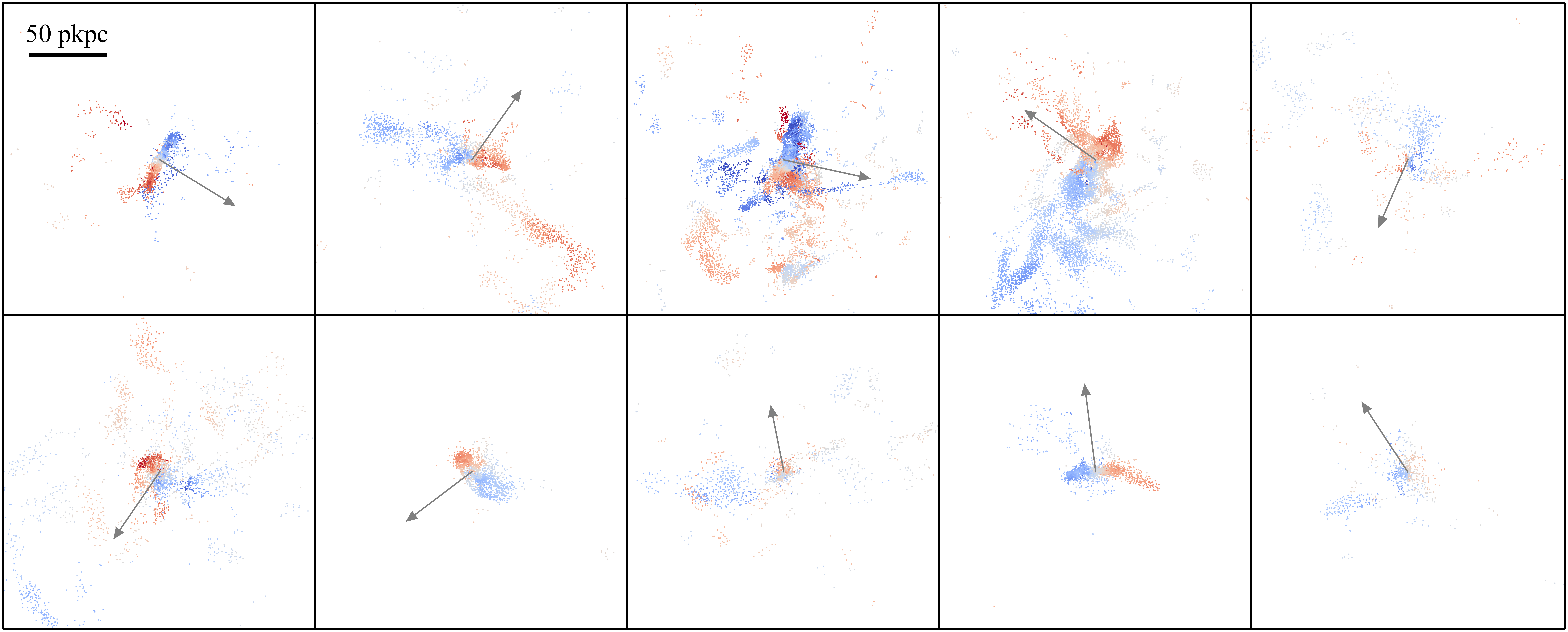}
    \caption{
        Projected LOS velocities of cold gas particles 
        with $T_\mathrm{gas} \leq 3\times10^4$ K.
        Each panel projects the cold gas particles 
        of each central galaxy onto a 2D plane. 
        Every particle is color-coded by its projected LOS velocity. 
        Red and blue colors represent 
        redshifted and blueshifted cold gas particles respectively, 
        and we use the same velocity scale as in the top row of
        Figure~\ref{fig:ppj_jr_example_2p5e5}.  
        The gray arrow shows the direction of the net angular momentum 
        of the cold gas, 
        which is measured within an aperture of 10-pkpc 
        from the galaxy center.
        More than half of the galaxies show signs of rotation.
        Each panel has the same spatial scale of 200 pkpc $\times$ 200 pkpc.
        }
    \label{fig:particle_pjplot_3e4_vlos}
\end{figure*}

While we use one \EAGLE\ galaxy to demonstrate
the possibility of detecting corotating cold gas 
using mock quasar sightlines, 
the presence of rotating structures,
which can give rise to corotating gas detection, 
is not unique to this galaxy.
Figure~\ref{fig:particle_pjplot_3e4_vlos} 
shows the cold gas distribution of the remaining 10 central galaxies, 
and we color each particle by its projected LOS velocity.  
Although the spatial distribution of particles indicates that
not all galaxies have thin-disk-like morphologies,
at least half of the galaxies have cold gas structures 
that show signs of rotation,
i.e., an approaching side and a receding side.  
\editR{Rotating gas structures are not only 
common among our selected galaxies.  
Simulated galaxies with halo mass $\gtrsim 10^{11} M_\odot$
from the \texttt{FIRE} project also show clear signs of rotation,
even though the gas distribution is not morphologically `disky'
\citep{ElBadry2018}.} 
Hence, sightlines through \editR{all} these galaxies 
may also detect corotating cold gas as in our example, 
and this may imply that the corotation detection 
is not a rare phenomenon.

%%%%%%%%%%%% FIG: Example Particle Plots (color by vlos) %%%%%%%%%%%%
%%%%%%%%%%%%%%%%%%%%%%%%%%%%%%%%%%%%%%%%%%%%%%%%

%%%%%%%%%%%%%%%%%%%%%%%%%
% Velocity Range
%%%%%%%%%%%%%%%%%%%%%%%%%
\subsection{Velocity Range Along LOS}
\label{ssec:vlos_range}

Quasar sightline observations often detect broad LOS velocity ranges
that cannot be explained by a thin rotating disk model.  
Therefore, using our example galaxy that has an extended rotating disk,
we investigate the velocity ranges 
that this galaxy can produce.\footnote{
    In spectroscopic observations, 
    whether or not the cold gas can be detected 
    and the measured velocity ranges 
    depend on 
    the equivalenth widths of the absorption systems
    and the spectroscopic sensitivity.  
    Our discussion focuses on the velocity ranges
    that can be produced by the cold gas, 
    and we implicitly assume that
    all cold gas can be detected.  
    } 
\editR{
By examining how LOS velocity varies along different sightlines, 
we discuss the resultant velocity ranges 
that can possibly be created by different gas structures.
}

\editR{We choose individual sightlines 
that intersect different components of the cold gas structure:
(1) the thin disk: at small $b$ and along the galaxy major axis
    ($\alpha=0$\deg) ,
(2) the thick disk: at small $b$ and at least several kpc 
    above the disk plane (e.g., $\alpha=20$\deg), and  
(3) a gas stream. 
Specifically for our edge-on galaxy example, 
a sightline that intersects the disk midplane 
represents the thin disk component.  
Since \EAGLE\ imposes a temperature floor of 8000 K 
to prevent metal-rich gas particles from cooling to 
very cold, interstellar gas \citep{Schaye2015, Crain2015}, 
this sets a minimum disk height,
considerably larger than the physical scale height 
for observed edge-on galaxies
\citep{deGrijs1998,Kregel2002}.
As a result, \EAGLE\ cannot produce very flat galaxies
(also see \citet{Lagos2018} for a related discussion).
We select our sightline that intersects the thick disk component
$\sim10$ pkpc above the disk midplane, 
avoiding the `unresolved' thin disk.  
We call this the `thick disk' component
also because the 10-kpc thickness 
is comparable to that of the modeled rotating gas disks 
(tens of kpc tall)
in explaining the measured CGM kinematics 
(e.g., \citealt{Steidel2002,Ho2017,Ho2019}).}

\begin{figure}[htb]
    \centering
    \includegraphics[width=0.84\linewidth]{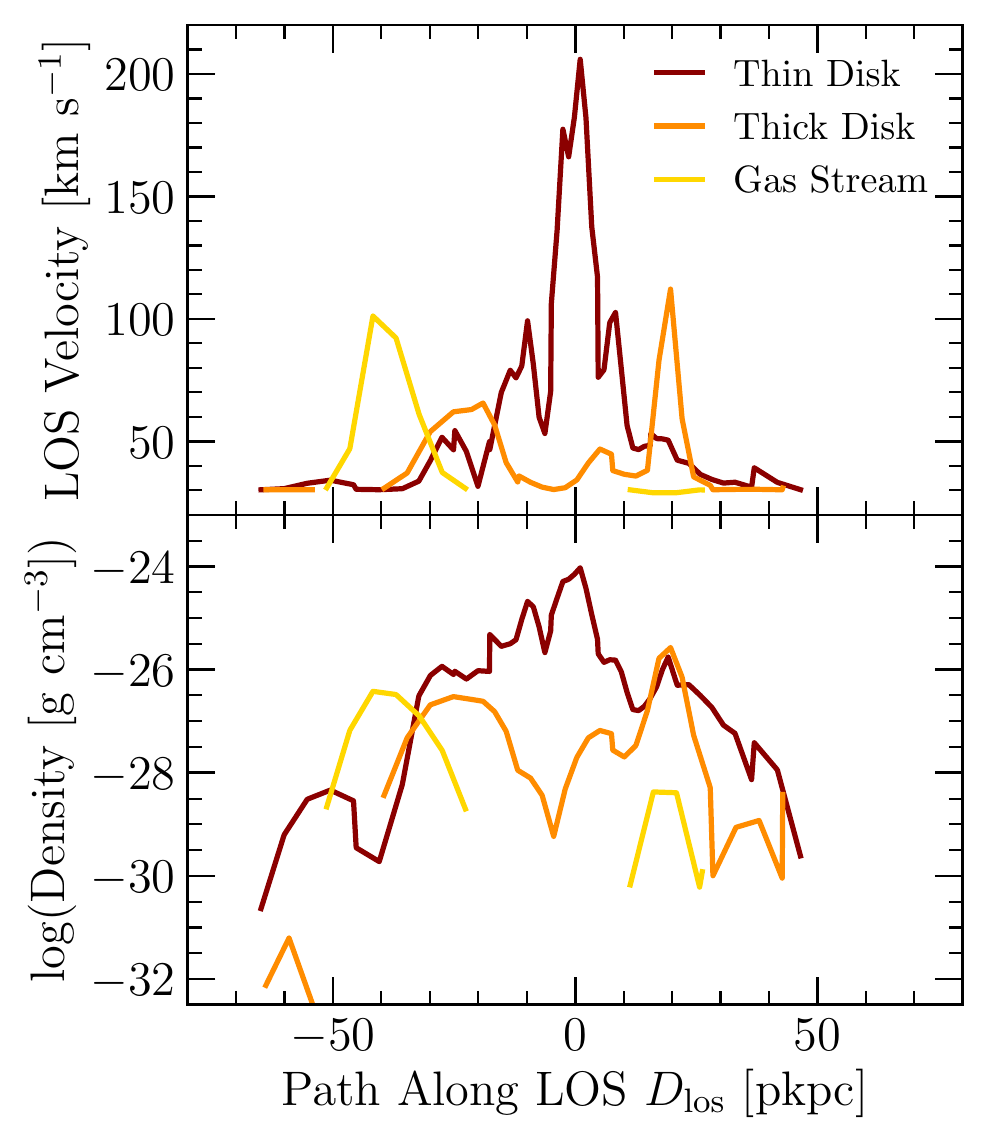}
    \caption{
        Variation of LOS velocity and cold gas density along 
        selected sightlines of the example edge-on galaxy 
        (galaxy ID: 37448).
        The top and the bottom panel show 
        how the LOS velocity and cold gas density vary 
        along sightlines that individually intersect 
        three structural components:
        a thin disk (red), a thick disk (orange), and a gas stream (yellow).  
        A positive LOS velocity indicates the LOS velocity 
        has the same sign as the disk angular momentum.
        $D_\mathrm{los} = 0$ separates the near side and the far side 
        of the galaxy disk with respect to the observer.
        Any discontinuity along $D_\mathrm{los}$ 
        indicates that cold gas is not intersected
        by that section of the sightline.  
        We only show $|D_\mathrm{los}| \leq 80$ pkpc along the path, 
        beyond which all three sightlines rarely intersect any cold gas.  
        }
    \label{fig:los_var} 
\end{figure}

Figure~\ref{fig:los_var} shows 
the variation of LOS velocity and cold gas density along 
each of the three sightlines.  
Along the path of each sightline, 
$D_\mathrm{los} = 0$ pkpc 
separates the near side and the far side of the edge-on disk.
Although our sightlines extend to $r_\mathrm{vir}$,
we only plot $|D_\mathrm{los}| \leq 80$ pkpc.  
Beyond this pathlength, 
the sightlines rarely intersect any cold gas within $r_\mathrm{vir}$.  
Even if cold gas is intersected, 
it has order-of-magnitudes lower density than 
the intersected cold gas within $|D_\mathrm{los}| \leq 80$ pkpc.
In addition, a positive LOS velocity indicates that 
it has the same sign as the disk angular momentum, 
i.e., the intersected cold gas corotates with the galaxy disk.  
For each of the three sightlines 
that intersect the three structures, 
the LOS velocities stay positive.  
This indicates that the intersected cold gas 
always shares the same sense of rotation as the inner galaxy disk.  
In the following, 
we briefly discuss the velocity variation along 
individual sightlines.

%%%%%%%%%%%% FIG: LOS velocity and cold gas density %%%%%%%%%%%%
%%%%%%%%%%%%%%%%%%%%%%%%%%%%%%%%%%%%%%%%%%%%%%%%

First, the sightline that intersects the thin disk 
detects the broadest LOS velocity range.  
The LOS velocity $v_\mathrm{los}$ reaches 210\kms\ 
at $D_\mathrm{los} = 0$, 
where the cold gas density also peaks 
(red line in Figure~\ref{fig:los_var}).
The large $v_\mathrm{los}$ is produced by 
the circular motion of the gas on the disk,
and the rotation velocity vector is tangent to the sightline
at $D_\mathrm{los} = 0$.  
Increasing $|D_\mathrm{los}|$ then
reduces the magnitude of $v_\mathrm{los}$ significantly, 
because the LOS no longer runs in parallel with the rotation velocity vector.  
Along this sightline, 
since the LOS velocity varies from around 30\kms\ to 210\kms,
this sightline can detect a broad LOS velocity range of 180\kms.

Second, the sightline that intersects the thick disk component 
produces a narrower velocity range. 
The LOS velocity reaches maxima of around 70\kms\ and 110\kms\
at the near and the far side of the disk respectively,
and the velocity minimum along the sightline is around  30\kms 
(orange line in Figure~\ref{fig:los_var}).
The density peaks at the two locations of the velocity maxima, 
which indicates that the thick disk component 
does not comprise a uniform disk.
This is evident from the gas distribution 
in Figure~\ref{fig:ppj_jr_example_2p5e5}, 
which shows the gas having spiral structures 
instead of being a solid disk.  
Since this sightline detects LOS velocity that varies 
from around 30\kms\ to a maximum of 110\kms, 
the velocity range reaches 80\kms,
which significantly exceeds the thermal linewidth.

As for the sightline that intersects a gas stream, 
only discrete sections along the sightline have intersected cold gas.  
The sightline intersects the gas stream at around 
$D_\mathrm{los} = -40$ pkpc (yellow line in Figure~\ref{fig:los_var}).  
At the far side of the disk of $D_\mathrm{los} = 20$ pkpc, 
the sightline intersects cold gas 
whose density is of 100 times lower.
The higher density gas stream produces LOS velocities
from around 30\kms\ to a maximum of 100\kms.  
Therefore, this sightline produces a LOS velocity range 
comparable to that of the thick disk component.

Using the three cases of our example edge-on galaxy, 
we have demonstrated that velocity ranges of over 100\kms\ 
in observations can 
be produced by a sightline that intersects 
a thin rotating disk edge-on.  
Sightlines that intersect 
the thick disk component or a gas stream
can detect LOS velocity ranges 
that are significantly broader than the thermal linewidth.  
All these gas structures likely contribute 
to the broad line profiles in real observations,
especially because the 
observed profiles broadened by instrumental resolution 
are expected to consist of multiple velocity components.

\editR{However, we emphasize that
while the three example sightlines can 
produce broad velocity ranges in principle,
such sightlines are extremely rare.  
As a result, they cannot explain
the large number of broad absorption systems observed.
Moreover, 
the observed velocity range of the absorption 
depends on the column density 
and thereby the ionization state of the gas,
which is beyond the scope of this paper.
Therefore, in the future,
we will conduct the ionization analysis
and use mock sightlines to 
study the absorption line profiles of different ionic species.  
This will provide us deeper insight into 
the circumgalactic gas kinematics measured
in quasar sightline observations.
}

%%%%%%%%%%%%%%%%%%%%%%%%%%%%%%%%%%%%%%%%
% Summary (Conclusion)
%%%%%%%%%%%%%%%%%%%%%%%%%%%%%%%%%%%%%%%%
\section{Summary and Conclusions}
\label{sec:summary}

Galaxies grow and fuel star formation by accreting gas, 
yet direct observation of gas inflows onto galaxies remains sparse.
In this pilot study, 
we have studied the properties of cold inflowing gas around 
\EAGLE\ galaxies at $z=0.27$.   
We have identified the cold inflowing gas using two methods: 
(i) 
a ballistic approximation that calculates 
the trajectories of particles moving under gravity,
and then predicts which particles accrete 
onto the inner galaxy
within a rotation period, 
and
(ii) 
tracking particles through the same time interval in \EAGLE,
which includes full hydrodynamic calculations.
We have compared the two sets of deduced inflow properties
and discussed the limitations of the ballistic approximation.  
To gain insight into understanding cold gas kinematics
measured from quasar sightline observations, 
we have probed the cold CGM ($\sim10^4$ K) using mock quasar sightlines.  
Our analysis has focused on the CGM corotation with the galactic disk,
as well as the velocity ranges produced by different galaxy structures.

Using either the ballistic approximation or particle tracking, 
we find that
the galaxies typically have low inflow speeds 
of around 20\kms\ to 60\kms.
The mass inflow rates deduced from the two methods  
agree to within a factor of two, 
and the ballistic approximation often over-estimates.
We have attributed the major cause of the discrepancy to feedback, 
which has removed the newly accreted gas
or can even disrupt the star-forming regions.  
In other words, 
feedback reduces the overall amount of cold gas accreted
and thereby suppresses the mass inflow rates.
Nevertheless, 
the mass inflow rates are 
generally comparable to the galaxy SFRs,
\editR{This suggests that
the cold inflowing gas plausibly sustains
the galaxy star formation activities
and thereby prolong the disk lifetimes.
}

Inspired by recent observational measurements 
of cold gas kinematics along quasar sightlines, 
we have used mock quasar sightlines 
to probe the cold gas within $r_\mathrm{vir}$ around 
a selected \EAGLE\ galaxy.  
This galaxy has an extended cold gas disk,
and the measurements of its cold gas kinematics
share similar characteristics as in sightline observations.  
This motivates future work
using larger simulation boxes with a higher resolution
to `observe' the cold gas kinematics in detail.

We have posed the question of
which sightlines can detect cold gas
that corotates with the inner galaxy disk.
By viewing the selected \EAGLE\ galaxy edge-on, 
we have found that sightlines with 
azimuthal angles of $\alpha=0$\deg\ and 10\deg\ can
detect corotating cold gas out to 60 pkpc and 40 pkpc respectively.  
Since quasar sightline observations often detect 
corotating cold gas near the galaxy major axes, 
\editR{our results suggest}
it is possible that the observed sightlines have 
intersected cold gas disks that extend beyond the optical disks.

Because quasar sightline observations 
also often detect broad velocity profiles, 
we have explored whether sightlines that individually intersect 
a thin disk component, 
a thick disk component (i.e., above the disk plane), and a gas stream 
can produce comparable velocity ranges. 
We have demonstrated that sightlines intersecting 
these three gas components can produce velocity ranges 
of over 70\kms, 
significantly broader than the thermal linewidth.  
All these structures possibly contribute to the broad 
line profiles in observational measurements,
especially because the observed line profiles 
are expected to consist of multiple velocity components.

In the future,
we will use \EAGLE\ simulations with large box sizes 
and higher resolutions to study cold gas kinematics.  
With larger cosmological volumes,
we can increase the sampling size of galaxies.
This will produce better statistics on the probabilities 
of detecting corotating cold gas along sightlines and 
finding galaxies with extended cold gas disks.
Using simulations with better resolution will allow us
to generate high resolution, mock absorption line spectra
and compare them to quasar sightline measurements.  
This will allow us to interpret quasar sightline observations 
using \EAGLE\ simulations and 
thereby gain insight into understanding 
the measured cold gas kinematics.

\acknowledgments

\editR{We thank the referee for the thoughtful and detailed
comments that improved the manuscript.}
We gratefully acknowledge Joop Schaye for the inspirational discussions on
both the technical and science aspects of this paper.  
We thank Marie Lau and Michael Lipatov 
for their suggestions on the manuscript.
We also thank Suoqing Ji and Nathan Goldbaum for the helpful discussion 
regarding the use of \texttt{yt}.
Material presented in this paper is based in part 
upon work supported by 
the National Science Foundation under Grant No.~1817125.  
We acknowledge the Virgo Consortium for making their simulation data available. 
The \EAGLE\ simulations were performed using the DiRAC-2 facility at Durham, 
managed by the ICC, 
and the PRACE facility Curie based in France 
at TGCC, CEA, Bruy\`{e}resle-Ch\^{a}tel.
This work has made use of the \texttt{yt} 
astrophysics analysis and visualization tool \citep{Turk2011}.

%%%%%%%%%%% REFERENCES %%%%%%%%%%%
\bibliography{master_eagle19}

%%%%%%%%%%%%%%%%%%%%%%%%%%%%%%%%%%%%%%%%%%
% Appendix (if needed)
%%%%%%%%%%%%%%%%%%%%%%%%%%%%%%%%%%%%%%%%%%
%\newpage
\appendix

%%%%%%%%%%%% FIG: Particle Plots (color by LOS velocity) for 10 centrals %%%%%%%%%%%%
\begin{figure*}[htb]
    \centering
    \includegraphics[width=0.89\linewidth]{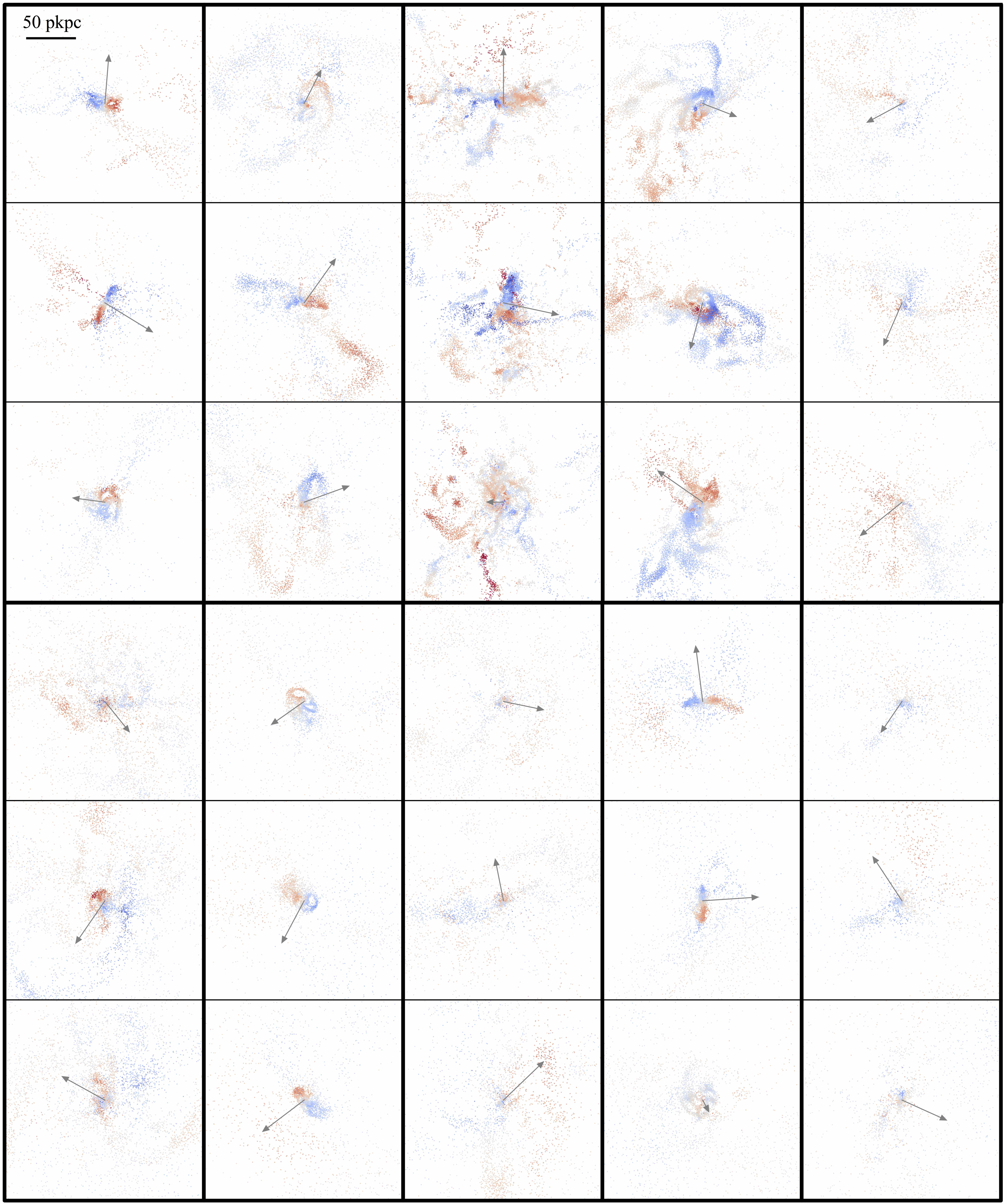}
    \caption{
        Particle projection plots showing the 
        projected LOS velocities of cold gas particles 
        ($T_\mathrm{gas} \leq 2.5\times10^5$ K).
        Each particle is color-coded by its projected velocity, 
        and we use the same velocity scale as in the top row of 
        Figure~\ref{fig:ppj_jr_example_2p5e5}.
        This figure follows the same format as in
        Figure~\ref{fig:particle_pjplot_2p5e5}.  
        Each galaxy occupies 3 rows $\times$ 1 column, 
        enclosed by the thick black lines.  
        For individual galaxies, from top to bottom, each panel
        represents the cut in $x$, $y$ and $z$ plane respectively.
        The side length of each panel is 200 pkpc.
        The plotted central galaxies 
        are ordered in decreasing stellar masses, 
        from left to right, and top to bottom.
        }
    \label{fig:particle_pjplot_2p5e5_vlos}
\end{figure*}

\end{document}

%% file: setdef.tex
% LAST UPDATE: Nov 12, 2015
% set short-hand definitions for convenience
%
%
%\def\etal{et al.\ }
\def\etali{{\it et al.\thinspace}}
\def\etns{{\rm et\thinspace al.}}   % et al.
\def\et{et al.\thinspace}

\def\EAGLE{\texttt{EAGLE}}
\def\AREPO{\texttt{AREPO}}
\def\GADGET3{\texttt{GADGET-3}}
\def\OWLS{\texttt{OWLS}}

%%%%%%%%%%%%%%%%%%%%
% Ions related
%%%%%%%%%%%%%%%%%%%%
%\def\mgIIdb{Mg II $\lambda\lambda$2796, 2803}
%\def\mgIIdbl{Mg II $\lambda$2796}
%\def\mgIIdbu{Mg II $\lambda$2796}
\def\mgIIdb{\ion{Mg}{2} $\lambda\lambda$2796, 2803}
\def\mgIIdbl{\ion{Mg}{2} $\lambda$2796}
\def\mgIIdbu{\ion{Mg}{2} $\lambda$2803}
\def\oI{[\ion{O}{1}] $\lambda$6300}
\def\nII{[\ion{N}{2}] $\lambda\lambda$6548, 6583}
\def\sII{[\ion{S}{2}] $\lambda\lambda$6716, 6731}
\def\oIII{[\ion{O}{3}] $\lambda$5007}
\def\hI{\mbox {\ion{H}{1}}}
%\def\mgII{\mbox {\ion{Mg}{2}}}

%%%%%%%%%%%%%%%%%%%%
% Units related
%%%%%%%%%%%%%%%%%%%%
\def\kms{\mbox{km s$^{-1}$}}
\def\kmstb{km s$^{-1}$}
\def\micron{\mbox{$\mu$m}}
\def\modotyr{\mbox {$\rm M_\odot$~yr$^{-1}$}}
\def\percm2{\mbox{cm$^{-2}$}}

%  Sept. 1993
%  CLM
%  Use ``\input mymac.tex'' in a tex file to use these definitions.
%
%\renewcommand{\deg}{\mbox{$^{\circ}$}}
%
\def \dlow {\mbox{$400 {\rm ~l~mm}^{-1}$}}
\def \dhigh {\mbox{$600 {\rm ~l~mm}^{-1}$}}
% PAPER SPECIFIC
% EVAN S.
\newcommand{\be}{\begin{equation}} \newcommand{\ba}{\begin{eqnarray}}
\newcommand{\ee}{\end{equation}} \newcommand{\ea}{\end{eqnarray}}
\def\etal{{\it et al.\thinspace}}
\def\-{{\em{---}}}
\def \mA {\mbox{${\rm m \AA} $} }
\def \rr {\mbox{${\rm RR}$} }
\def \rarb {\mbox{${\rm R_AR_B}$} }
\def \rara {\mbox{${\rm R_AR_A}$} }
\def \dd {\mbox{${\rm DD}$} }
\def \dada {\mbox{${\rm D_AD_A}$} }
\def \dadb {\mbox{${\rm D_AD_B}$} }
\def \dr {\mbox{${\rm DR}$} }
\def \darb {\mbox{${\rm D_AR_B}$} }
\def \dara {\mbox{${\rm D_AR_A}$} }
\def \dbra {\mbox{${\rm D_BR_A}$} }
\def \hMpc      {h^{-1}{\rm\ Mpc}}
\def \hkpc      {h^{-1}{\rm\ kpc}}
\def \h         {\hbox{$\, h$} }
\def \hinv      {\hbox{$\, h^{-1}$} }
\def \hinvseven    {\hbox{$\, h_{70}^{-1}$} }
\def\ewr{\mbox {EW$_r$}}
\def\ewo{\mbox {EW$_o$}}
\def\H7{\mbox {$h_{0.7}$}}
\def\naI{\mbox {\ion{Na}{1}}}
\def\mgI{\mbox {\ion{Mg}{1}}}
\def\feI{\mbox {\ion{Fe}{1}}}
\def\oVI{\mbox {\ion{O}{6}}}
\def\znII{\mbox {\ion{Zn}{2}}}
\def\crII{\mbox {\ion{Cr}{2}}}
\def\alI{\mbox {\sc Al~I~}}
\def\alII{\mbox {\sc Al~II~}}
\def\alIII{\mbox {\sc Al~III~}}
\def\mgII{\mbox {\ion{Mg}{2}}}
\def\mnII{\mbox {\ion{Mn}{2}}}
\def\niII{\mbox {\ion{Ni}{2}}}
\def\feII{\mbox {\ion{Fe}{2}}}
\def\feIII{\mbox {\ion{Fe}{3}}}
\def\cIV{\mbox {\ion{C}{4}}}
\def\sV{\mbox {\ion{S}{5}}}
\def\siIV{\mbox {\ion{Si}{4}}}
\def\siIII{\mbox {\ion{Si}{3}}}
\def\siII{\mbox {\ion{Si}{2}}}
\def\siI{\mbox {\ion{Si}{1}}}
\def\cII{\mbox {\ion{C}{2}}}
\def\cIII{\mbox {\ion{C}{3}}}
\def\llambda{\mbox {$\lambda$}}
\def\mstar{\mbox {$M_*$}}
\def\hlen{\mbox {$h_{0.7}^{-1}$}}
\def\lstarlya{\mbox {$L^*_{Ly\alpha}$}}
\def\IZw18{I~Zw~18}
\def\m82{M82}
\def\Ab{Abell~}
\def\gi{\mbox {\rm g-i}}
\def\ug{\mbox {\rm u-g}}
\def\br{\mbox {\rm b-r}}
\def\eqn{equation}
\def\vesc{\mbox {$v_{\rm esc}$}}
% HE PAPER
\def\heha{\mbox {He~I~$\lambda 5876$ / H$\alpha$}}
\def\xhe{\mbox {$\chi({\rm He}) / \chi({\rm H})$} }
\def\heii{\mbox {${\rm He}^+$}}
\def\he{\mbox {\rm He}}
\def\hii{\mbox {${\rm H}^+$}}
\def\h{\mbox {\rm H}}
\def\mab{\mbox {$\rm m_{AB}$}}
\def\ssp{\baselineskip=13pt plus 1pt minus 1pt}
\def\tsp{\baselineskip=5pt plus 1pt minus 1pt}
%
% ASTRO SYMBOLS (revised to work in/out mathmode).
%
\def\deg{\mbox {$^{\circ}$}}
\def\msun{\mbox {${\rm ~M_\odot}$}}
\def\zsun{\mbox {${\rm ~Z_{\odot}}$}}
\def\lsun{\mbox {${~\rm L_\odot}$}}
\def\msunyr{\mbox {$~{\rm M_\odot}$~yr$^{-1}$}}
\def\angs{\mbox {~\AA}}
\def\lya{\mbox {Ly$\alpha$}}
\def\lyb{\mbox {Ly$\beta$}}
\def\Ha{\mbox {H$\alpha$}}
\def\Hb{\mbox {H$\beta$}}
\def\Hg{\mbox {H$\gamma$}}
\def\tion{\mbox {$T_{\rm ion}$~}}
\def\ch{\mbox {$\bigtriangleup$}}
\def\grad{\mbox {$\bigtriangledown$}}
\def\lstar{\mbox {$L^*$}}
\def\line{\mbox {~$\lambda$}}
\def\lines{\mbox {~$\lambda\lambda$~}}
\def\h0{\mbox {~H$_0$}}
\def\q0{\mbox {~q$_0$}}
%
% **** LINE RATIOS ****
%
\def\auroral{[OIII]~$\lambda4363$~}
\def\auroral{[OIII]~$\lambda4363$~}
\def\ohsun{\mbox {(O/H)$_{\odot}$~}}

\def\O1ha{[OI]$\lambda6300$~/~H$\alpha$~}
\def\Ru{[OII]$\lambda\lambda3727$~/~[OIII]$\lambda5007$~}
\def\s2ha{[SII]$\lambda\lambda6717,31$~/~H$\alpha$~}
\def\2z2{HeII~$\lambda4686$~}
\def\z7{[NII]~$\lambda6583$ }
\def\N2{[NII]~$\lambda6583$~/~H$\alpha$~}
\def\16z2{[SII]~$\lambda\lambda6717, 6731$ }
\def\HgI{HgI~$\lambda4358$~}
\def\Sdensity{[SII]~$\lambda6717 / \lambda6731$}
\def\Temp{[OIII]~$\lambda\lambda4959 + 5007 ~{\rm to}~ \lambda4363$~}
%
% **** end ****
%
\def\j{J}
\def\n{NGC~}
\def\asec{\ifmmode {'' }\else $''~$\fi}  % arc sec
\def\amin{\ifmmode {' }\else $'~$\fi}    % arc min
\def\arcsper{\ifmmode \rlap.{'' }\else $\rlap{.}'' $\fi} % '' %Arcsec period
\def\arcmper{\ifmmode \rlap.{' }\else $\rlap{.}' $\fi} % '  %Arcmin period
\def\sles{\lesssim}
\def\sgreat{\gtrsim}
%
%\def\sgreat{\lower2pt\hbox{$\buildrel {\scriptstyle >}
%    \over {\scriptstyle\sim}$}} % approximately greater than
%
%These are smaller.
%\def\gapp{$_>\atop{^\sim}$}  % approximately greater than
%\def\lapp{$_<\atop{^\sim}$}  % approximately less than
\def\gapp{\mbox {$_>\atop{^\sim}$}}  % approximately greater than
\def\lapp{\mbox {$_<\atop{^\sim}$}}  % approximately less than
%
% UNITS
\def\kms{\mbox {~km~s$^{-1}$}}
\def\ergsec{~ergs~s$^{-1}$~}
\def\sb{~ergs~s$^{-1}$~cm$^{-2}$~arcsec$^{-2}$}
\def\flux{~ergs~s$^{-1}$~cm$^{-2}$}
\def\flam{~ergs~s$^{-1}$~cm$^{-2}$ \AA$^{-1}$}
\def\cm3{~cm$^{-3}$}
\def\col{\mbox {~cm$^{-2}$}}
\def\mpc3{~Mpc$^{3}$}
\def\mpc-3{~Mpc$^{-3}$}
\def\rate{~sec$~{-1}$}
\def\um{~${\mu}$m~}
% ABBREVIATIONS
\def\fig{{Figure}}
\def\figs{{Figures}}
\def\tbl{{Table}~}
\def\sec{{Sec.}~}
\def\x{{X-ray}~}
\def\xs{{X-rays}~}
\def\X{{X-Ray}~}

%
% REFERENCES
\def\et{{\rm et\thinspace al.}\ }   % et al.
\def\ets{{\rm et\thinspace al.'s}\ }   % et al.'s
\def\reff{\par\noindent\parskip=1pt\hangindent=3pc\hangafter=1}
%
%\def\apjl{ApJL}
%\def\apj{ApJ}
%\def\apjs{ApJS}
%\def\pasp{PASP}
%\def\aj{AJ}
%\def\mn{MNRAS}
%\def\nat{Nature}
%\def\aa{A\&A}
%\def\aasup{A\&AS}
%\def\baas{BAAS}
%\def\annrev{ARA\&R}
%\def\aar{A\&AR}
%\def\pasj{PASJ}
%

% USE THIS FOR TABLE REFERENCES
%
\def\beginrefs{
         {\normalsize}
         {\noindent}
         \small
        \baselineskip=11pt
        \parindent=0pt
        \frenchspacing
        \parskip=1pt plus 1pt
%        \interlinepenalty=1000\tolerance=400
        \everypar={\hangindent=0.42in}}

%% file: table1.tex
% Target basic info
\begin{deluxetable}{lll}
\tablecaption{Characteristics of the Ref-L012N0188 simulation used in this paper.  
\label{tb:simulation}  }
%\tablecolumns{8}
%\tabletypesize{\normalsize}
%\tabletypesize{\scriptsize}
\tabletypesize{\footnotesize}
\tablewidth{1\linewidth}
\tablehead{
\colhead{} & \colhead{Simulation Property} & \colhead{Value}
}
\startdata
(1) & Boxsize $L$ (cMpc)      & 12.5\\
(2) & Number of particles $N$          & $188^3$\\
(3) & Initial baryonic particle mass $m_\mathrm{g}$ ($M_\odot$)    & $1.81 \times 10^6$\\
(4) & Dark matter particle mass $m_\mathrm{dm}$ ($M_\odot$)   & $9.70 \times 10^6$\\
(5) & Gravitational softening length $\epsilon_\mathrm{com}$ (ckpc) & 2.66\\
(6) & Maximum softening length $\epsilon_\mathrm{prop}$ (pkpc)         & 0.70
%(1) & $L$ (cMpc)      & 12.5\\
%(2) & $N$          & $188^3$\\
%(3) & $m_\mathrm{g}$ ($M_\odot$)    & $1.81 \times 10^6$\\
%(4) & $m_\mathrm{dm}$ ($M_\odot$)   & $9.70 \times 10^6$\\
%(5) & $\epsilon_{com}$ (ckpc) & 2.66\\
%(6) & $\epsilon_{prop}$ (pkpc)         & 0.70
\enddata
\tablecomments{
(1) Comoving boxsize. 
(2) Number of dark matter particles (initially there is an equal number of baryonic particles).
(3) Initial baryonic particle mass.
(4) Dark matter particle mass. 
(5) Comoving Plummer-equivalent gravitational softening length. 
(6) Maximum proper softening length.  
}
%\label{tb:galfit_result}  
\end{deluxetable}

%% file: table2.tex
% Target basic info
\begin{deluxetable*}{cccccccccc}
\tablecaption{Mass inflow rate and mass-weighted average inflow speed of cold gas ($T_\mathrm{gas} \leq 2.5\times10^5$ K) at central galaxies}
%\tablecolumns{8}
%\tabletypesize{\normalsize}
%\tabletypesize{\scriptsize}
\tabletypesize{\footnotesize}
\tablewidth{0pt}
\tablehead{
\colhead{Galaxy ID} & 
\colhead{$\log(M_\star/M_\odot)$} &
\colhead{$r_\mathrm{vir}$} &
\colhead{SFR} &
\colhead{$\dot{M}_\mathrm{in}^\mathrm{ballistic}$} &
\colhead{$\langle v_r \rangle_\mathrm{in}^\mathrm{ballistic}$} &
\colhead{$\langle v_r \rangle_\mathrm{in, top 10\%}^\mathrm{ballistic}$} & 
\colhead{$\dot{M}_\mathrm{in}^\mathrm{tracking}$} &
\colhead{$\langle v_r \rangle_\mathrm{in}^\mathrm{tracking}$} & 
\colhead{$\langle v_r \rangle_\mathrm{in, top 10\%}^\mathrm{tracking}$}
\\
\colhead{} &
\colhead{} &
\colhead{(pkpc)} &
\colhead{(\modotyr)} &
\colhead{(\modotyr)} &
\colhead{(\kmstb)} &
\colhead{(\kmstb)} &
\colhead{(\modotyr)} &
\colhead{(\kmstb)} &
\colhead{(\kmstb)}
% \\
% \colhead{(1)} &
% \colhead{(2)} &
% \colhead{(3)} &
% \colhead{(4)} &
% \colhead{(5)} &
% \colhead{(6)} &
% \colhead{(7)} &
% \colhead{(8)} &
% \colhead{(9)} &
% \colhead{(10)}
}
\decimalcolnumbers
\startdata
37448 & 10.36 & 238 & 1.34 & 1.49 & 34 & 132 & 0.982 & 17 & 67\\ % 486/4_0
43166 & 10.25 & 197 & 1.72 & 2.23 & 62 & 161 & 1.67 & 42 & 164\\ % 573/6_0
32523 & 10.06 & 190 & 1.26 & 2.12 & 37 & 123 & 1.14 & 34 & 103\\ % 585/7_0
30497 & 10.02 & 219 & 2.34 & 11.5 & 61 & 206 & 6.57 & 48 & 140\\ % 416/3_0
40127 & 9.95 & 227 & 0.99 & 6.00 & 35 & 143 & 3.29 & 39 & 117\\ % 525/5_0
48386 & 9.87 & 186 & 0.41 & 0.863 & 30 & 100 & 0.0380 & 20 & 26\\ % 698/10_0
44548 & 9.85 & 185 & 1.11 & 4.14 & 40 & 155 & 2.89 & 40 & 148\\ % 617/8_0
57650 & 9.68 & 156 & 0.53 & 2.75 & 36 & 117 & 1.10 & 18 & 55\\ % 818/15_0
46369 & 9.59 & 177 & 0.34 & 1.85 & 45 & 102 & 1.08 & 59 & 99\\ % 661/9_0
51643 & 9.54 & 168 & 0.56 & 1.74 & 39 & 128 & 1.63 & 22 & 61\\ % 723/11_0
53907 & 9.50 & 166 & 0.40 & 1.73 & 17 & 49 & 1.15 & 25 & 57\\ % 739/12_0
\enddata
\tablecomments{
    (1) Galaxy ID.
    (2) Stellar mass.  
    (3) Virial radius. 
    (4) Star formation rate.
    (5) Average mass inflow rate from inflow particles 
        in the ballistic calculation.
    (6) Mass-weighted average inflow speed from inflow particles 
        in the ballistic calculation.
    (7) Mass-weighted average inflow speed 
        (10\% mass with highest inflow speeds) 
        from inflow particles in the ballistic calculation.  
    (8) Average mass inflow rate from tracking inflow particles.
    (9) Mass-weighted average inflow speed from tracking inflow particles.
    (10) Mass-weighted average inflow speed 
         (10\% mass with highest inflow speeds)
         from tracking inflow particles.
}
\label{tb:inflow_avgv_result}  
\end{deluxetable*}

%% file: ms_x1.bbl
\begin{thebibliography}{}
\expandafter\ifx\csname natexlab\endcsname\relax\def\natexlab#1{#1}\fi

\bibitem[{{Bah{\'e}} {et~al.}(2016){Bah{\'e}}, {Crain}, {Kauffmann}, {Bower},
  {Schaye}, {Furlong}, {Lagos}, {Schaller}, {Trayford}, {Dalla Vecchia}, \&
  {Theuns}}]{Bahe2016}
{Bah{\'e}}, Y.~M., {Crain}, R.~A., {Kauffmann}, G., {et~al.} 2016, \mnras, 456,
  1115

\bibitem[{{Bigiel} {et~al.}(2008){Bigiel}, {Leroy}, {Walter}, {Brinks}, {de
  Blok}, {Madore}, \& {Thornley}}]{Bigiel2008}
{Bigiel}, F., {Leroy}, A., {Walter}, F., {et~al.} 2008, \aj, 136, 2846

\bibitem[{{Bigiel} {et~al.}(2011){Bigiel}, {Leroy}, {Walter}, {Brinks}, {de
  Blok}, {Kramer}, {Rix}, {Schruba}, {Schuster}, {Usero}, \&
  {Wiesemeyer}}]{Bigiel2011}
{Bigiel}, F., {Leroy}, A.~K., {Walter}, F., {et~al.} 2011, \apjl, 730, L13

\bibitem[{{Boselli} {et~al.}(2001){Boselli}, {Gavazzi}, {Donas}, \&
  {Scodeggio}}]{Boselli2001}
{Boselli}, A., {Gavazzi}, G., {Donas}, J., \& {Scodeggio}, M. 2001, \aj, 121,
  753

\bibitem[{{Bouch{\'e}} {et~al.}(2012){Bouch{\'e}}, {Hohensee}, {Vargas},
  {Kacprzak}, {Martin}, {Cooke}, \& {Churchill}}]{Bouche2012}
{Bouch{\'e}}, N., {Hohensee}, W., {Vargas}, R., {et~al.} 2012, \mnras, 426, 801

\bibitem[{{Bouch{\'e}} {et~al.}(2013){Bouch{\'e}}, {Murphy}, {Kacprzak},
  {P{\'e}roux}, {Contini}, {Martin}, \& {Dessauges-Zavadsky}}]{Bouche2013}
{Bouch{\'e}}, N., {Murphy}, M.~T., {Kacprzak}, G.~G., {et~al.} 2013, Science,
  341, 50

\bibitem[{{Bouch{\'e}} {et~al.}(2010){Bouch{\'e}}, {Dekel}, {Genzel}, {Genel},
  {Cresci}, {F{\"o}rster Schreiber}, {Shapiro}, {Davies}, \&
  {Tacconi}}]{Bouche2010}
{Bouch{\'e}}, N., {Dekel}, A., {Genzel}, R., {et~al.} 2010, \apj, 718, 1001

\bibitem[{{Bouch{\'e}} {et~al.}(2016){Bouch{\'e}}, {Finley}, {Schroetter},
  {Murphy}, {Richter}, {Bacon}, {Contini}, {Richard}, {Wendt}, {Kamann},
  {Epinat}, {Cantalupo}, {Straka}, {Schaye}, {Martin}, {P{\'e}roux},
  {Wisotzki}, {Soto}, {Lilly}, {Carollo}, {Brinchmann}, \&
  {Kollatschny}}]{Bouche2016}
{Bouch{\'e}}, N., {Finley}, H., {Schroetter}, I., {et~al.} 2016, \apj, 820, 121

\bibitem[{{Bowen} {et~al.}(2016){Bowen}, {Chelouche}, {Jenkins}, {Tripp},
  {Pettini}, {York}, \& {Frye}}]{Bowen2016}
{Bowen}, D.~V., {Chelouche}, D., {Jenkins}, E.~B., {et~al.} 2016, \apj, 826, 50

\bibitem[{{Brinchmann} {et~al.}(2004){Brinchmann}, {Charlot}, {White},
  {Tremonti}, {Kauffmann}, {Heckman}, \& {Brinkmann}}]{Brinchmann2004}
{Brinchmann}, J., {Charlot}, S., {White}, S.~D.~M., {et~al.} 2004, \mnras, 351,
  1151

\bibitem[{{Brook} {et~al.}(2011){Brook}, {Governato}, {Ro{\v s}kar}, {Stinson},
  {Brooks}, {Wadsley}, {Quinn}, {Gibson}, {Snaith}, {Pilkington}, {House}, \&
  {Pontzen}}]{Brook2011}
{Brook}, C.~B., {Governato}, F., {Ro{\v s}kar}, R., {et~al.} 2011, \mnras, 415,
  1051

\bibitem[{{Bryan} \& {Norman}(1998)}]{Bryan1998}
{Bryan}, G.~L., \& {Norman}, M.~L. 1998, \apj, 495, 80

\bibitem[{{Coil} {et~al.}(2011){Coil}, {Blanton}, {Burles}, {Cool},
  {Eisenstein}, {Moustakas}, {Wong}, {Zhu}, {Aird}, {Bernstein}, {Bolton}, \&
  {Hogg}}]{Coil2011}
{Coil}, A.~L., {Blanton}, M.~R., {Burles}, S.~M., {et~al.} 2011, \apj, 741, 8

\bibitem[{{Cool} {et~al.}(2013){Cool}, {Moustakas}, {Blanton}, {Burles},
  {Coil}, {Eisenstein}, {Wong}, {Zhu}, {Aird}, {Bernstein}, {Bolton}, {Hogg},
  \& {Mendez}}]{Cool2013}
{Cool}, R.~J., {Moustakas}, J., {Blanton}, M.~R., {et~al.} 2013, \apj, 767, 118

\bibitem[{{Correa} {et~al.}(2018{\natexlab{a}}){Correa}, {Schaye}, {van de
  Voort}, {Duffy}, \& {Wyithe}}]{Correa2018a}
{Correa}, C.~A., {Schaye}, J., {van de Voort}, F., {Duffy}, A.~R., \& {Wyithe},
  J.~S.~B. 2018{\natexlab{a}}, \mnras, 478, 255

\bibitem[{{Correa} {et~al.}(2018{\natexlab{b}}){Correa}, {Schaye}, {Wyithe},
  {Duffy}, {Theuns}, {Crain}, \& {Bower}}]{Correa2018b}
{Correa}, C.~A., {Schaye}, J., {Wyithe}, J.~S.~B., {et~al.} 2018{\natexlab{b}},
  \mnras, 473, 538

\bibitem[{{Crain} {et~al.}(2015){Crain}, {Schaye}, {Bower}, {Furlong},
  {Schaller}, {Theuns}, {Dalla Vecchia}, {Frenk}, {McCarthy}, {Helly},
  {Jenkins}, {Rosas-Guevara}, {White}, \& {Trayford}}]{Crain2015}
{Crain}, R.~A., {Schaye}, J., {Bower}, R.~G., {et~al.} 2015, \mnras, 450, 1937

\bibitem[{{Crain} {et~al.}(2017){Crain}, {Bah{\'e}}, {Lagos}, {Rahmati},
  {Schaye}, {McCarthy}, {Marasco}, {Bower}, {Schaller}, {Theuns}, \& {van der
  Hulst}}]{Crain2017}
{Crain}, R.~A., {Bah{\'e}}, Y.~M., {Lagos}, C.~d.~P., {et~al.} 2017, \mnras,
  464, 4204

\bibitem[{{Dalla Vecchia} \& {Schaye}(2008)}]{DallaVecchia2008}
{Dalla Vecchia}, C., \& {Schaye}, J. 2008, \mnras, 387, 1431

\bibitem[{{Dalla Vecchia} \& {Schaye}(2012)}]{DallaVecchia2012}
---. 2012, \mnras, 426, 140

\bibitem[{{Danovich} {et~al.}(2015){Danovich}, {Dekel}, {Hahn}, {Ceverino}, \&
  {Primack}}]{Danovich2015}
{Danovich}, M., {Dekel}, A., {Hahn}, O., {Ceverino}, D., \& {Primack}, J. 2015,
  \mnras, 449, 2087

\bibitem[{{Danovich} {et~al.}(2012){Danovich}, {Dekel}, {Hahn}, \&
  {Teyssier}}]{Danovich2012}
{Danovich}, M., {Dekel}, A., {Hahn}, O., \& {Teyssier}, R. 2012, \mnras, 422,
  1732

\bibitem[{{Dav{\'e}} {et~al.}(2012){Dav{\'e}}, {Finlator}, \&
  {Oppenheimer}}]{Dave2012}
{Dav{\'e}}, R., {Finlator}, K., \& {Oppenheimer}, B.~D. 2012, \mnras, 421, 98

\bibitem[{{Davis} {et~al.}(1985){Davis}, {Efstathiou}, {Frenk}, \&
  {White}}]{Davis1985}
{Davis}, M., {Efstathiou}, G., {Frenk}, C.~S., \& {White}, S.~D.~M. 1985, \apj,
  292, 371

\bibitem[{{de Grijs}(1998)}]{deGrijs1998}
{de Grijs}, R. 1998, \mnras, 299, 595

\bibitem[{{Dekel} \& {Birnboim}(2006)}]{Dekel2006}
{Dekel}, A., \& {Birnboim}, Y. 2006, \mnras, 368, 2

\bibitem[{{Diamond-Stanic} {et~al.}(2016){Diamond-Stanic}, {Coil}, {Moustakas},
  {Tremonti}, {Sell}, {Mendez}, {Hickox}, \& {Rudnick}}]{DiamondStanic2016}
{Diamond-Stanic}, A.~M., {Coil}, A.~L., {Moustakas}, J., {et~al.} 2016, \apj,
  824, 24

\bibitem[{{Dolag} {et~al.}(2009){Dolag}, {Borgani}, {Murante}, \&
  {Springel}}]{Dolag2009}
{Dolag}, K., {Borgani}, S., {Murante}, G., \& {Springel}, V. 2009, \mnras, 399,
  497

\bibitem[{{El-Badry} {et~al.}(2018){El-Badry}, {Quataert}, {Wetzel}, {Hopkins},
  {Weisz}, {Chan}, {Fitts}, {Boylan-Kolchin}, {Kere{\v s}},
  {Faucher-Gigu{\`e}re}, \& {Garrison-Kimmel}}]{ElBadry2018}
{El-Badry}, K., {Quataert}, E., {Wetzel}, A., {et~al.} 2018, \mnras, 473, 1930

\bibitem[{{Fall} \& {Efstathiou}(1980)}]{Fall1980}
{Fall}, S.~M., \& {Efstathiou}, G. 1980, \mnras, 193, 189

\bibitem[{{Fox} {et~al.}(2014){Fox}, {Wakker}, {Barger}, {Hernandez},
  {Richter}, {Lehner}, {Bland-Hawthorn}, {Charlton}, {Westmeier}, {Thom},
  {Tumlinson}, {Misawa}, {Howk}, {Haffner}, {Ely}, {Rodriguez-Hidalgo}, \&
  {Kumari}}]{Fox2014}
{Fox}, A.~J., {Wakker}, B.~P., {Barger}, K.~A., {et~al.} 2014, \apj, 787, 147

\bibitem[{{Furlong} {et~al.}(2015){Furlong}, {Bower}, {Theuns}, {Schaye},
  {Crain}, {Schaller}, {Dalla Vecchia}, {Frenk}, {McCarthy}, {Helly},
  {Jenkins}, \& {Rosas-Guevara}}]{Furlong2015}
{Furlong}, M., {Bower}, R.~G., {Theuns}, T., {et~al.} 2015, \mnras, 450, 4486

\bibitem[{{Gehrels}(1986)}]{Gehrels1986}
{Gehrels}, N. 1986, \apj, 303, 336

\bibitem[{{Heald} {et~al.}(2011){Heald}, {J{\'o}zsa}, {Serra}, {Zschaechner},
  {Rand}, {Fraternali}, {Oosterloo}, {Walterbos}, {J{\"u}tte}, \&
  {Gentile}}]{Heald2011}
{Heald}, G., {J{\'o}zsa}, G., {Serra}, P., {et~al.} 2011, \aap, 526, A118

\bibitem[{{Heckman} {et~al.}(2015){Heckman}, {Alexandroff}, {Borthakur},
  {Overzier}, \& {Leitherer}}]{Heckman2015}
{Heckman}, T.~M., {Alexandroff}, R.~M., {Borthakur}, S., {Overzier}, R., \&
  {Leitherer}, C. 2015, \apj, 809, 147

\bibitem[{{Ho} \& {Martin}(2019)}]{Ho2019}
{Ho}, S.~H., \& {Martin}, C.~L. 2019, arXiv e-prints, arXiv:1901.11182

\bibitem[{{Ho} {et~al.}(2017){Ho}, {Martin}, {Kacprzak}, \&
  {Churchill}}]{Ho2017}
{Ho}, S.~H., {Martin}, C.~L., {Kacprzak}, G.~G., \& {Churchill}, C.~W. 2017,
  \apj, 835, 267

\bibitem[{{Kacprzak} {et~al.}(2011){Kacprzak}, {Churchill}, {Barton}, \&
  {Cooke}}]{Kacprzak2011ApJ}
{Kacprzak}, G.~G., {Churchill}, C.~W., {Barton}, E.~J., \& {Cooke}, J. 2011,
  \apj, 733, 105

\bibitem[{{Kacprzak} {et~al.}(2010){Kacprzak}, {Churchill}, {Ceverino},
  {Steidel}, {Klypin}, \& {Murphy}}]{Kacprzak2010}
{Kacprzak}, G.~G., {Churchill}, C.~W., {Ceverino}, D., {et~al.} 2010, \apj,
  711, 533

\bibitem[{{Kacprzak} {et~al.}(2012){Kacprzak}, {Churchill}, \&
  {Nielsen}}]{Kacprzak2012}
{Kacprzak}, G.~G., {Churchill}, C.~W., \& {Nielsen}, N.~M. 2012, \apjl, 760, L7

\bibitem[{{Kacprzak} {et~al.}(2015){Kacprzak}, {Muzahid}, {Churchill},
  {Nielsen}, \& {Charlton}}]{Kacprzak2015}
{Kacprzak}, G.~G., {Muzahid}, S., {Churchill}, C.~W., {Nielsen}, N.~M., \&
  {Charlton}, J.~C. 2015, \apj, 815, 22

\bibitem[{{Kacprzak} {et~al.}(2014){Kacprzak}, {Martin}, {Bouch{\'e}},
  {Churchill}, {Cooke}, {LeReun}, {Schroetter}, {Ho}, \&
  {Klimek}}]{Kacprzak2014}
{Kacprzak}, G.~G., {Martin}, C.~L., {Bouch{\'e}}, N., {et~al.} 2014, \apjl,
  792, L12

\bibitem[{{Kamphuis} {et~al.}(2013){Kamphuis}, {Rand}, {J{\'o}zsa},
  {Zschaechner}, {Heald}, {Patterson}, {Gentile}, {Walterbos}, {Serra}, \& {de
  Blok}}]{Kamphuis2013}
{Kamphuis}, P., {Rand}, R.~J., {J{\'o}zsa}, G.~I.~G., {et~al.} 2013, \mnras,
  434, 2069

\bibitem[{{Kennicutt}(1998)}]{Kennicutt1998}
{Kennicutt}, Jr., R.~C. 1998, \araa, 36, 189

\bibitem[{{Kennicutt} {et~al.}(1994){Kennicutt}, {Tamblyn}, \&
  {Congdon}}]{Kennicutt1994}
{Kennicutt}, Jr., R.~C., {Tamblyn}, P., \& {Congdon}, C.~E. 1994, \apj, 435, 22

\bibitem[{{Kere{\v s}} {et~al.}(2009){Kere{\v s}}, {Katz}, {Fardal},
  {Dav{\'e}}, \& {Weinberg}}]{Keres2009}
{Kere{\v s}}, D., {Katz}, N., {Fardal}, M., {Dav{\'e}}, R., \& {Weinberg},
  D.~H. 2009, \mnras, 395, 160

\bibitem[{{Kere{\v s}} {et~al.}(2005){Kere{\v s}}, {Katz}, {Weinberg}, \&
  {Dav{\'e}}}]{Keres2005}
{Kere{\v s}}, D., {Katz}, N., {Weinberg}, D.~H., \& {Dav{\'e}}, R. 2005,
  \mnras, 363, 2

\bibitem[{{Kimm} {et~al.}(2011){Kimm}, {Devriendt}, {Slyz}, {Pichon}, {Kassin},
  \& {Dubois}}]{Kimm2011}
{Kimm}, T., {Devriendt}, J., {Slyz}, A., {et~al.} 2011, ArXiv e-prints,
  arXiv:1106.0538

\bibitem[{{Kregel} {et~al.}(2002){Kregel}, {van der Kruit}, \& {de
  Grijs}}]{Kregel2002}
{Kregel}, M., {van der Kruit}, P.~C., \& {de Grijs}, R. 2002, \mnras, 334, 646

\bibitem[{{Lagos} {et~al.}(2018){Lagos}, {Schaye}, {Bah{\'e}}, {Van de Sande},
  {Kay}, {Barnes}, {Davis}, \& {Dalla Vecchia}}]{Lagos2018}
{Lagos}, C.~d.~P., {Schaye}, J., {Bah{\'e}}, Y., {et~al.} 2018, \mnras, 476,
  4327

\bibitem[{{Lagos} {et~al.}(2017){Lagos}, {Theuns}, {Stevens}, {Cortese},
  {Padilla}, {Davis}, {Contreras}, \& {Croton}}]{Lagos2017}
{Lagos}, C.~d.~P., {Theuns}, T., {Stevens}, A.~R.~H., {et~al.} 2017, \mnras,
  464, 3850

\bibitem[{{Lagos} {et~al.}(2015){Lagos}, {Crain}, {Schaye}, {Furlong}, {Frenk},
  {Bower}, {Schaller}, {Theuns}, {Trayford}, {Bah{\'e}}, \& {Dalla
  Vecchia}}]{Lagos2015}
{Lagos}, C.~d.~P., {Crain}, R.~A., {Schaye}, J., {et~al.} 2015, \mnras, 452,
  3815

\bibitem[{{Leroy} {et~al.}(2008){Leroy}, {Walter}, {Brinks}, {Bigiel}, {de
  Blok}, {Madore}, \& {Thornley}}]{Leroy2008}
{Leroy}, A.~K., {Walter}, F., {Brinks}, E., {et~al.} 2008, \aj, 136, 2782

\bibitem[{{Leroy} {et~al.}(2013){Leroy}, {Walter}, {Sandstrom}, {Schruba},
  {Munoz-Mateos}, {Bigiel}, {Bolatto}, {Brinks}, {de Blok}, {Meidt}, {Rix},
  {Rosolowsky}, {Schinnerer}, {Schuster}, \& {Usero}}]{Leroy2013}
{Leroy}, A.~K., {Walter}, F., {Sandstrom}, K., {et~al.} 2013, \aj, 146, 19

\bibitem[{{Martin}(1999)}]{Martin1999}
{Martin}, C.~L. 1999, \apj, 513, 156

\bibitem[{{Martin} {et~al.}(2019){Martin}, {Ho}, {Kacprzak}, \&
  {Churchill}}]{Martin2019}
{Martin}, C.~L., {Ho}, S.~H., {Kacprzak}, G.~G., \& {Churchill}, C.~W. 2019,
  arXiv e-prints, arXiv:1901.09123

\bibitem[{{Martin} {et~al.}(2012){Martin}, {Shapley}, {Coil}, {Kornei},
  {Bundy}, {Weiner}, {Noeske}, \& {Schiminovich}}]{Martin2012}
{Martin}, C.~L., {Shapley}, A.~E., {Coil}, A.~L., {et~al.} 2012, \apj, 760, 127

\bibitem[{{McAlpine} {et~al.}(2016){McAlpine}, {Helly}, {Schaller}, {Trayford},
  {Qu}, {Furlong}, {Bower}, {Crain}, {Schaye}, {Theuns}, {Dalla Vecchia},
  {Frenk}, {McCarthy}, {Jenkins}, {Rosas-Guevara}, {White}, {Baes}, {Camps}, \&
  {Lemson}}]{McAlpine2016}
{McAlpine}, S., {Helly}, J.~C., {Schaller}, M., {et~al.} 2016, Astronomy and
  Computing, 15, 72

\bibitem[{{Mo} {et~al.}(1998){Mo}, {Mao}, \& {White}}]{Mo1998}
{Mo}, H.~J., {Mao}, S., \& {White}, S.~D.~M. 1998, \mnras, 295, 319

\bibitem[{{Moustakas} {et~al.}(2013){Moustakas}, {Coil}, {Aird}, {Blanton},
  {Cool}, {Eisenstein}, {Mendez}, {Wong}, {Zhu}, \& {Arnouts}}]{Moustakas2013}
{Moustakas}, J., {Coil}, A.~L., {Aird}, J., {et~al.} 2013, \apj, 767, 50

\bibitem[{{Nelson} {et~al.}(2015){Nelson}, {Genel}, {Vogelsberger}, {Springel},
  {Sijacki}, {Torrey}, \& {Hernquist}}]{Nelson2015}
{Nelson}, D., {Genel}, S., {Vogelsberger}, M., {et~al.} 2015, \mnras, 448, 59

\bibitem[{{Nelson} {et~al.}(2013){Nelson}, {Vogelsberger}, {Genel}, {Sijacki},
  {Kere{\v s}}, {Springel}, \& {Hernquist}}]{Nelson2013}
{Nelson}, D., {Vogelsberger}, M., {Genel}, S., {et~al.} 2013, \mnras, 429, 3353

\bibitem[{{Nielsen} {et~al.}(2015){Nielsen}, {Churchill}, {Kacprzak}, {Murphy},
  \& {Evans}}]{Nielsen2015}
{Nielsen}, N.~M., {Churchill}, C.~W., {Kacprzak}, G.~G., {Murphy}, M.~T., \&
  {Evans}, J.~L. 2015, \apj, 812, 83

\bibitem[{{Oosterloo} {et~al.}(2007){Oosterloo}, {Fraternali}, \&
  {Sancisi}}]{Oosterloo2007}
{Oosterloo}, T., {Fraternali}, F., \& {Sancisi}, R. 2007, \aj, 134, 1019

\bibitem[{{Oppenheimer} \& {Schaye}(2013)}]{Oppenheimer2013}
{Oppenheimer}, B.~D., \& {Schaye}, J. 2013, \mnras, 434, 1043

\bibitem[{{Peng} {et~al.}(2010){Peng}, {Lilly}, {Kova{\v c}}, {Bolzonella},
  {Pozzetti}, {Renzini}, {Zamorani}, {Ilbert}, {Knobel}, {Iovino}, {Maier},
  {Cucciati}, {Tasca}, {Carollo}, {Silverman}, {Kampczyk}, {de Ravel},
  {Sanders}, {Scoville}, {Contini}, {Mainieri}, {Scodeggio}, {Kneib}, {Le
  F{\`e}vre}, {Bardelli}, {Bongiorno}, {Caputi}, {Coppa}, {de la Torre},
  {Franzetti}, {Garilli}, {Lamareille}, {Le Borgne}, {Le Brun}, {Mignoli},
  {Perez Montero}, {Pello}, {Ricciardelli}, {Tanaka}, {Tresse}, {Vergani},
  {Welikala}, {Zucca}, {Oesch}, {Abbas}, {Barnes}, {Bordoloi}, {Bottini},
  {Cappi}, {Cassata}, {Cimatti}, {Fumana}, {Hasinger}, {Koekemoer},
  {Leauthaud}, {Maccagni}, {Marinoni}, {McCracken}, {Memeo}, {Meneux}, {Nair},
  {Porciani}, {Presotto}, \& {Scaramella}}]{Peng2010}
{Peng}, Y.-j., {Lilly}, S.~J., {Kova{\v c}}, K., {et~al.} 2010, \apj, 721, 193

\bibitem[{{Pichon} {et~al.}(2011){Pichon}, {Pogosyan}, {Kimm}, {Slyz},
  {Devriendt}, \& {Dubois}}]{Pichon2011}
{Pichon}, C., {Pogosyan}, D., {Kimm}, T., {et~al.} 2011, \mnras, 418, 2493

\bibitem[{{Prochaska} {et~al.}(2017){Prochaska}, {Werk}, {Worseck}, {Tripp},
  {Tumlinson}, {Burchett}, {Fox}, {Fumagalli}, {Lehner}, {Peeples}, \&
  {Tejos}}]{Prochaska2017}
{Prochaska}, J.~X., {Werk}, J.~K., {Worseck}, G., {et~al.} 2017, \apj, 837, 169

\bibitem[{{Putman} {et~al.}(2012){Putman}, {Peek}, \& {Joung}}]{Putman2012}
{Putman}, M.~E., {Peek}, J.~E.~G., \& {Joung}, M.~R. 2012, \araa, 50, 491

\bibitem[{{Rahman} {et~al.}(2012){Rahman}, {Bolatto}, {Xue}, {Wong}, {Leroy},
  {Walter}, {Bigiel}, {Rosolowsky}, {Fisher}, {Vogel}, {Blitz}, {West}, \&
  {Ott}}]{Rahman2012}
{Rahman}, N., {Bolatto}, A.~D., {Xue}, R., {et~al.} 2012, \apj, 745, 183

\bibitem[{{Roberts}(1963)}]{Roberts1963}
{Roberts}, M.~S. 1963, \araa, 1, 149

\bibitem[{{Robertson} \& {Kravtsov}(2008)}]{Robertson2008}
{Robertson}, B.~E., \& {Kravtsov}, A.~V. 2008, \apj, 680, 1083

\bibitem[{{Ro{\v s}kar} {et~al.}(2010){Ro{\v s}kar}, {Debattista}, {Brooks},
  {Quinn}, {Brook}, {Governato}, {Dalcanton}, \& {Wadsley}}]{Roskar2010}
{Ro{\v s}kar}, R., {Debattista}, V.~P., {Brooks}, A.~M., {et~al.} 2010, \mnras,
  408, 783

\bibitem[{{Rubin} {et~al.}(2012){Rubin}, {Prochaska}, {Koo}, \&
  {Phillips}}]{Rubin2012}
{Rubin}, K.~H.~R., {Prochaska}, J.~X., {Koo}, D.~C., \& {Phillips}, A.~C. 2012,
  \apjl, 747, L26

\bibitem[{{Rudie} {et~al.}(2012){Rudie}, {Steidel}, {Trainor}, {Rakic},
  {Bogosavljevi{\'c}}, {Pettini}, {Reddy}, {Shapley}, {Erb}, \&
  {Law}}]{Rudie2012}
{Rudie}, G.~C., {Steidel}, C.~C., {Trainor}, R.~F., {et~al.} 2012, \apj, 750,
  67

\bibitem[{{Rupke} {et~al.}(2005){Rupke}, {Veilleux}, \& {Sanders}}]{Rupke2005}
{Rupke}, D.~S., {Veilleux}, S., \& {Sanders}, D.~B. 2005, \apjs, 160, 115

\bibitem[{{Saintonge} {et~al.}(2011){Saintonge}, {Kauffmann}, {Kramer},
  {Tacconi}, {Buchbender}, {Catinella}, {Fabello}, {Graci{\'a}-Carpio}, {Wang},
  {Cortese}, {Fu}, {Genzel}, {Giovanelli}, {Guo}, {Haynes}, {Heckman},
  {Krumholz}, {Lemonias}, {Li}, {Moran}, {Rodriguez-Fernandez}, {Schiminovich},
  {Schuster}, \& {Sievers}}]{Saintonge2011}
{Saintonge}, A., {Kauffmann}, G., {Kramer}, C., {et~al.} 2011, \mnras, 415, 32

\bibitem[{{Sancisi} \& {Allen}(1979)}]{Sancisi1979}
{Sancisi}, R., \& {Allen}, R.~J. 1979, \aap, 74, 73

\bibitem[{{Schaye} \& {Dalla Vecchia}(2008)}]{Schaye2008}
{Schaye}, J., \& {Dalla Vecchia}, C. 2008, \mnras, 383, 1210

\bibitem[{{Schaye} {et~al.}(2010){Schaye}, {Dalla Vecchia}, {Booth}, {Wiersma},
  {Theuns}, {Haas}, {Bertone}, {Duffy}, {McCarthy}, \& {van de
  Voort}}]{Schaye2010}
{Schaye}, J., {Dalla Vecchia}, C., {Booth}, C.~M., {et~al.} 2010, \mnras, 402,
  1536

\bibitem[{{Schaye} {et~al.}(2015){Schaye}, {Crain}, {Bower}, {Furlong},
  {Schaller}, {Theuns}, {Dalla Vecchia}, {Frenk}, {McCarthy}, {Helly},
  {Jenkins}, {Rosas-Guevara}, {White}, {Baes}, {Booth}, {Camps}, {Navarro},
  {Qu}, {Rahmati}, {Sawala}, {Thomas}, \& {Trayford}}]{Schaye2015}
{Schaye}, J., {Crain}, R.~A., {Bower}, R.~G., {et~al.} 2015, \mnras, 446, 521

\bibitem[{{Schiminovich} {et~al.}(2010){Schiminovich}, {Catinella},
  {Kauffmann}, {Fabello}, {Wang}, {Hummels}, {Lemonias}, {Moran}, {Wu},
  {Giovanelli}, {Haynes}, {Heckman}, {Basu-Zych}, {Blanton}, {Brinchmann},
  {Budav{\'a}ri}, {Gon{\c c}alves}, {Johnson}, {Kennicutt}, {Madore}, {Martin},
  {Rich}, {Tacconi}, {Thilker}, {Wild}, \& {Wyder}}]{Schiminovich2010}
{Schiminovich}, D., {Catinella}, B., {Kauffmann}, G., {et~al.} 2010, \mnras,
  408, 919

\bibitem[{{Schmidt}(1963)}]{Schmidt1963}
{Schmidt}, M. 1963, \apj, 137, 758

\bibitem[{{Sommer-Larsen}(1991)}]{SommerLarsen1991}
{Sommer-Larsen}, J. 1991, \mnras, 249, 368

\bibitem[{{Springel}(2005)}]{Springel2005}
{Springel}, V. 2005, \mnras, 364, 1105

\bibitem[{{Springel} {et~al.}(2001){Springel}, {White}, {Tormen}, \&
  {Kauffmann}}]{Springel2001}
{Springel}, V., {White}, S.~D.~M., {Tormen}, G., \& {Kauffmann}, G. 2001,
  \mnras, 328, 726

\bibitem[{{Steidel} {et~al.}(2002){Steidel}, {Kollmeier}, {Shapley},
  {Churchill}, {Dickinson}, \& {Pettini}}]{Steidel2002}
{Steidel}, C.~C., {Kollmeier}, J.~A., {Shapley}, A.~E., {et~al.} 2002, \apj,
  570, 526

\bibitem[{{Steidel} {et~al.}(2014){Steidel}, {Rudie}, {Strom}, {Pettini},
  {Reddy}, {Shapley}, {Trainor}, {Erb}, {Turner}, {Konidaris}, {Kulas}, {Mace},
  {Matthews}, \& {McLean}}]{Steidel2014}
{Steidel}, C.~C., {Rudie}, G.~C., {Strom}, A.~L., {et~al.} 2014, \apj, 795, 165

\bibitem[{{Stevens} {et~al.}(2017){Stevens}, {Lagos}, {Contreras}, {Croton},
  {Padilla}, {Schaller}, {Schaye}, \& {Theuns}}]{Stevens2017}
{Stevens}, A.~R.~H., {Lagos}, C.~d.~P., {Contreras}, S., {et~al.} 2017, \mnras,
  467, 2066

\bibitem[{{Stewart} {et~al.}(2013){Stewart}, {Brooks}, {Bullock}, {Maller},
  {Diemand}, {Wadsley}, \& {Moustakas}}]{Stewart2013}
{Stewart}, K.~R., {Brooks}, A.~M., {Bullock}, J.~S., {et~al.} 2013, \apj, 769,
  74

\bibitem[{{Stewart} {et~al.}(2011){Stewart}, {Kaufmann}, {Bullock}, {Barton},
  {Maller}, {Diemand}, \& {Wadsley}}]{Stewart2011ApJ}
{Stewart}, K.~R., {Kaufmann}, T., {Bullock}, J.~S., {et~al.} 2011, \apj, 738,
  39

\bibitem[{{Stewart} {et~al.}(2017){Stewart}, {Maller}, {O{\~n}orbe}, {Bullock},
  {Joung}, {Devriendt}, {Ceverino}, {Kere{\v s}}, {Hopkins}, \&
  {Faucher-Gigu{\`e}re}}]{Stewart2017}
{Stewart}, K.~R., {Maller}, A.~H., {O{\~n}orbe}, J., {et~al.} 2017, \apj, 843,
  47

\bibitem[{{Stocke} {et~al.}(2013){Stocke}, {Keeney}, {Danforth}, {Shull},
  {Froning}, {Green}, {Penton}, \& {Savage}}]{Stocke2013}
{Stocke}, J.~T., {Keeney}, B.~A., {Danforth}, C.~W., {et~al.} 2013, \apj, 763,
  148

\bibitem[{{Teklu} {et~al.}(2015){Teklu}, {Remus}, {Dolag}, {Beck}, {Burkert},
  {Schmidt}, {Schulze}, \& {Steinborn}}]{Teklu2015}
{Teklu}, A.~F., {Remus}, R.-S., {Dolag}, K., {et~al.} 2015, \apj, 812, 29

\bibitem[{{Trayford} {et~al.}(2016){Trayford}, {Theuns}, {Bower}, {Crain},
  {Lagos}, {Schaller}, \& {Schaye}}]{Trayford2016}
{Trayford}, J.~W., {Theuns}, T., {Bower}, R.~G., {et~al.} 2016, \mnras, 460,
  3925

\bibitem[{{Trayford} {et~al.}(2015){Trayford}, {Theuns}, {Bower}, {Schaye},
  {Furlong}, {Schaller}, {Frenk}, {Crain}, {Dalla Vecchia}, \&
  {McCarthy}}]{Trayford2015}
---. 2015, \mnras, 452, 2879

\bibitem[{{Tumlinson} {et~al.}(2017){Tumlinson}, {Peeples}, \&
  {Werk}}]{Tumlinson2017}
{Tumlinson}, J., {Peeples}, M.~S., \& {Werk}, J.~K. 2017, \araa, 55, 389

\bibitem[{{Turk} {et~al.}(2011){Turk}, {Smith}, {Oishi}, {Skory}, {Skillman},
  {Abel}, \& {Norman}}]{Turk2011}
{Turk}, M.~J., {Smith}, B.~D., {Oishi}, J.~S., {et~al.} 2011, The Astrophysical
  Journal Supplement Series, 192, 9

\bibitem[{{Turner} {et~al.}(2017){Turner}, {Schaye}, {Crain}, {Rudie},
  {Steidel}, {Strom}, \& {Theuns}}]{Turner2017}
{Turner}, M.~L., {Schaye}, J., {Crain}, R.~A., {et~al.} 2017, \mnras, 471, 690

\bibitem[{{Turner} {et~al.}(2014){Turner}, {Schaye}, {Steidel}, {Rudie}, \&
  {Strom}}]{Turner2014}
{Turner}, M.~L., {Schaye}, J., {Steidel}, C.~C., {Rudie}, G.~C., \& {Strom},
  A.~L. 2014, \mnras, 445, 794

\bibitem[{{van de Voort} {et~al.}(2011){van de Voort}, {Schaye}, {Booth},
  {Haas}, \& {Dalla Vecchia}}]{vandeVoort2011}
{van de Voort}, F., {Schaye}, J., {Booth}, C.~M., {Haas}, M.~R., \& {Dalla
  Vecchia}, C. 2011, \mnras, 414, 2458

\bibitem[{{van den Bergh}(1962)}]{vandenBergh1962}
{van den Bergh}, S. 1962, \aj, 67, 486

\bibitem[{{Werk} {et~al.}(2014){Werk}, {Prochaska}, {Tumlinson}, {Peeples},
  {Tripp}, {Fox}, {Lehner}, {Thom}, {O'Meara}, {Ford}, {Bordoloi}, {Katz},
  {Tejos}, {Oppenheimer}, {Dav{\'e}}, \& {Weinberg}}]{Werk2014}
{Werk}, J.~K., {Prochaska}, J.~X., {Tumlinson}, J., {et~al.} 2014, \apj, 792, 8

\bibitem[{{Worthey} {et~al.}(1996){Worthey}, {Dorman}, \&
  {Jones}}]{Worthey1996}
{Worthey}, G., {Dorman}, B., \& {Jones}, L.~A. 1996, \aj, 112, 948

\bibitem[{{Zheng} {et~al.}(2015){Zheng}, {Putman}, {Peek}, \&
  {Joung}}]{Zheng2015}
{Zheng}, Y., {Putman}, M.~E., {Peek}, J.~E.~G., \& {Joung}, M.~R. 2015, \apj,
  807, 103

\bibitem[{{Zschaechner} {et~al.}(2012){Zschaechner}, {Rand}, {Heald},
  {Gentile}, \& {J{\'o}zsa}}]{Zschaechner2012}
{Zschaechner}, L.~K., {Rand}, R.~J., {Heald}, G.~H., {Gentile}, G., \&
  {J{\'o}zsa}, G. 2012, \apj, 760, 37

\end{thebibliography}
